%% file: lhc_smqcd.tex
\documentclass[epj]{svjour}
\usepackage{graphics}
\usepackage{hepunits2007}
\usepackage{lhc_defs}
\usepackage{calc}
\usepackage{bm}
\usepackage{makeidx}
\usepackage{subfigure}
\usepackage{xspace}
\usepackage[noprefix]{nomencl}
\usepackage{marvosym}
\usepackage{wasysym}
\usepackage{delarray}
\usepackage{footmisc}
\usepackage{graphicx}
\usepackage[sectionbib,numbers,sort,compress]{natbib}
\usepackage{color}
\usepackage[hyperfootnotes=false]{hyperref}
\graphicspath{{.}}

\begin{document}

\title{Studies of Quantum Chromodynamics at the LHC}
\label{sec:smqcd}
\author{Tancredi Carli$^{\dagger}$, Klaus Rabbertz$^{\ddagger}$, and Steffen Schumann$^{\amalg}$}
\institute{%
$^{\dagger}$CERN, 1211~Geneva~23, Switzerland\\
$^{\ddagger}$Institut f\"ur Experimentelle Kernphysik, KIT, Wolfgang-Gaede-Str.~1, 76131~Karlsruhe, Germany\\
$^{\amalg}$II. Physikalisches Institut, Georg-August-Universit\"at G\"ottingen, Friedrich-Hund-Platz~1, 37077~G\"ottingen, Germany}

\abstract{A successful description of hadron-hadron collision data
  demands a profound understanding of quantum
  chromodynamics. Inevitably, the complexity of strong-interaction
  phenomena requires the use of a large variety of theoretical
  techniques---from perturbative cross-section calculations up to the
  modelling of exclusive hadronic final states. Together with the
  unprecedented precision of the data provided by the experiments in
  the first running period of the LHC, a solid foundation of
  hadron-hadron collision physics at the \TeV\ scale could be
  established that allowed the discovery of the Higgs boson and that
  is vital for estimating the background in searches for new
  phenomena. This chapter on studies of quantum chromodynamics at the
  LHC is part of a recent book on the results of LHC
  Run~1~~\cite{thisbook} and presents the advances in theoretical
  methods side-by-side with related key measurements in an integrated
  approach.}

\authorrunning{Carli, Rabbertz, Schumann}
\titlerunning{Studies of Quantum Chromodynamics at the LHC}

\maketitle
\tableofcontents
\blfootnote{Originally published in ``The Large Hadron Collider --- Harvest of Run~1'', edited by T.~Sch\"orner-Sadenius, Springer, 2015, pp. 139--194~\cite{thisbook}.}

\input{introduction}

\input{qcdbasics}

\input{pqcd}

\input{partonshowers}

\newpage
\input{nlo}

\input{nnlo}

\newpage
\input{nlops}

\input{Vplusb}

\input{resummation-fragmentation}

\input{beyond-pqcd}

\newpage
\input{summary}

\bibliography{lhc_smqcd}

\end{document}

%% file: introduction.tex
\section{Introduction}\label{sec:qcd:intro}

Quantum chromodynamics (QCD) is the well-established quantum field
theory of the strong interaction~\cite{Gross:1973id,Politzer:1973fx}
and one cornerstone of the Standard Model (SM) of particle physics.
Like the electromagnetic and the weak force, QCD belongs to the class
of gauge field theories. The assumption that the corresponding gauge
symmetry is an exact symmetry of nature results in truly massless
force carriers of the strong force, the
gluons.
However, as a consequence of the non-Abelian character
of the $\text{SU}(3)$
QCD gauge group, the gluons carry a strong or so-called ``colour''
charge
and interact amongst
themselves. This is a striking difference to the electromagnetic
force, which is mediated by electrically uncharged photons, and it
induces---amongst other effects---the confinement
of strongly interacting particles at low energies. In the following, the
colour-charged constituents of hadrons, i.e.\ quarks, antiquarks, and
gluons, will generically be denoted as ``partons''.

In hadronic collisions like at the LHC, QCD effects are omnipresent,
and their detailed understanding is indispensable for the
interpretation of collider data, whether to search for new phenomena
or to perform precision studies of model parameters. Despite
complications through the gluon self-interactions, the perturbative
expansion of QCD (pQCD)
that
describes interactions with large momentum exchanges (``hard
interactions'')
in terms of parton-parton
scatterings remains the most powerful theoretical technique. This
technique allows quantitative predictions at
parton level
for observables ranging from
inclusive production rates to shape observables, which are sensitive
to details of the QCD radiation pattern. Nevertheless the theory's
confining nature obliges us to address non-perturbative
aspects.
This includes a reliable understanding of the short-distance parton structure of the
initial-state protons, the fragmentation of final-state partons into
hadrons, or the modelling of soft proton interactions.

The unprecedented experimental precision achieved with the new
detectors at the LHC requires equally accurate theoretical predictions
and has sparked rapid progress in the field of perturbative
calculations using both analytical methods and modern Monte Carlo (MC)
event generators.
The interplay between experiment and
theory enforced the development of new observables and novel
techniques to match the challenges arising on both sides. In summary,
the improved theoretical understanding of the QCD dynamics including
the ability to precisely predict even complicated high-multiplicity
final states and the excellent performance and understanding of the LHC
machine and detectors together with sophisticated analysis techniques
reveal a more refined and detailed picture of QCD than ever before.

For the presentation of QCD-related experimental results obtained
during LHC Run~1 and the underlying theoretical developments, an
integrative approach is chosen---reflecting the productive and
fruitful interplay of the two communities. As a consequence,
compromises on the content had to be taken, and only a selection of
the most important measurements is discussed---omissions in the
presentation of theoretical methods were unavoidable.

The chapter at hands is structured as follows: After a brief reminder
of the basics of the QCD theory and the central aspects of
perturbative QCD, the discussion focuses on various approximations in
the modelling of scattering processes, including parton-shower
simulations and parton-level predictions at next-to-leading and
next-to-next-to-leading order in the strong coupling. Then the
discussion turns to the observation of multi-jet final states, the
successful description of which requires the combination of both
fixed-order calculations and all-order (i.e.\ parton-shower)
techniques. This is followed by a section on analytical methods for
the resummation of large logarithms; these are exemplified using
gap-fraction and jet-substructure observables. The chapter closes with
a presentation of various phenomena and measurements sensitive to
non-perturbative aspects of QCD\@.

%% file: qcdbasics.tex
\section{Basic Elements of QCD}
\label{sec:smqcd:qcdbasics}

The discussion shall begin with a brief reminder of the ingredients of
the QCD Lagrangian that defines the Feynman rules required for a
perturbative analysis of QCD\@.
The classical QCD Lagrangian
is composed out of the free Dirac Lagrangians for the six quark fields
and the kinetic and self-interaction terms for the gluon fields
$A^a_\mu$, labelled by a colour index $a=1,\dots,8$. These two parts
get minimally coupled through a gauge covariant derivative $D_\mu$:
  \begin{equation*} {\mathcal{L}}_{\text{QCD}} =
    {\mathcal{L}}_{\text{gauge}} +{\mathcal{L}}_{\text{quarks}}\, ,
  \end{equation*}
where
  \begin{equation*} {\mathcal{L}}_{\text{gauge}} = -\frac14
    F^a_{\mu\nu}F^{\mu\nu}_a\,,\quad\text{with}\quad
    F^a_{\mu\nu}=\partial_\mu A^a_\nu - \partial_\nu A^a_\mu -\gs
    f_{abc}A^b_\mu A^c_\nu
  \end{equation*}
the gluon field-strength tensor, and
%
  \begin{equation*} {\mathcal{L}}_{\rm quarks} = \sum\limits_{\qq\in
      \{\uq,\dq,\sq,\cq,\bq,\tq\} }\qbar\left(i\gamma^\mu
      D_\mu-m_\qq\right)\qq\,,\quad\text{with}\quad D_\mu=\partial_\mu
    + i\gs t^aA^a_\mu
  \end{equation*}
the QCD covariant derivative (see also the introduction to the SM Lagrangian in
Chap.~4 of Ref.~\cite{thisbook}).
Quark masses are denoted by $m_q$.
The $\text{SU}(3)$ generator matrices
introduced here obey the algebra
  \begin{equation*} [t^a,t^b] = if_{abc}t^c\, ,
  \end{equation*}
defining the QCD structure constants $f_{abc}$.
The classical QCD Lagrangian
exhibits the property of local gauge invariance,
 i.e.\ invariance under a
simultaneous redefinition of the quark and gluon fields.  As a
consequence of this internal symmetry, it is impossible to define the
gluon field propagator without explicitly specifying a choice of
gauge. A Lorentz-covariant way to fix the gauge is given by the class
of $R_\xi$ gauges, imposed by adding a term
  \begin{equation*} {\mathcal{L}}_{\text{gauge-fixing}} =
    -\frac{1}{2\xi}(\partial^\mu A^a_\mu)^2
  \end{equation*}
to the classical Lagrangian. Because of the non-Abelian character of
the QCD gauge group, the full Lagrangian of the quantum field theory
features a further contribution, the ghost Lagrangian
%
  \begin{equation*} {\mathcal{L}}_{\text{ghost}} =
    \partial_\mu\eta^{a\dagger}\left(D^\mu_{ab}\eta^b\right)\, ,
  \end{equation*}
that represents the field-dependent Faddeev--Popov determinant. The
ghost fields $\eta^a$ are represented by anti-commuting scalar fields.
This completes the Lagrangian for a consistent version of a quantum
field theory of the strong interaction. Accordingly one can read off
the QCD Feynman rules, featuring three-point quark-quark-gluon and
ghost-ghost-gluon interactions as well as triple and quartic gluon
self-interactions.
All of these interaction vertices are proportional to the strong
charge $\gs$.
This is also
the relevant parameter when applying the method of perturbation theory
to QCD\@. Defining the QCD counterpart of the QED fine-structure
constant
$\alpS=\gs\squared/4\pi$, one can expect a truncation of the
power-series expansion for a given observable ${\mathcal{O}}$, i.e.\
  \begin{equation*} {\mathcal{O}} = {\mathcal{O}}_0 +
    {\mathcal{O}}_1\alpS + {\mathcal{O}}_2\alpS^2+\dots\,,
  \end{equation*}
to yield meaningful estimates as long as $\alpS\ll 1$.

A prime example of a quantity evaluated in perturbation theory is the
QCD $\beta$ function.
It determines the running of the coupling constant $\alpS$ through the
renormalisation group equation
%
  \begin{equation}
    Q^2\frac{\partial\alpS}{\partial Q^2} =
    \beta(\alpS)\,,\quad\text{with}\quad
    \beta(\alpS)=-\alpS^2(b_0+b_1\alpS+b_2\alpS^2+\mathcal{O}(\alpS^3))\, ,
    \label{eq:qcd:rge}
  \end{equation}
and
\begin{equation}
  b_0= \frac{33-2\NF}{12\pi}\,,\;\; b_1=\frac{153-19\NF}{24\pi^2}\,,\;\; b_2=\frac{77139-15099\NF+325\NF^2}{3456\pi^3}\, .
\end{equation}

\begin{figure}[ht!]
  \begin{center}
    \includegraphics[width=0.8\textwidth]{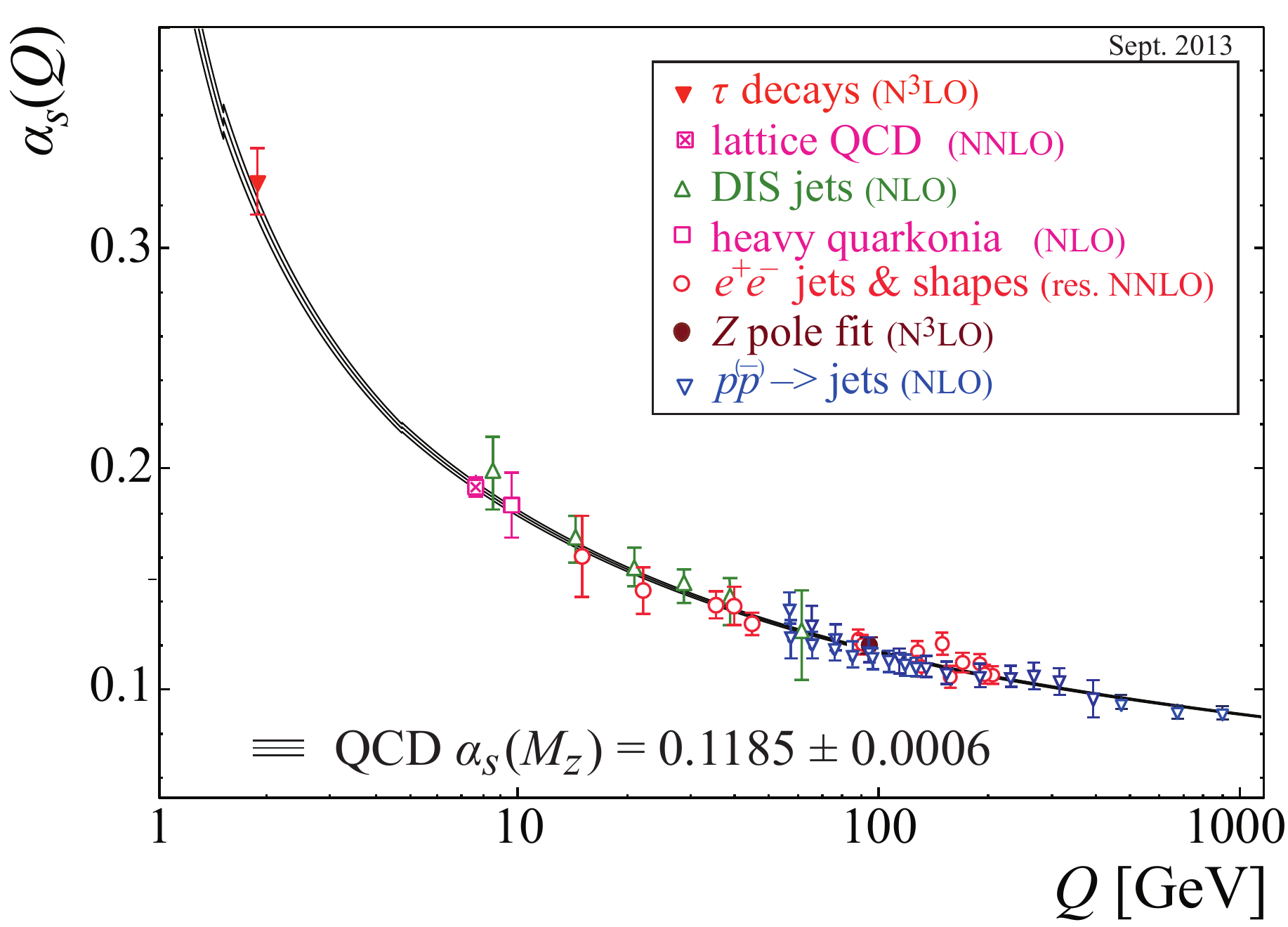}
  \caption{Summary of measurements of the strong coupling \alpS as a
    function of the respective energy scale $Q$. By including new LHC
    data, the range in $Q$ could be extended to the \TeV\ scale.
    \textit{(Adapted from Ref.~\cite{Agashe:2014kda}.)}
    \label{fig:qcd:smqcd_as}}
\end{center}
\end{figure}

$\NF$ denotes the number of quark flavours with masses $m_q$ smaller
than the scale $Q$. Note that the higher coefficients $b_2$ and
$b_3$ (see~Ref.~\cite{vanRitbergen:1997va}) are
renormalisation-scheme dependent. Here $b_2$ is quoted in the \MSbar
scheme. Retaining only the leading term $b_0$,
\eqn{\eqref{eq:qcd:rge}} is solved by
\begin{equation}
  \alpS(Q^2) = \frac{\alpS(\mu^2)}{1+b_0\ln\left(Q^2/\mu^2\right)\alpS(\mu^2)}\, ,
  \label{eq:qcd:runningas}
\end{equation}
which relates the strength of the coupling at a scale $Q$ to the one
at scale $\mu$, assuming both scales to be in the perturbative regime. The
non-Abelian nature of QCD
manifests itself in the negative sign of the $\beta$
function.
Thus, as long as $\NF< 17$, the coupling becomes weaker at
higher scales $Q$, or, in other words, the QCD colour charge decreases
when the distance decreases. For high scales $Q$, QCD becomes almost a
free theory---a property known as ``asymptotic
freedom''.
It is this weakly coupled regime where perturbative methods can successfully
be applied and quantitative predictions for hard scattering processes
can be made. The world average value of the strong
coupling
as of 2014, quoted at the scale of the
$\Zb$-boson mass $\MZ$, is given by
  \begin{equation*}
    \alpS(\MZ) = 0.1185 \pm 0.0006\, ,
  \end{equation*}
derived from hadronic $\tau$-lepton decays, lattice QCD calculations,
deep-inelastic scattering data, electron-positron annihilation
processes, and electroweak precision fits~\cite{Agashe:2014kda}.
Figure~\ref{fig:qcd:smqcd_as} shows a summary of measurements of the
strong coupling at energy scales ranging from the mass of the
$\tau$-lepton of $M_\tau \approx \unit{1.8}{\GeV}$ up to the \TeV\
scale thanks to newly included LHC data.
The historical development of \alpS determinations is discussed in
Chap.~12, Fig.~12.2, of Ref.~\cite{thisbook}.

The dynamical behaviour of $\alpS(Q^2)$ implies an increase of the QCD
coupling at small momentum transfer, i.e.\ large distances.
When the coupling approaches unity, perturbation theory is not valid
anymore. The parameter \LambdaQCD is defined as the scale, where
$\alpS(Q^2)$ formally diverges. With this definition,
\eqn{\eqref{eq:qcd:runningas}} can be rewritten as $\alpS(Q^2) =
\left(b_0\ln\left(Q^2/\LambdaQCD^2\right)\right)^{-1}$. For $\NF=5$
flavours in the \MSbar scheme, \LambdaQCD roughly amounts to
$\unit{214}{\MeV}$ and represents the dividing line below which one is
in the manifestly non-perturbative regime of QCD\@.
It is the growth of the coupling at small scales that makes QCD a
theory of the strong interaction---the fundamental force that confines
the quarks and gluons into ordinary hadronic matter, e.g.\ the protons
and neutrons. For the purpose of LHC physics, one has to account for
this phenomenon of ``confinement''
when modelling the transition from free quarks and gluons to the bound-state hadrons
observed in the detectors. Lacking a first-principles understanding of
this process, one mostly has to rely on MC models for this aspect.
Even more
fundamentally, the partonic content of the colliding protons needs to
be parametrised in order to allow for a description of LHC collision
events through partonic scattering processes.

%% file: pqcd.tex
\section{Perturbative QCD}\label{sec:qcd:pqcd}

Quantitative predictions based on the non-Abelian QCD Lagrangian
can be obtained either with lattice methods employing a
discretised space-time or using perturbation theory.
Given the complexity of the final states produced in proton-proton collisions
with high momentum transfer, lattice techniques are of no practical
importance for the prediction of LHC events at present. Instead one
has to rely on perturbative methods, which are possibly supplemented by
models for the transition of partons to hadrons.

\subsection{Cross-Section Predictions}

In perturbative QCD, the cross section for a hard scattering
process
at a hadron-hadron collider can be written in the following factorised form
  \begin{eqnarray}
    \sigma_{H_1H_2\to X} &=& \sum\limits_{i,j}\int \dif{x_1} \dif{x_2}\; f_{i/{H_1}}(x_1,\mufs) f_{j/{H_2}}(x_2,\mufs)\nonumber\\
    &&\times \hat{\sigma}_{ij\to X}\left(x_1P_1,x_2P_2,\alpS(\murs),\frac{Q}{\mufs}\right)\, ,\label{eq:qcd:QCDxsec}
  \end{eqnarray}
where the sum extends over all contributing initial-state partons
$i,j\in\{\qq,\qbar,\glue\}$. It is assumed here that the scale
associated with the hard process, $Q$, is much larger than \LambdaQCD,
the delimiting scale for the applicability of perturbative methods to
QCD\@.  In this high-energy limit, effects related to the binding of
the partons in the initial-state protons can be neglected.  As a
consequence, the cross section simply
factorises
into a product of
parton distribution functions (PDFs) $f_{i/{H}}(x,\mufs)$ of non-perturbative
origin and the perturbatively calculable partonic cross section.
The parton distribution functions model the probability to find a parton
of flavour $i$ in the incoming hadron $H$ (protons at the LHC) with a
fraction $x$ of the hadron's momentum $P$. The resulting squared
partonic centre-of-mass energy is given by $\hat{s}=x_1x_2s$, with
$s=(P_1+P_2)^2$ the squared hadronic centre-of-mass energy.

The collinear factorisation
ansatz underlying
\eqn{\eqref{eq:qcd:QCDxsec}} is the key to quantitative predictions in
the framework of QCD that can be compared to actual LHC collision
data. Based on the property of asymptotic freedom of QCD, the desired
cross section can be expanded as a power series of the coupling
constant \alpS. The lowest-order coefficient is denoted as leading
order (LO), the subsequent ones as next-to-leading order (NLO) and
next-to-next-to-leading order (NNLO), respectively.

When calculating the LO, NLO or NNLO estimate for a partonic cross
section, the appropriate QCD evolution of the parton distribution
functions has to be used.

Leading-order cross-section calculations are fully automated by means
of tree-level matrix-element generator programs
such as
\ALPGEN~\cite{Mangano:2002ea}, \AMEGIC~\cite{Krauss:2001iv},
\COMIX~\cite{Gleisberg:2008fv},
\HELAC/\PHEGAS~\cite{Cafarella:2007pc}, \MADGRAPH~\cite{Alwall:2011uj}
or \WHIZARD~\cite{Kilian:2007gr}. These codes are capable of providing
integrated cross sections and parton-level events for almost arbitrary
Standard Model final states, with multiplicities ranging up to ten
particles. In particular for high-multiplicity final states,
implementations relying on recursive algorithms for the generation of
the expressions for the amplitudes, e.g.\ Berends--Giele
recursion~\cite{Berends:1987me}, prove most
efficient~\cite{Duhr:2006iq,Dinsdale:2006sq}.

Over the past years there has been enormous progress in the evaluation
of processes at NLO and NNLO in the strong coupling. These
developments and related precision measurements will be addressed in
detail in later sections of this chapter.

\subsection{Fragmentation and Hadronic Jets}


One entity from \eqn{\eqref{eq:qcd:QCDxsec}} that has not yet been
discussed is the final state $X$ of a collision. The simplest reaction
that can be considered is the Drell--Yan process~\cite{Drell:1970yt},
where a quark and an antiquark annihilate to produce a lepton pair:
$\hat{\sigma}\left(\qqbar\to \lplm \right)$. In this case there are no
strongly interacting particles in the final state, and the theory
prediction can directly be compared to the measured leptons. Merely
the proton remnants, which fragment into hadrons along the beam lines,
have to be described by non-perturbative models in MC event
generators. At high transverse momenta, the two leptons are well
separated from any such proton debris and high-precision comparisons
with theory even at NNLO become possible. This is discussed in more
detail later in this chapter.

However, in the vast majority of reactions at least some
colour-charged partons are produced so that a further step covering
the transition from the partonic final state to measurable particles,
the so-called ``particle level'',
is needed. Here, ``measurable'' refers to colour-neutral particles with
mean decay lengths such that $c\tau>\unit{10}{mm}$, where $c$ is the
speed of light and $\tau$ the lifetime of a particle. One possibility
to account for this transition is to reuse the concept underlying the
PDFs that describe the partonic content of a hadron, only in an
inverted sense. The necessary functions $D_{k\to h}(z,\mufs)$ are
called fragmentation functions (FFs) and are the final-state analogues
of the PDFs\@. They parametrise the probability of finding a hadron
$h$ within the fragmentation products of parton $k$, carrying the
fraction $z$ of the parton momentum. Like the
PDFs, fragmentation functions depend on a non-physical resolution or
fragmentation scale \mufs. Again, these functions can currently not be
determined by first principles in QCD, but once they have been
measured (for example under the experimentally more favourable
conditions of \epem collisions at the LEP collider), they are
universally valid.

\begin{figure}[tb]
  \begin{center}
    \includegraphics[width=0.6\textwidth]{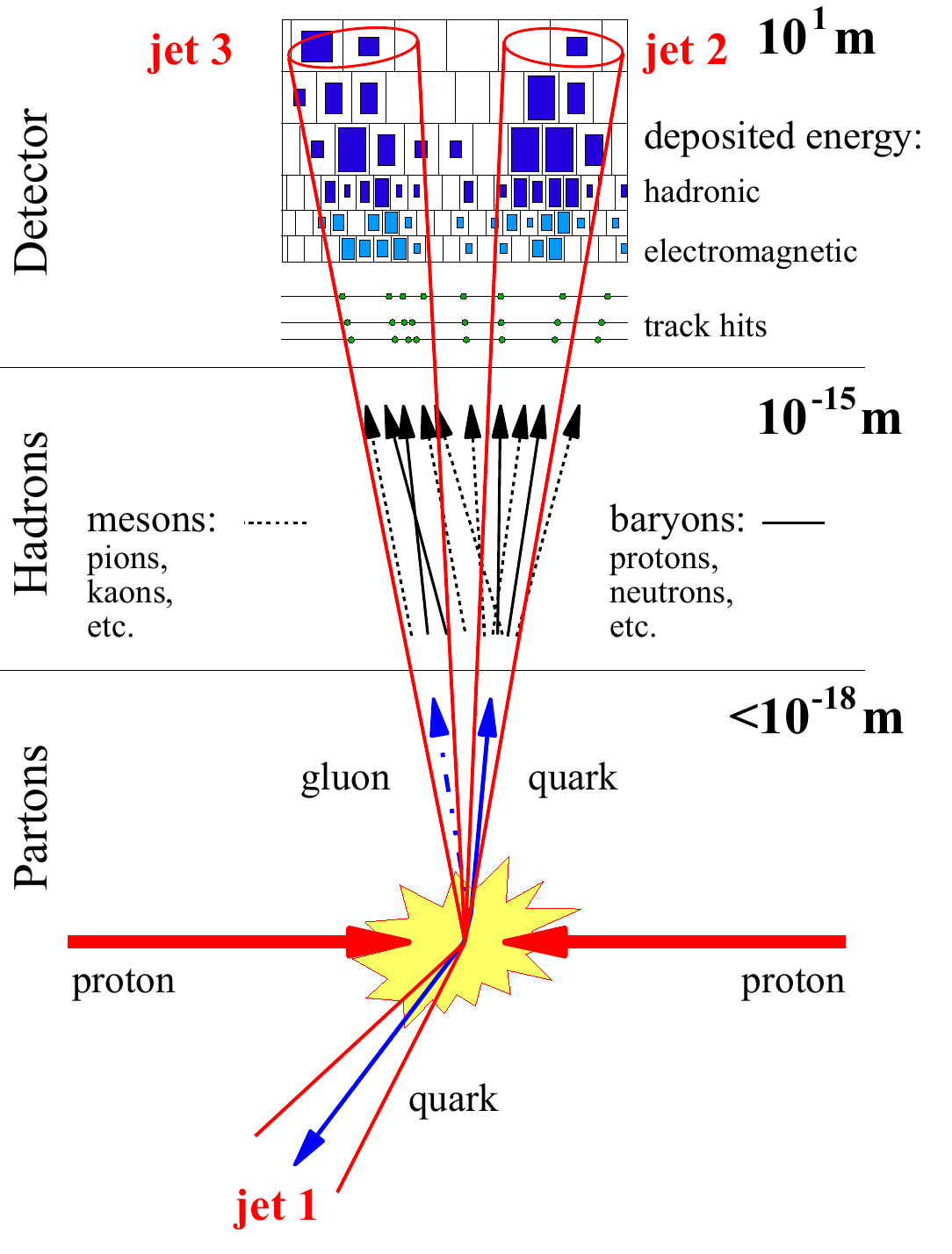}
  \end{center}
  \caption{Illustration of a jet to which bundles of partons, hadrons,
    or detector measurements are grouped
    together.
    \label{fig:qcd:jetlevels}
  }
\end{figure}

A second possibility to account for the transition to measurable particles
makes use of the concept of hadronic
jets. Instead of looking into the detailed production of identified
particles---an experimentally very challenging endeavour---for the
majority of processes it is sufficient to know how much energy and
momentum is carried away by hadrons. QCD predicts that large-distance
non-perturbative (NP) effects are mostly decoupled from the hard
reaction so that highly energetic partons fragment into a collimated
stream or ``jet'' of hadrons, which inherits energy and momentum from
its parent parton. To define what ``collimated'' means, a prescription
is required that, given some distance measure, unambiguously decides
which objects belong to a jet. As one wants to compare predictions by
pQCD with measured particles, tracks, or energy depositions as
illustrated in \fig{\ref{fig:qcd:jetlevels}}, a jet algorithm is
needed that is applicable to theoretical calculations as well as to
measurements from different experiments.

Most importantly, to deal with the cancellation of collinear and soft
singularities appearing in pQCD, a jet algorithm must be collinear-
and infrared-safe. This means that the outcome of a jet-clustering
procedure depends neither on the splitting or merging of collinear
parton four-vectors nor on the addition of arbitrarily soft partons to
the list of objects to be clustered. The first description of a
collinear- and infrared-safe jet algorithm, which grouped partons or
particles together that are inside an angular cone around a specific
direction, was given by G.~Sterman and S.~Weinberg in
1977~\cite{Sterman:1977wj}.

The most important requirements for jet algorithms from the experimental side are i)
independence of detector details, ii) maximal reconstruction
efficiency, iii) minimal resolution smearing, iv) computational
efficiency, and v) ease of calibration. The extension of the original
jet definition, which was specialised to 2-jet events in \epem
collisions, to all kind of reactions and the partially conflicting
requirements lead to various new propositions that were tried and
discussed in several
workshops~\cite{Huth:238477,Blazey:2000qt,Buttar:2008jx}. Two classes
of jet algorithms emerged:
\begin{enumerate}
\item cone algorithms that geometrically assign objects to the leading
  energy-flow objects in an event;
\item sequential-recombination algorithms that iteratively combine the
  closest pairs of objects~\cite{Bartel:1986ua}.
\end{enumerate}

Many cone algorithms need starting points with a minimum energy or
momentum for the cone directions---so-called ``seeds''. These spoil
the condition of collinear safety. A seedless infrared-safe cone
algorithm exists in the form of the \SISCONE
algorithm~\cite{Salam:2007xv}, which avoids this problem. For reasons
of computational efficiency, though, the method of choice employed at
the LHC is the anti-\ktalgo clustering algorithm~\cite{Cacciari:2008gp} as
implemented in the \FASTJET package~\cite{Cacciari:2011ma}. For the
jet size parameter $R$ (the equivalent to the cone size in cone
algorithms) the LHC collaborations chose the values 0.4 and 0.6 for ATLAS
respectively 0.5 and 0.7 for CMS\@.  An extensive overview of jet
definitions in QCD and their history is presented in
Ref.~\cite{Salam:2009jx}.

\begin{figure}[tb]
  \begin{center}
    \includegraphics[width=0.95\textwidth]{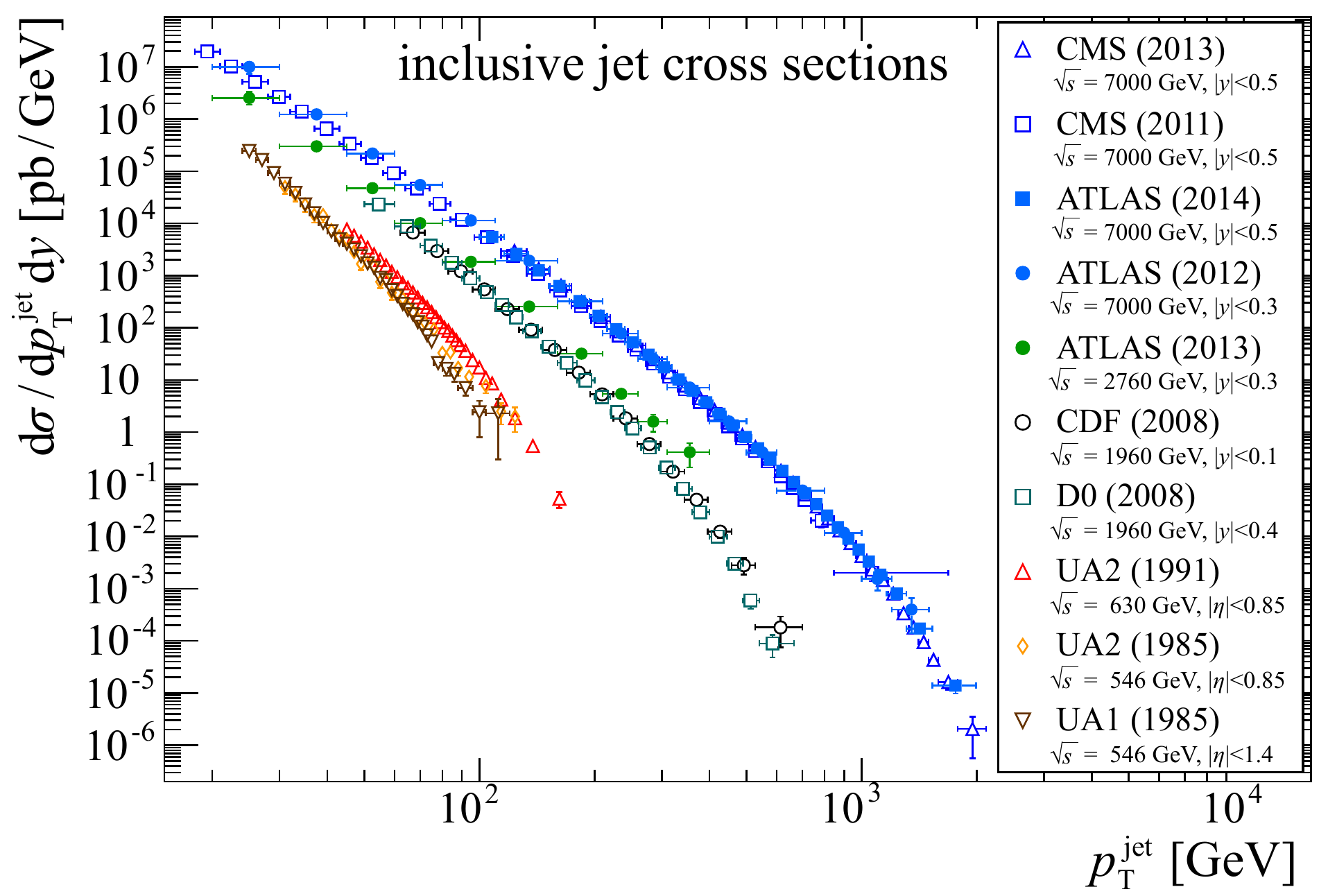}
    \caption{Summary of inclusive jet cross-section measurements at
      \pp or \ppbar colliders.
Data taken from Refs~.\cite{
Arnison:1985rv,Appel:1985rm,Alitti:1990aa,Abazov:2008ae,Aaltonen:2008eq,
Aad:2011fc,Chatrchyan:2011ab,Chatrchyan:2012bja,Aad:2014vwa,
Chatrchyan:2014gia,Aad:2013lpa}.
    \label{fig:qcd:jet_history}}
  \end{center}
\end{figure}

The most fundamental
quantity of jet production that can be investigated is the inclusive
jet cross section as a function of the jet \pT and rapidity $y$,
$\dif^2\sigma/\dif\pTjet{}\dif y$, where every jet of an event
contributes. Figure~\ref{fig:qcd:jet_history} presents a summary of
inclusive jet cross section measurements, performed with various jet
algorithms, from \pp or \ppbar collisions at centre-of-mass energies
from \unit{546}{\GeV} up to \unit{7}{\TeV} at central rapidity. The
measured cross sections stretch over eleven orders of magnitude and
range up to \unit{200}{\GeV} at \sps, \unit{600}{\GeV} at the
\Tevatron, and \unit{2}{\TeV} at the LHC.

LO predictions (not shown) correctly describe the steep drop with
increasing \pTjet, but fail to predict the absolute values more
precisely than in an order-of-magnitude estimation. More accurate
calculations at NLO are able to improve this situation.


%% file: partonshowers.tex
\section{Parton Showers: The Bulk of the Emissions}
\label{sec:qcd:ps}

The strength of the collinear factorisation
ansatz allowing us to
define universal PDFs and FFs lies in its repeated applicability.
It enables us to define evolution equations such as the DGLAP
equations~\cite{Gribov:1972ri,Lipatov:1974qm,Dokshitzer:1977sg,Altarelli:1977zs},
which determine the variation of the PDFs and FFs when the
factorisation scale $\mufs$ is changed. For the case of the initial
state PDFs one derives
\begin{equation}
  \mufs^2\frac{\partial f_i(x,\mufs)}{\partial \mufs^2} = \sum\limits_{j=\{\qq,\qbar,\glue\}}\int\limits^1_x\frac{dz}{z}\frac{\alpS}{2\pi} P_{ij}(z)f_{j/H}(x/z,\mufs)\, ,
  \label{eqn:qcd:pdfevolution}
\end{equation}
with $P_{ij}$ the (plus-prescription) regularised Altarelli--Parisi
splitting functions
%
\begin{eqnarray}
  P_{gq}(z) &=& \CF\left(\frac{1+(1-z)^2}{z}\right)\,,\, P_{qq}(z) = \CF\left(\frac{1+z^2}{(1-z)_+}+\frac32\delta(1-z)\right)\,,\nonumber\\
  P_{qg}(z) &=& \TR \left(z^2+(1-z^2)\right)\,,\\
  P_{gg}(z) &=& 2\CA\left(\frac{z}{(1-z)_+}+\frac{1-z}{z}+z(1-z)\right) +\delta(1-z)\frac{11\CA-4\NF \TR}{6}\,. \nonumber
\end{eqnarray}
Furthermore, it should be noted that $P_{\glue\qbar}=P_{\glue\qq}$ and
$P_{\qbar\qbar}=P_{\qq\qq}$. While the LO, i.e.\ one-loop,
approximation is quoted here, the QCD splitting functions are
known up-to NNLO, i.e.\ three-loop accuracy~\cite{Moch:2004pa,Vogt:2004mw}.
The LO DGLAP evolution allows for a interpretation by means of
simple branching processes. A parton $i$ resolved at scale $\mufs$ may
have originated from a branching of parton $j$ resolved at some higher
scale. This transition of parton $j$ to $i$ is accompanied by the
emission of an additional QCD parton. When applying the DGLAP
equations to solve for the scale evolution of PDFs or FFs,
these emitted particles get ignored, by considering inclusive
processes only.

However, in parton-shower MC programs these emissions are
made explicit, and the subsequent branching of initial-state and final-state
partons results in a cascade-like picture, modelling the
initial- and final-state evolution of the hard process' parton
configuration into an ensemble of many QCD partons.
In essence these parton-shower emissions approximate higher-order
real-emission corrections to the hard scattering process. Shower
simulations form an integral part of MC event generators---simulation
tools to describe the fully exclusive hadronic final states
of individual scattering events~\cite{Buckley:2011ms}.
They link the
hard scattering at some high scale $Q^2$ to an exclusive set of
partons with typical separation scales of order
$Q^2_0\approx\unit{1}{\GeV\squared}$, the cut-off scale of the shower
evolution. Through this parton-shower link it is possible to invoke
universal models for the hadronisation of partons into hadrons,
independent of the hard process scale $Q^2$. Physically, parton
showers account for the intra-jet evolution of jets by accounting for
the emission of almost collinear partons, and they model the inter-jet QCD
activity through the emission of soft, wide-angle partons.

\subsection{Colour Coherence}

In particular the correct incorporation of soft-gluon emissions needs
special consideration. In contrast to collinear emissions, which
factorise at the cross-section level, soft-gluon emissions factorise
at the level of individual QCD amplitudes. One should therefore consider
soft gluons to be emitted by the scattering process as a whole, given
by the squared sum of all contributing amplitudes.  At first glance
this approach seems to spoil the parton-shower picture of independent
subsequent emissions. However, soft colour-coherence effects
originating from the interference of individual amplitudes can be
incorporated into parton-shower simulations by suitable choices for
the shower-evolution variable. Most prominently this can be achieved
by using the emission's opening angle as ordering parameter, as
employed in the \HERWIG and \HERWIGPP
generators~\cite{Marchesini:1989yk,Gieseke:2003rz}.

However, angular ordering is not the only option to include colour-coherence
effects. A lot of effort went into the construction of new parton-shower
algorithms based on subtraction formalisms as they are used in NLO QCD
calculations. These new shower schemes, either based on
Catani--Seymour dipole
factorisation~\cite{Schumann:2007mg,Dinsdale:2007mf,Platzer:2009jq} or
antenna subtraction~\cite{Winter:2007ye,Ritzmann:2012ca},
implement soft-gluon coherence through a partial fractioning of the QCD radiator
antennas, thereby introducing the notion of an emitter, i.e.\ splitter,
and an associated spectator parton that accompanies the splitting
process.

\begin{figure}[t!]
  \begin{center}
    \includegraphics[width=0.95\textwidth]{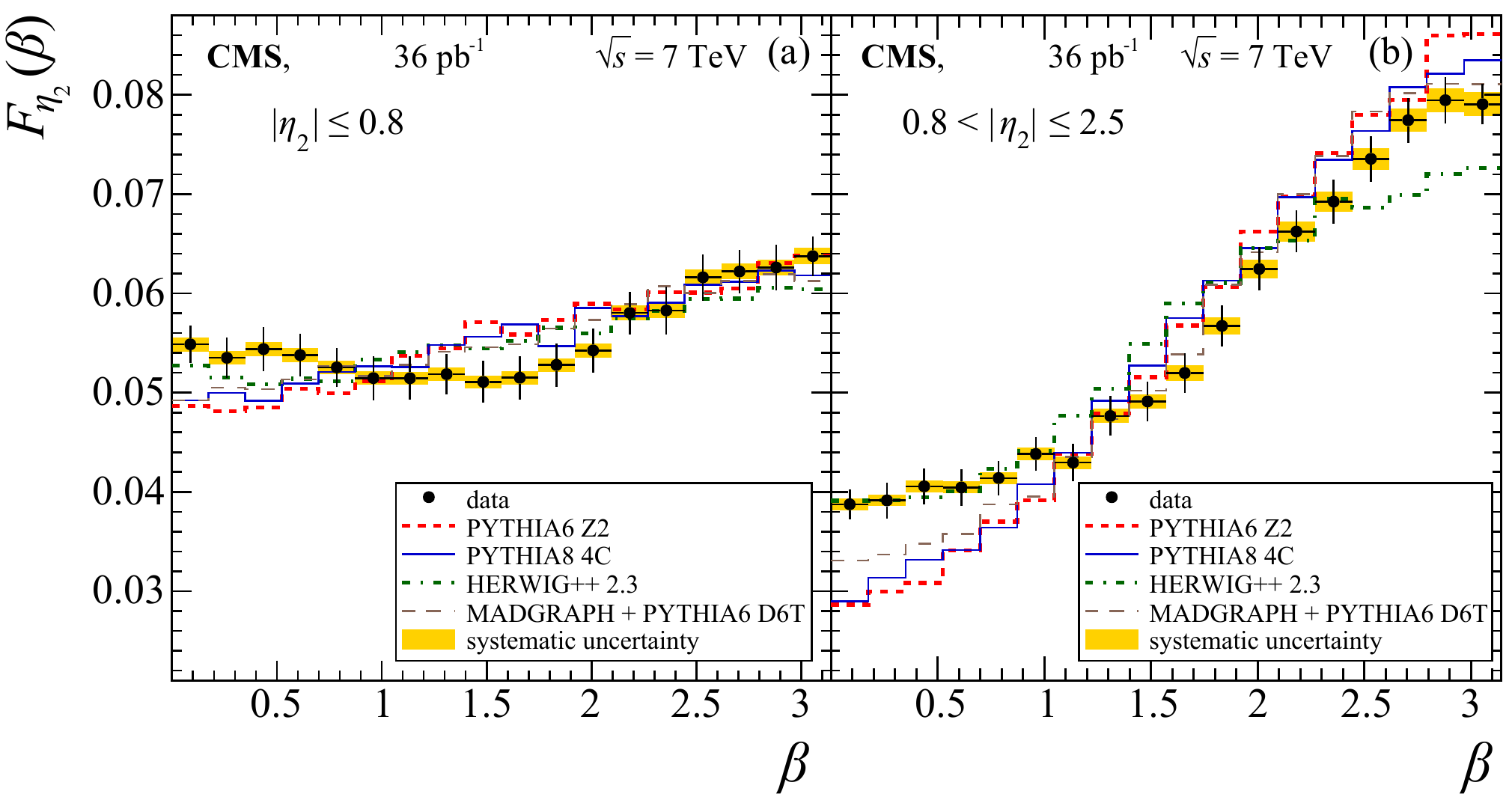}
  \end{center}
  \caption{Measured normalised $\beta$ distribution, corrected for
    detector effects, in comparison to various MC predictions. The
    error bars indicate the statistical uncertainties, while the
    shaded bands correspond to the total systematic
    uncertainty.
    \textit{(Adapted from Ref.~\cite{Chatrchyan:2013fha}.)}
  \label{fig:qcd:colourcoherence}
    }
\end{figure}

In order to study QCD coherence effects one needs to devise
observables sensitive to rather soft emissions. This can be achieved
by selecting final states with at least three jets exhibiting a
sizeable spread in transverse momentum between the hardest and the
softest jet. One such analysis was presented by the CMS collaboration
in Ref.~\cite{Chatrchyan:2013fha}, based on a data set with
an integrated luminosity of $\unit{36}{\invpb}$ collected in 2010. The
analysis inspects events with at least three anti-\ktalgo jets using a
distance parameter of $R=0.5$, ordered in transverse momentum such
that $\pTone>\pTtwo>\pTthree$. The event selection criteria are given
by
\begin{eqnarray*}
  &&\pTone > \unit{100}{\GeV}\,,\,\, \pTthree>\unit{30}{\GeV}\,,\,\,|\eta_{1,2}| < 2.5\\
  && M_{12}>\unit{220}{\GeV}\,,\,\,0.5<\Delta R_{23}<1.5\, .
\end{eqnarray*}
The observable considered to probe colour-coherence effects, called
$\beta$, is defined as the azimuthal angle of the third jet with
respect to the second jet, i.e.\
\begin{equation*}
  \tan\beta = \frac{|\Delta \phi_{23}|}{\Delta \eta_{23}}\, .
\end{equation*}
In the presence of QCD coherence, the emission of the parton initiating the
third jet is expected to preferentially lie in the event plane
defined by the emitting parton and the beam
axis. Figure~\ref{fig:qcd:colourcoherence} shows the normalised $\beta$
distributions measured for two regions of the second-jet
pseudo-rapidity, i.e.\ central $|\eta_2|\leq 0.8$ and forward
$0.8<|\eta_2|\leq 2.5$. The observable is thereby defined as
$F_{\eta_2,i}(\beta)=N_{\eta,i}/N_\eta$, with $N_\eta$ the total number of
events in the respective $\eta_2$ region and $N_{\eta,i}$ the number of
events in the $i$th $\beta$ bin. The data are compared to various particle-level
MC predictions, i.e.\ specific tunes of \PYTHIASIX,
\PYTHIAEIGHT~\cite{Sjostrand:2007gs} and \HERWIGPP, all based on LO
$2\to2$ matrix elements plus parton showers, and a combination of
\MADGRAPH using exact $2\to 2$ and $2\to 3$ tree-level matrix elements
supplemented by \PYTHIASIX parton showers. It has been shown in
Ref.~\cite{Chatrchyan:2013fha} that the data clearly support the
inclusion of coherence effects in the simulation; the
regions of small $\beta$ and $\beta\approx \pi/2$ are particularly
sensitive. However, none of the generators used in the analysis
describes the data satisfactorily over the entire phase space.
\HERWIGPP models the data best but the agreement is rather poor in the
forward region around $\beta \approx \pi$. The inclusion of the exact
$2\to 3$ matrix element from \MADGRAPH slightly improves the pure
\PYTHIASIX parton-shower prediction.

\subsection{Azimuthal Decorrelation}

Clearly, the inclusion of parton showers is indispensable when
attempting to simulate event configurations sensitive to the emission
of multiple partons. Exemplary quantities are event
shapes~\cite{Khachatryan:2011dx, Aad:2012np, Khachatryan:2014ika}, or
sub- or intra-jet-related observables~\cite{Aad:2012meb, ATLAS:2012am,
  Aad:2013ueu, Aad:2014haa}.
  Here, the measurement of dijet azimuthal
decorrelations
as performed by  ATLAS and CMS~\cite{daCosta:2011ni,Khachatryan:2011zj} shall briefly
be discussed. The analyses evaluate $\unit{36}{\invpb}$ and
$\unit{2.9}{\invpb}$ of \pp collision data at
$\sqrt{s}=\unit{7}{\TeV}$ by ATLAS and CMS, respectively. The observable
considered is the azimuthal angle $\Delta\phi$ between the two jets
leading in jet \pT. For events with exactly two high-\pT jets and
nothing else, the azimuthal angle between the two jets is fully correlated
through momentum conservation in the transverse plane and
$\Delta\phi \approx \pi$. Multi-jet production disturbs this balance
so that the two leading jets become decorrelated in azimuthal angle
and $\Delta \phi < \pi$. Hence, the observable $\Delta\phi$ is
indicative of multi-jet production, while at the same time only the
azimuthal angles of the two leading jets are measured, avoiding the
large uncertainties associated to the jet energy calibration.

\begin{figure}[t!]
  \begin{center}
    \includegraphics[width=1.0\textwidth]{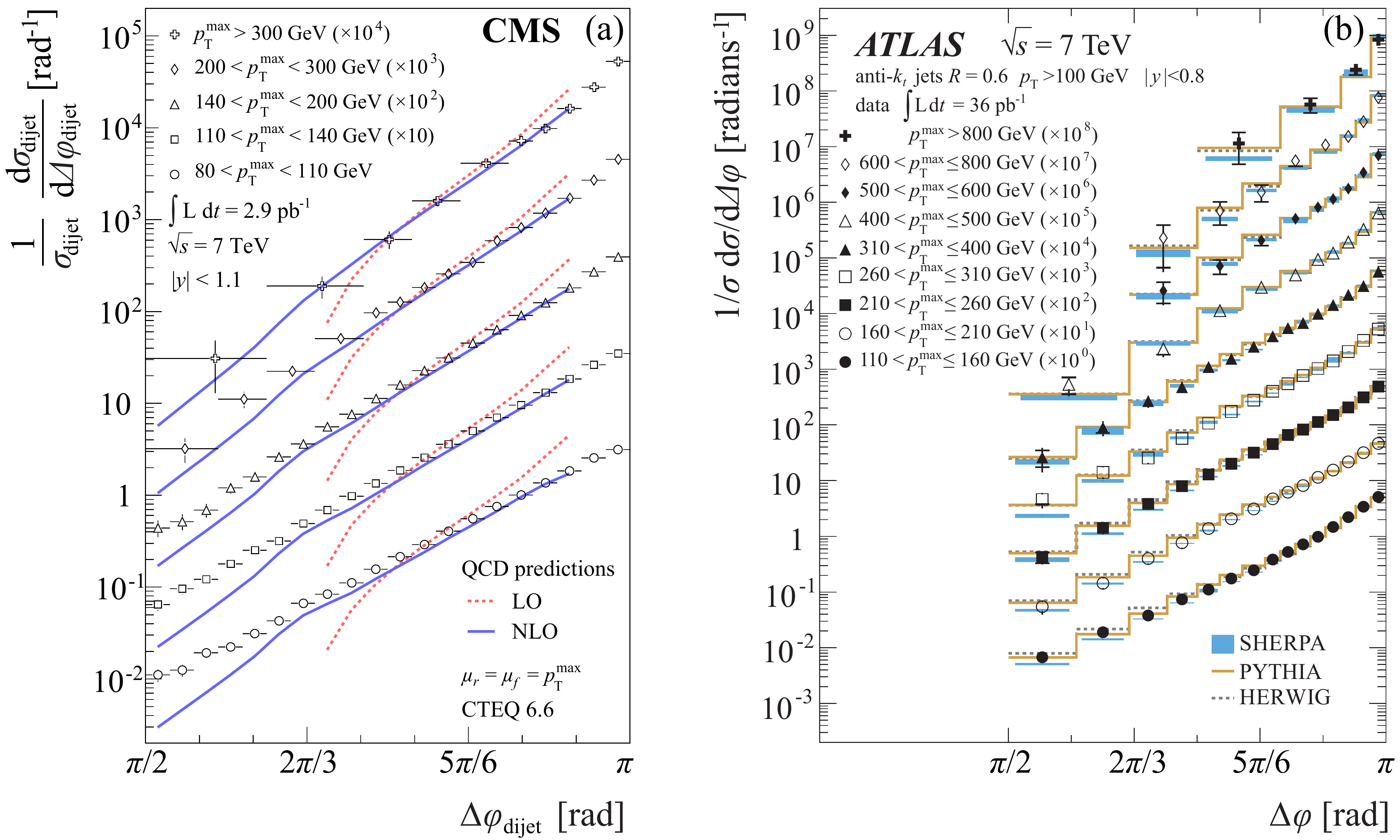}%
  \caption{Normalised differential cross section
    $(1/\sigma)(\dif\sigma/\dif\Delta\phi)$ for multiple slices in leading-jet \pT:
    (a) CMS data compared to fixed-order predictions in pQCD;
    (b) ATLAS data compared to predictions by the MC event generators
    \PYTHIASIX, \HERWIG, and \SHERPA.
    \textit{(Adapted from
    auxiliary material provided with Refs.~\cite{Khachatryan:2011zj, daCosta:2011ni}.)}
  \label{fig:qcd:dphi}
    }
  \end{center}
\end{figure}

With $\Delta\phi$ it is possible to probe complementary aspects of
perturbative QCD\@. The region $\Delta\phi \approx \pi$ is sensitive
to multiple soft and collinear emissions and thus requires a proper
resummation of soft and collinear logarithms. The region of large
azimuthal decorrelations, i.e.\ $\Delta\phi \ll \pi$, indirectly
probes the production mechanism for additional hard jets and requires
the inclusion of higher-order matrix
elements. Figure~\ref{fig:qcd:dphi}(a) shows a comparison of the
normalised distribution of $\Delta\phi$ for five slices in leading-jet
\pT as measured by CMS~\cite{Khachatryan:2011zj} to perturbative
predictions for three-parton production at LO and NLO, respectively,
in the region $2\pi/3 < \Delta\phi < \pi$. The leading order has predictive power
only in a very narrow range of $\Delta\phi$. The NLO calculation improves
this substantially, but fails when approaching $\Delta\phi \approx \pi$, and
below $2\pi/3$ matrix elements for $4$-jet production are required.

In \fig{\ref{fig:qcd:dphi}}(b) the normalised $\Delta\phi$
distributions as measured by ATLAS~\cite{daCosta:2011ni} are compared
for nine slices in leading-jet \pT to predictions from the MC event
generators \PYTHIASIX, \HERWIG, and \SHERPA. \PYTHIASIX and \HERWIG are
based on LO $2 \to 2$ QCD matrix elements matched with parton
showers. \SHERPA additionally includes tree-level matrix elements for
$2 \to$ 3--6 jet production properly matched to its parton-shower
algorithm~\cite{Schumann:2007mg,Hoeche:2009rj}.  All three generators
describe the data well over the measured range of $\pi/2 < \Delta\phi
< \pi$ and all
$\pTmax$ slices. This includes both the region $\pi/2 < \Delta\phi <
2\pi/3$ where multi-jet contributions are significant and
$\Delta\phi\approx \pi$ that cannot be described by a
fixed-order calculation.

%% file: nlo.tex
\section{NLO: The New Standard}\label{sec:qcd:nlo}

For most observables, NLO predictions in the strong coupling represent
the first accurate theoretical estimate that allows an assessment of
associated theoretical uncertainties. The evaluation of NLO cross
sections, however, is more involved and requires to consider
real-emission and virtual one-loop corrections that are individually
singular in the infrared region. While these divergences cancel in the
sum of both contributions for sufficiently inclusive observables, they
render the numerical evaluation of such cross sections
difficult. Several state-of-the-art techniques exist that allow the
exact cancellation of the divergences separately for the real and
virtual contributions through the introduction of suitable subtraction
terms. Most widely used is the dipole-factorisation method by Catani
and Seymour~\cite{Catani:1996vz};
  alternatives are provided by the
Frixione--Kunszt--Signer (FKS)
method~\cite{Frixione:1995ms},
antenna subtraction~\cite{Kosower:1997zr},
or the recently developed Nagy--Soper
formalism~\cite{Chung:2010fx,Bevilacqua:2013iha}.

The enormous progress recently experienced in the field of NLO QCD
calculations was sparked by two important developments: i) the
introduction of fast and efficient methods for the calculation of
virtual amplitudes, see for instance
Refs.~\cite{Denner:2005nn,Ossola:2006us,Anastasiou:2006jv,Ellis:2007br,Giele:2008ve,Berger:2008sj};
ii) the organisation and implementation of complete NLO calculations
in the framework of parton-level Monte Carlo event generators such as
\HELAC/\PHEGAS~\cite{Bevilacqua:2011xh},
\MADGRAPHFIVE~\cite{Alwall:2014hca}, or
\SHERPA~\cite{Gleisberg:2003xi,Gleisberg:2008ta}.  All these
approaches employ automated subtraction-term generators, see
Refs.~\cite{Gleisberg:2007md,Czakon:2009ss,Hasegawa:2009tx,Frederix:2010cj,
  Bevilacqua:2013iha}, that implement the Catani--Seymour or FKS
subtraction formalism. The real-emission corrections as well as the
phase-space integration are handled by tree-level matrix-element
generators such as \AMEGIC~\cite{Krauss:2001iv},
\COMIX~\cite{Gleisberg:2008fv}, \MADGRAPH~\cite{Alwall:2011uj},
\MADGRAPHFIVE~\cite{Frederix:2009yq,Alwall:2014hca},
or
\HELAC~\cite{Cafarella:2007pc}.  Virtual amplitudes, typically
provided by specialised one-loop generators such as
\BLACKHAT~\cite{Berger:2008sj}, \GOSAM~\cite{Cullen:2011ac},
\HELACONELOOP~\cite{vanHameren:2009dr}, \MADLOOP~\cite{Hirschi:2011pa},
\NJET~\cite{Badger:2012pg},
\OPENLOOPSPLUSCOLLIER~\cite{Cascioli:2011va} or
\RECOLA~\cite{Actis:2012qn} can be incorporated via the universal BLHA
interface~\cite{Binoth:2010xt,Alioli:2013nda} or dedicated
solutions. Examples of recent NLO calculations that have been
performed using a combination of the tools listed above include:
$\Wb+4,5$ jets~\cite{Berger:2010zx,Bern:2013gka}, $\Zb+4$
jets~\cite{Ita:2011wn}, $4$-jet and $5$-jet
production~\cite{Bern:2011ep,Badger:2012pf,Badger:2013yda}, $\ttbar+2$
jets~\cite{Bevilacqua:2010ve} and $\gamma\gamma+3$
jets~\cite{Badger:2013ava}. Most of these new tools are readily
available to perform NLO QCD event generation for use in LHC data
analyses.

\subsection{Jet Counting}

\begin{figure}[htbp]
  \begin{center}
    \includegraphics[width=0.95\textwidth]{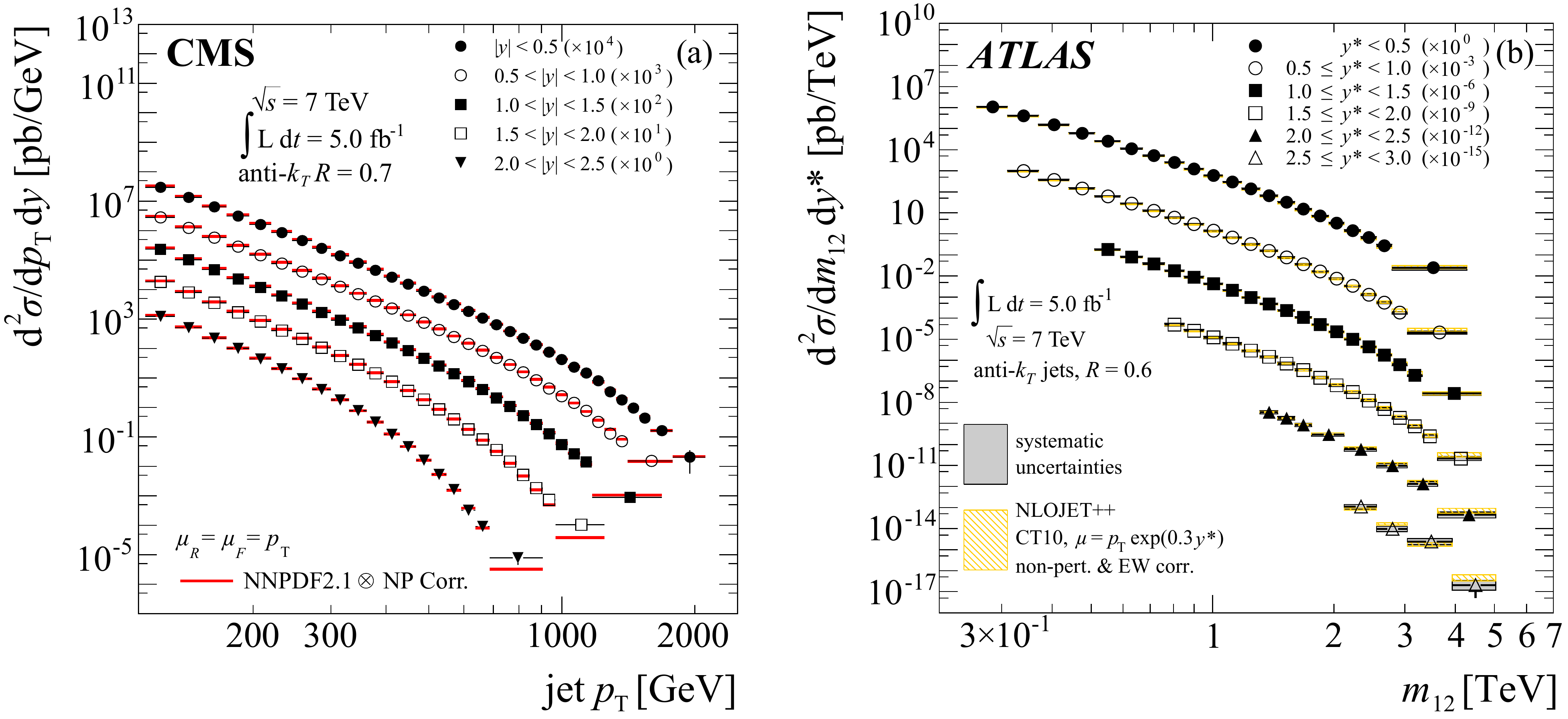}%
    \caption{(a) Inclusive jet cross section from CMS and (b) dijet
      cross section from ATLAS, both at \unit{7}{\TeV}
      centre-of-mass energy, in comparison to theory at NLO including
      non-perturbative (CMS) or non-perturbative and electroweak corrections (ATLAS). The anti-\ktalgo
      algorithm has been applied with the jet-size parameter $R$ being
      0.7 for CMS and 0.6 for ATLAS\@.
      \textit{(Adapted from Refs.~\cite{Chatrchyan:2012bja, Aad:2013tea}.)}
    \label{fig:qcd:jet_crosssections}
  }
\end{center}
\end{figure}

With NLO predictions at hand, more precise comparisons of data and
theory become possible. The production of single inclusive jets and of
dijets constitute the most basic QCD processes at hadron colliders.
Figure~\ref{fig:qcd:jet_crosssections} shows a comparison between data
and theory at NLO, i.e.\ $\mathcal{O}(c_1 \alpS^2 + c_2 \alpS^3)$, for
(a) inclusive jets from CMS~\cite{Chatrchyan:2012bja} and (b) dijets
from ATLAS~\cite{Aad:2013tea}, both at \unit{7}{\TeV}
centre-of-mass energy.
Recently, also trijet cross sections were measured~\cite{Aad:2014rma,CMSPaper12027}.
The inclusive jet cross section has been
measured as a function of \pTjet and $y$ in ranges of $0.1 \leq \pT <
\unit{2.0}{\TeV}$ and $|y| < 2.5$
(see Refs.~\cite{Chatrchyan:2012bja,Aad:2014vwa}).
The dijet cross section is
presented as a function of the dijet mass, $m_{12}$, and rapidity
separation, $y^* = |y_1-y_2|/2$, of the two highest-\pT jets in
regions of $0.3 \leq m_{12} < \unit{5.0}{\TeV}$ and $y^* < 3.0$,
respectively. In both cases the NLO theory has been
corrected for non-perturbative effects, and for the more recent ATLAS
publication also electroweak
(EW) corrections have been considered. While EW effects start being
sizeable
at high \pTjet and $|y|$ or large $m_{12}$~\cite{Dittmaier:2012kx},
non-perturbative effects, estimated via their modelling in MC event
generators,
become negligible. The data are well described by theory over many
orders of magnitude in cross section and in wide kinematic ranges so
that fits of parameters entering the pQCD calculations become
feasible.

\subsection{Jets and the Gluon PDF}

In the following, PDF fits are considered as a first example of
determining parameters entering pQCD calculations.
Many choices are possible in terms of data set
selection, theoretical ingredients and approximations,
parameterisation, or fitting techniques and criteria. There is no
unique approach to PDF fits. The PDF sets in use at the LHC are
produced by six PDF fitting groups: ABM, CTEQ, (G)JR, HERAPDF, MSTW,
and NNPDF~\cite{Alekhin:2012ig, Lai:2010vv, Gluck:2007ck,
  Aaron:2009aa, Martin:2009iq, Ball:2011mu}, where the quoted
publications refer to pre-LHC determinations. The fundamental
experimental input central to all PDF determinations are
deep-inelastic scattering (DIS) data collected by the H1 and ZEUS
collaborations at the \ep collider HERA\@. The HERA collaborations'
own PDF fits based solely on their data are presented as HERAPDF
sets. The ABM group extends the choice of data to other DIS
measurements and fixed-target Drell--Yan data, while the (G)JR group
also considers high-\pT jet production, which cannot yet be described
at NNLO, to better fix the gluon PDF\@. The trio of CTEQ, MSTW, and
NNPDF finally try to incorporate as many data sets as possible and to
provide truly global fits. New PDF sets including LHC data became
available recently or are in preparation for the next start-up of the
LHC with \unit{13}{\TeV} center-of-mass energy.

It is possible to judge the potential impact of new data, e.g.\ from
the LHC, on a given PDF set without a complete refit. The technique of
``Bayesian reweighting''
was first implemented in the context of the NNPDF
approach~\cite{Ball:2010gb}. The NNPDF
collaboration~\cite{Ball:2008by} provides PDF sets in the form of an
ensemble of replicas, which sample variations in the PDF parameter
space within allowed uncertainties. Their method fits a PDF to smeared
data points, where each data point is sampled from a multi-Gaussian
distribution. The central value of the distribution is equal to the
measurement point and the covariance is taken from the experimental
uncertainties. This procedure is repeated $N$ times and provides an
ensemble of $N$ PDFs representing the uncertainty of the PDF
fit. Hence, the NNPDF prediction for an observable is given by the
mean and standard deviation as estimators for the true value and its
uncertainty. Via the Bayesian reweighting, the impact of new data can
be approximately taken into account by giving according weights
different from unity to each replica. If this leads to many replicas
with zero or very small weights, this technique cannot be applied and
refits must be performed. Otherwise the re-evaluation of an observable
while using the re-weighted set of replicas approximately accounts for
changes in the PDFs due to the additional data.

\begin{figure}[htb]
  \begin{center}
    \includegraphics[width=0.95\textwidth]{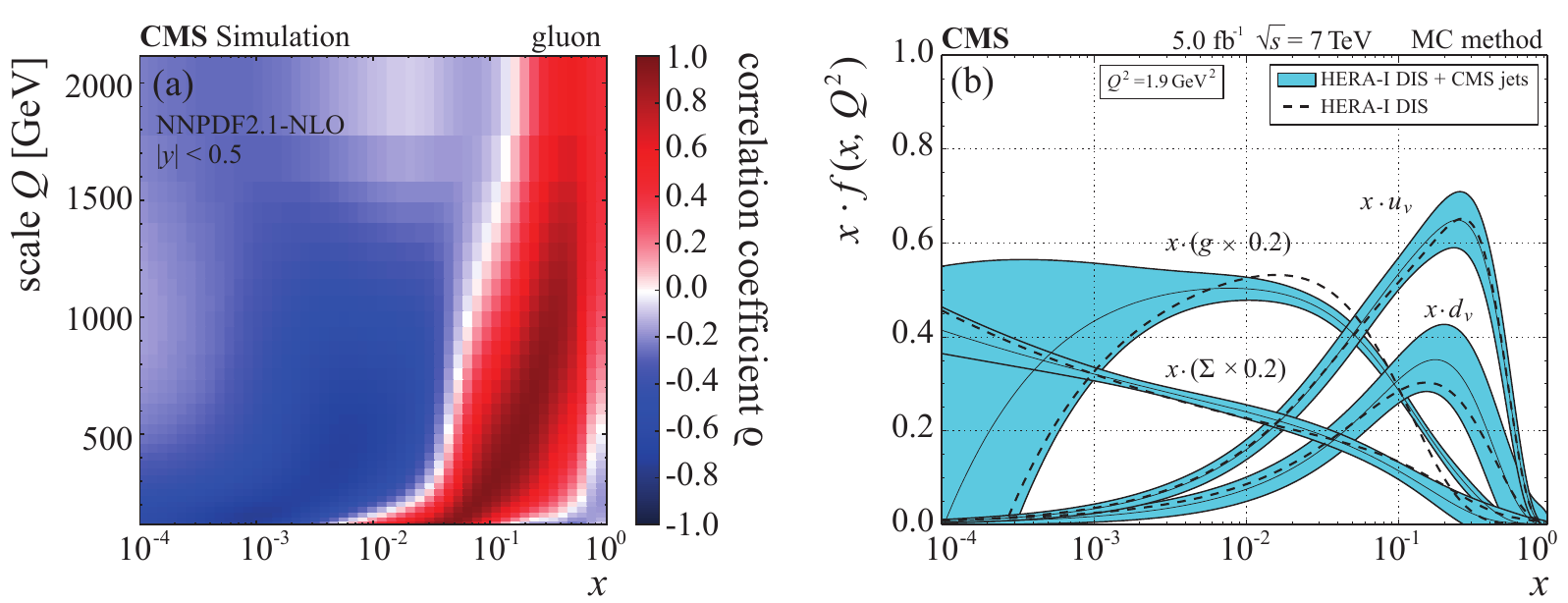}%
    \caption{(a) Correlation coefficient $\rho$ between the inclusive jet
      cross section and the gluon PDF as a function of energy scale
      $Q$ and momentum fraction $x$. %
      (b) PDFs as determined from fits to HERA DIS data alone or in
      combination with CMS jet data. The
      uncertainties on the gluon PDF at high $x$ are reduced compared
      to a fit with DIS data alone (uncertainties not
      shown).
      \textit{(Adapted from Ref.~\cite{CMS:2013yua} and auxiliary material.)}
      \label{fig:qcd:pdfs_incljets}
    }
  \end{center}
\end{figure}

In a similar way to the mean and standard deviation as estimators for
the main prediction and uncertainty of an observable, the NNPDF
ensemble of replicas can be used to investigate the correlation
between an observable and the PDFs as a function of energy scale $Q$
and momentum fraction $x$. As an example,
\fig{\ref{fig:qcd:pdfs_incljets}}(a) shows the correlation coefficient
$\rho$ between the inclusive jet cross section as measured by
CMS~\cite{CMS:2013yua} and the gluon PDF\@. For $x > 0.01$ a
large positive correlation is exhibited between the two.

In case of the other five PDF sets, the method of diagonalisation of
the Hessian matrix~\cite{Stump:2001gu}
(also called ``eigenvector method'')
is employed to express uncertainty estimates for their respective choices of PDF
parameterisation. Their prediction for an observable in pQCD can thus
be evaluated from one central PDF set with an uncertainty given by
quadratic addition of all deviations obtained while using additional
PDFs that correspond to variations
along the directions of the eigenvectors.
A Bayesian reweighting was shown to be feasible here as
well~\cite{Watt:2012tq}, and an alternative approach is studied in
Ref.~\cite{Paukkunen:2014zia}.

Employing an open-source tool like the \HERAFITTER
project~\cite{Alekhin:2014rma, Aaron:2009kv, Aaron:2009aa}, which
provides fitting code, data sets, and theory predictions in a common
framework, it is even possible to perform complete PDF fits.
The CMS collaboration used this tool to study the impact of their
inclusive jet measurements on the gluon PDF~\cite{CMS:2013yua}
following the fitting setup and using the same DIS data as described
in Ref.~\cite{Aaron:2009aa}. The shape of the PDFs at a starting scale
$Q_0$ is assumed to be:

  \begin{align*}
    xg(x) &= A_g x^{B_g} (1-x)^{C_g} - A'_g x^{B'_g} (1-x)^{C'_g}\, , \\
    xu_v(x) &= A_{u_{v}} x^{B_{u_{v}}} (1-x)^{C_{u_{v}}} (1 + E_{u_{v}}x^2)\, ,\\
    xd_v(x) &= A_{d_v} x^{B_{d_v}} (1-x)^{C_{d_{v}}}\, ,\\
    x\overline U(x) &= A_{\overline U} x^{B_{\overline U}} (1-x)^{C_{\overline U}}\, ,\\
    x\overline D(x) &= A_{\overline D} x^{B_{\overline D}}
    (1-x)^{C_{\overline D}}\,.
  \end{align*}

Here, $xg(x)$ is the gluon distribution, $xu_v(x)$ and $xd_v(x)$
represent the valence quarks with $u_v(x) = u(x) - \overline{u}(x)$,
$d_v(x) = d(x) - \overline{d}(x)$, and $\overline{U}(x)$,
$\overline{D}(x)$ are the up- and down-type antiquark distributions
$\overline{U}(x) = \overline{u}(x)$, $\overline{D}(x) =
\overline{d}(x) + \overline{s}(x)$.  The indexed symbols $A_i$, $B_i$,
$C_i$ with $i \in \left\{g,u_v,d_v,\overline U,\overline D\right\}$,
$A'_g$, $B'_g$, $C'_g$, and $E_{u_{v}}$ stand for 19 parameters in the
definition of the PDFs. Note that there are no heavy quarks at a
starting scale that is chosen to be $Q_0^2 =
\unit{1.9}{\GeV\squared}$. The evolution of PDFs in $Q^2$ follows from
QCD according to \eqn{\eqref{eqn:qcd:pdfevolution}}, where in the
\HERAFITTER setup a generalised-mass variable flavour number
scheme~\cite{Thorne:1997ga,Thorne:2006qt} is used.

$A_g$, $A_{u_{v}}$, and $A_{d_{v}}$ are normalisation parameters that
are constrained by QCD sum rules. The $B$ and $C$ parameters describe
the limiting behaviour when very small, $x\to 0$, or very large, $x\to
1$, proton momentum fractions are approached. Additional terms with
parameters for the gluon and \uq-valence distribution allow for some
more flexibility in shape, while constraints of $B_{\overline
  U}=B_{\overline D}$ and $A_{\overline U} = A_{\overline D}(1-f_s)$
are applied to ensure the same normalisation for the \ubarq and
\dbarq densities at high $x$. The strangeness fraction is set
to $f_s = 0.31$, as obtained from neutrino-induced dimuon
production~\cite{Mason:2007zz}. The parameter $C'_g$ is fixed to
25~\cite{Martin:2009iq,Thorne:2006qt} and the strong coupling constant
to $\alpS(\MZ)= 0.1176$ so that in total 13 free parameters remain to
be determined.

The PDF fit result using HERA DIS data alone or in combination with
CMS jet data is presented in \fig{\ref{fig:qcd:pdfs_incljets}}(b). As
expected, a clear impact on the gluon PDF for momentum fractions
$x>0.01$ and a reduction of uncertainties (not shown) is exhibited.

\begin{figure}[htb!]
  \begin{center}
    \includegraphics[width=0.95\textwidth]{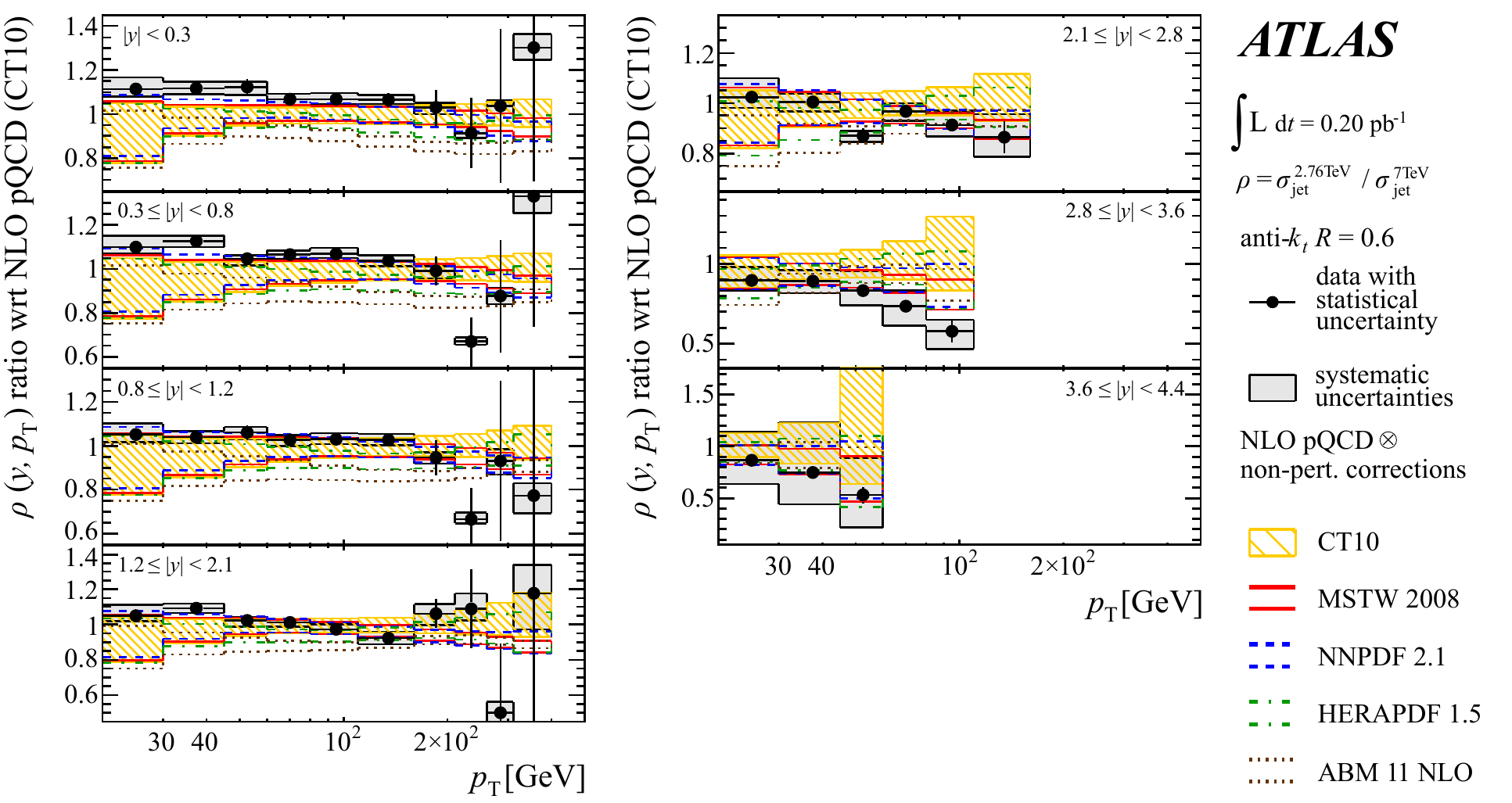}
    \caption{The ratio of the inclusive jet cross sections at $\sqrt{s} = \unit{2.76
      and 7}{\TeV}$ as measured by ATLAS divided by
      a pQCD prediction at NLO (with CT10 PDFs)  of the same ratio. Experimental
      uncertainties are indicated by error bars (statistical) and gray
      boxes (systematic). The uncertainties inherent to the PDFs are
      represented by bands.
      \textit{(Adapted from Ref.~\cite{Aad:2013lpa}.)}
      \label{fig:qcd:incljets_ecms_ratio}}
  \end{center}
\end{figure}

\subsection{Jets, Cross-Section Ratios, and the Strong Coupling}

Absolute jet cross sections as used in the above CMS example are
subject to many experimental sources of uncertainty, and in particular
to details of the jet energy calibration. Since the jet-\pT spectrum
is steeply falling (see \fig{\ref{fig:qcd:jet_crosssections}}), an
uncertainty in the jet energy scale (JES) of 5\% would translate into
an uncertainty on the measured cross section of about 25--30\%.  A
means to at least partially cancel such experimental uncertainties was
investigated by ATLAS in the form of the ratio of inclusive jet cross
sections measured for two different center-of-mass energies
(\unit{2.76 and 7}{\TeV})~\cite{Aad:2013lpa}.
Figure~\ref{fig:qcd:incljets_ecms_ratio} shows the considerably
reduced experimental uncertainties of this ratio compared to the cross
sections themselves, thus demonstrating the possibility to provide
stronger constraints on PDFs. For comparison, the PDF uncertainties
for various PDF sets are indicated as well.

\begin{figure}[ht!]
  \begin{center}
    \includegraphics[width=0.95\textwidth]{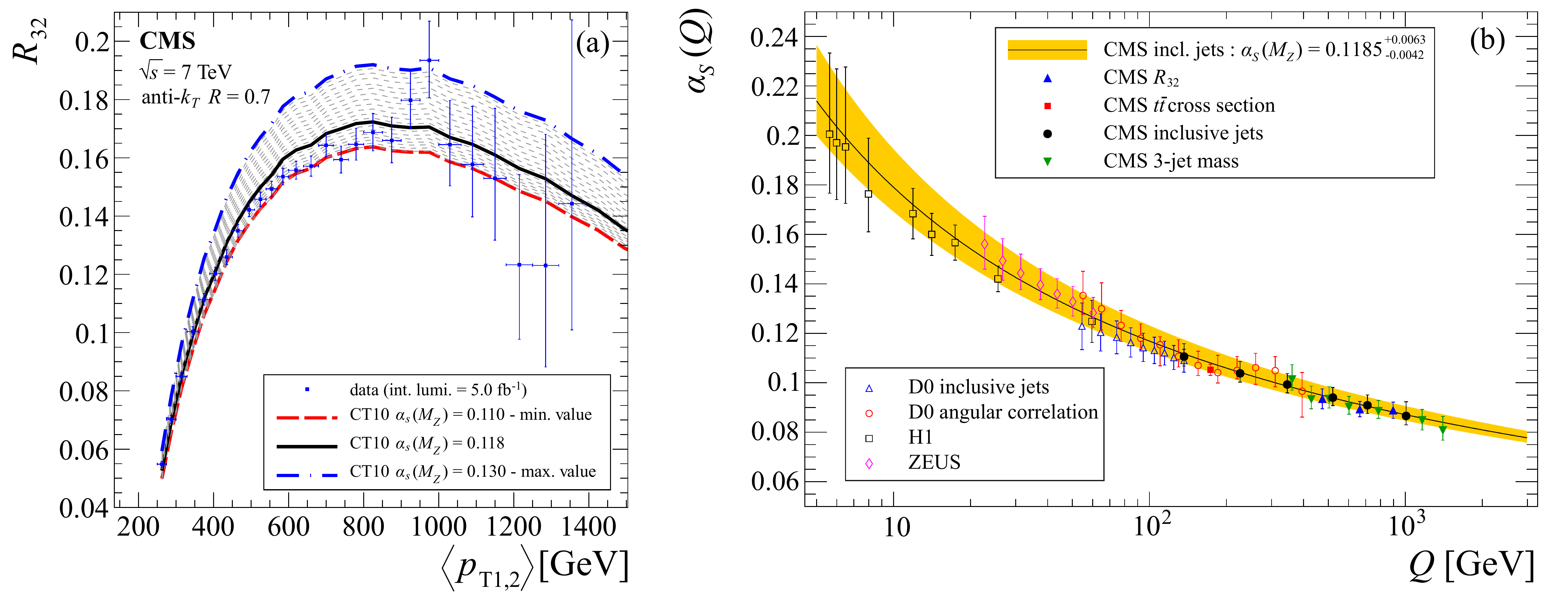}%
  \caption{(a) \Ratio as measured by CMS in comparison to predictions at
    NLO with the CT10 PDF for a series of assumptions on $\alpS(\MZ)$.
    (b) Summary of extractions of $\alpS(Q)$ from low to very high
    scales $Q$ from various experiments including the latest results
    achieved at the LHC up to $Q\approx\unit{1}{\TeV}$.
    \textit{(Adapted from Refs.~\cite{Chatrchyan:2013txa,CMSPaper12027} and auxiliary material.)}
    \label{fig:qcd:running_as}}
\end{center}
\end{figure}

Finally, the possibility to extract the strong coupling constant
$\alpS(\MZ)$ from inclusive jet production data is studied. As can be
seen from \fig{\ref{fig:qcd:pdfs_incljets}} and the DGLAP
equations~\eqref{eqn:qcd:pdfevolution}, the appearance of gluons and
the strength of $\alpS(Q)$ are coupled. Therefore, a simultaneous fit
of PDFs and $\alpS(\MZ)$ with DIS data alone is not possible, which is
also the reason why the value of $\alpS(\MZ)$ is fixed in the
above-mentioned PDF fit of Ref.~\cite{Aaron:2009aa}. The inclusion of
jet data facilitates such a simultaneous fit of PDFs and the strong
coupling as demonstrated by CMS~\cite{CMS:2013yua}. In the
following, the focus shall be, however, on the running behaviour of
$\alpS(Q)$ according to \eqn{\eqref{eq:qcd:runningas}}. To eliminate
or reduce numerous experimental sources of uncertainty---such as
those related to the luminosity measurement or the
jet energy scale---again a ratio is used: that of the inclusive $3$-jet over the
inclusive $2$-jet production cross section, \Ratio. In this quantity,
also theoretical uncertainties like those stemming from the PDFs are
reduced. Since $3$-jet production requires one real emission more than
the usual dijets, the numerator is approximately proportional to
$\alpS^3$, while the denominator is $\propto \alpS\squared$. \Ratio is
therefore a direct measure for the strength of the strong
coupling. The CMS collaboration has investigated this ratio with the
result $\alpS(\MZ) = 0.1148 \pm 0.0014\,\text{(exp)}\pm
0.0018\text{(PDF)}\pm 0.0050
\text{(theo)}$~\cite{Chatrchyan:2013txa}. By performing the analysis
separately for different regions of the average \pT of the two leading
jets, the running could be tested up to scales of \unit{1.39}{\TeV}.
Figure~\ref{fig:qcd:running_as}(a) shows the ratio \Ratio measured by
CMS in comparison to predictions at NLO obtained with the CT10 PDF set
for a series of assumptions on $\alpS(\MZ)$.
Figure~\ref{fig:qcd:running_as}(b) presents a summary of extractions
of $\alpS(Q)$ including the latest results achieved at the LHC with
scales $Q$ reaching beyond \unit{1}{\TeV}.

%% file: nnlo.tex
\section{NNLO: The Quest for Precision}\label{sec:qcd:nnlo}

For a few benchmark processes and observables, the experimental
accuracy of QCD measurements at the LHC is aiming for the few-percent
level or even better. Examples are inclusive single-jet or dijet production,
vector boson (pair) production, and top-quark or Higgs-boson production.
Clearly, for these channels theoretical predictions at NNLO in the strong
coupling are needed, as otherwise the extraction of Standard Model parameters
will be limited by theoretical uncertainties.

The evaluation of NNLO corrections for a given process requires the
calculation of three contributions: i) generic two-loop graphs; ii)
one-loop graphs with one extra final-state parton; iii) real-emission
tree diagrams with two additional partons. All of these contributions
exhibit infrared singularities, and only their sum yields finite results.
In explicit calculations, the divergences are typically treated in
dimensional regularisation. Similar to what was discussed for the treatment of
infrared singularities appearing in NLO corrections before, there
exist now several subtraction schemes to handle these divergences
consistently~\cite{Anastasiou:2003gr,GehrmannDeRidder:2005cm,Catani:2007vq,Czakon:2010td}.

Concerning the above-mentioned processes in need of precision calculations,
full NNLO results exist for inclusive $\Wb/\Zb$-boson
production~\cite{Anastasiou:2003yy,Melnikov:2006kv,Catani:2009sm},
for diboson~\cite{Grazzini:2013bna,Cascioli:2014yka,Gehrmann:2014fva}
and diphoton production~\cite{Catani:2011qz}, for top-quark
pair production~\cite{Czakon:2013goa} and for various Higgs-boson production
channels~\cite{Harlander:2002wh,Anastasiou:2002yz,Harlander:2003ai,Brein:2003wg,Pak:2009dg,Harlander:2009my,Bolzoni:2010xr,Ferrera:2011bk}. For the NNLO evaluation of dijet production,
results have been achieved for two-gluon final states,
initiated by gluon annihilation~\cite{Currie:2013dwa}, and for
quark-antiquark initial states~\cite{Currie:2014upa}.
For the inclusive jet cross section at $\sqrt{s}=\unit{8}{\TeV}$,
considering anti-\ktalgo jets with $R=0.7$, transverse
jet momenta of $\pT > \unit{80}{\GeV}$ and jet rapidities $|y|<4.4$, the authors
find an increase of the NNLO prediction with respect to the NLO estimate of about
27--16\%.
The corrections are largest for small jet transverse
momenta and decrease for larger values of jet \pT. The given
calculation relies on the MSTW08-NNLO PDF set~\cite{Martin:2009iq}
and uses a dynamical scale choice of $\mu=\murs=\mufs$ equal to
the transverse momentum of the leading jet. The size of these
corrections highlights the importance of higher-order calculations
for LHC precision observables. In what follows, two standard
``candles'' (i.e.\ well-measured processes or processes with
theoretically well-understood cross sections) at the LHC---the
Drell--Yan processes and diphoton production---shall be discussed
in some more detail.

\input{wincl}
\input{zpt}
\input{charge}

\subsection{Production of Photon Pairs with Large Invariant Mass}

At leading order in pQCD, the production of photon pairs can rather easily be
described by quark-antiquark annihilation,
$\left(\qqbar\to\gamma\gamma\right)$, as for the Drell--Yan process
before. However, in contrast to the massive vector bosons \Wb and \Zb,
which are predominantly produced in hard scattering processes and can
easily be identified via their leptonic decay modes, massless photons
can also be radiated off initial-state or final-state partons or may be
created in decays. In particular the $\pi^0$ and $\eta$ mesons, and to some
extent also the $\omega$, all of which are copiously produced within
jets, have neutral decay modes. Since these particles are boosted
within the jets, their decay photons are collimated and difficult to distinguish from a
single highly energetic photon.
Without applying further selection criteria, the mode of ``photon''
pair production at large invariant mass via meson decays within two
(different) jets is even dominating by several orders of magnitude.

To distinguish the ``non-prompt'' decay photons within jets
from ``prompt'' photons stemming from the hard scattering, isolation
criteria are applied. Typically, isolation demands that the transverse
energy not associated to a photon candidate but deposited within a
cone of $\Delta R = \sqrt{(\Delta\eta)\squared+(\Delta\phi)\squared} =
0.4$ around it may not exceed a few \GeV. In addition, a minimal
separation of the two photons is required to avoid overlap: $\Delta
R_{\gamma\gamma} > 0.40,\, 0.45$ for ATLAS and CMS, respectively. In
practice, the experimental selections are more complicated (see e.g.\
Refs.~\cite{Aad:2011mh,Aad:2012tba,Chatrchyan:2011qt,Chatrchyan:2014fsa}),
and great care has to be taken to ensure that they can be matched
properly to selection criteria applicable in theory (see
Ref.~\cite{Butterworth:2014efa} for a recent discussion on photon
isolation).

Finely segmented electromagnetic calorimeters are of great advantage
in such photon studies. Investigating the shower shape of energy
deposits in these calorimeters helps to differentiate between
electromagnetic showers that are caused by a single high-energetic
signal photon and those from multiple photons like in
$\pi^0 \to \gamma\gamma$ decays. Photon candidates are separated from
electron (or positron) candidates by using reconstructed tracks.
Depending on the amount of material to be traversed (e.g.\ the beampipe or the
silicon-pixel and silicon-strip detectors), however, highly energetic photons
may convert into \epem pairs before reaching the electromagnetic
calorimeter. Dedicated reconstruction methods are applied to avoid
losing these converted photons, which can make up to
50\%~\cite{Chatrchyan:2014fsa} of the total photon yield.

\begin{figure}[htb]
  \begin{center}
    \includegraphics[width=0.95\textwidth]{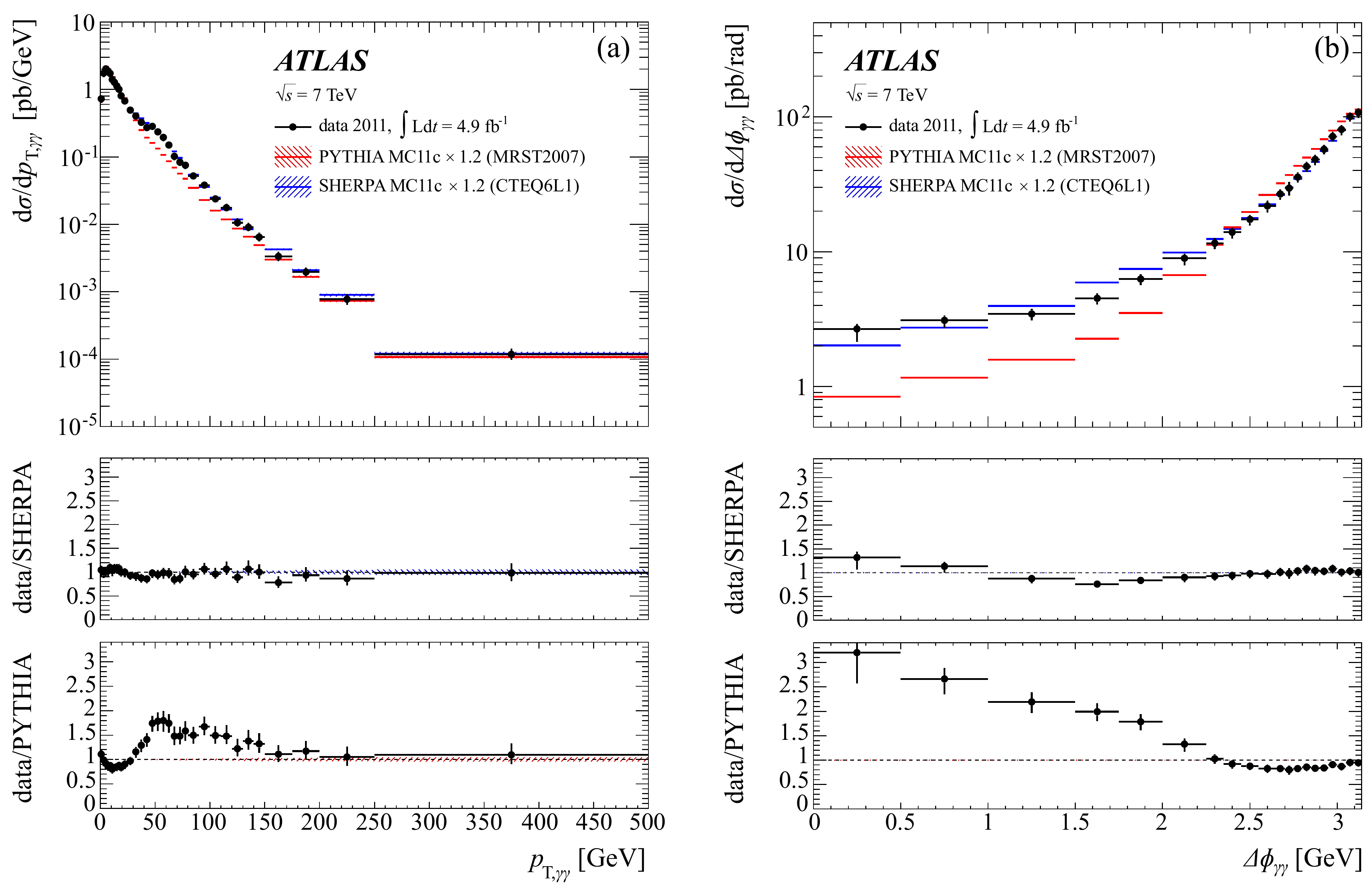}%
    \caption{Photon-pair production cross section at \unit{7}{\TeV}
      centre-of-mass energy as a function of (a) the diphoton transverse
      momentum \pTgamgam and (b) the azimuthal
      angular separation $\Delta\phi_{\gamma\gamma}$ as
      measured by ATLAS~\cite{Aad:2012tba}. The data are compared to
      parton-shower-improved LO predictions by \PYTHIASIX and \SHERPA.
      The MC predictions are rescaled by a factor of $1.2$ to match the total
      cross section observed in data.
      \textit{(Adapted from Ref.~\cite{Aad:2012tba}.)}
    \label{fig:qcd:atlas_diphoton}}
  \end{center}
\end{figure}

For background subtraction, techniques relying on data are
preferred.
Detailed studies, necessary to further quantify detector effects on
the photon reconstruction and isolation and to unfold the signal
yields, are performed with simulated events provided by the LO event
generator \PYTHIA~\cite{Sjostrand:2006za,Sjostrand:2007gs} or by
multi-jet-improved event generators that include additional real
emissions (extra jets) like \SHERPA~\cite{Gleisberg:2008ta,
  Hoeche:2009xc} or \MADGRAPH~\cite{Alwall:2011uj}. The latter is
combined with \PYTHIA for parton showering, hadronisation, and for the
modelling of the underlying event, see \sect{\ref{sec:qcd:npqcd}}. %

At next-to-leading order, the gluon-initiated processes
$\left(\glue\qq(\qbar)\to\gamma\gamma \qq(\qbar)\right)$ join the
annihilation process $\left(\qqbar\to\gamma\gamma \glue\right)$
together with corresponding virtual corrections. Here, a further
complication arises from the collinear fragmentation of a hard
outgoing parton into a photon.
Theoretically, this process is described using fragmentation functions
$D_{\qq\to \gamma}(z,\mufs)$ and $D_{\glue\to \gamma}(z,\mufs)$.  In
the limit of $\mufs\approx Q$, where $Q$ is the scale of the hard
scattering, this fragmentation process results in a contribution to
the cross section that is similar in size to the one of the LO
annihilation process~\cite{Binoth:1999qq}.

Prompt photons not emerging via fragmentation from the hard process
are also called ``direct''.
The effect of the above-mentioned isolation conditions, which aim to
suppress fragmentation photons in favour of direct ones, have to be
properly estimated not only experimentally for mesons decaying within
jets, but also theoretically, e.g.\ via FFs in the perturbative
calculations. The NLO parton-level program
\RESBOS~\cite{Balazs:1997hv,Balazs:2007hr} effectively includes the
fragmentation of one quark/gluon to a single photon at leading order
and additionally features soft and collinear gluon resummation (see
\sect{\ref{sec:qcd:resumfrag}}). \DIPHOX~\cite{Binoth:1999qq} provides
parton-level results at NLO taking fully into account up to two
fragmentation photons. Although formally an NNLO box graph, the
process $\left(\glue\glue\to\gamma\gamma\right)$ is drastically
enhanced at the LHC through the large gluon luminosity. It is
comparable in size to the LO terms, and therefore \DIPHOX includes this
contribution at NLO precision, i.e.\ up to $\text{N}^3\mathrm{LO}$
corrections in the strong coupling \alpS, via
\GAMMATWOMC~\cite{Bern:2002jx}. Finally, a full NNLO calculation is
available in the form of the \TWOGAMMANNLO~\cite{Catani:2011qz}
program, however without consideration of fragmentation photons.

\begin{figure}[htb]
  \begin{center}
    \includegraphics[width=0.95\textwidth]{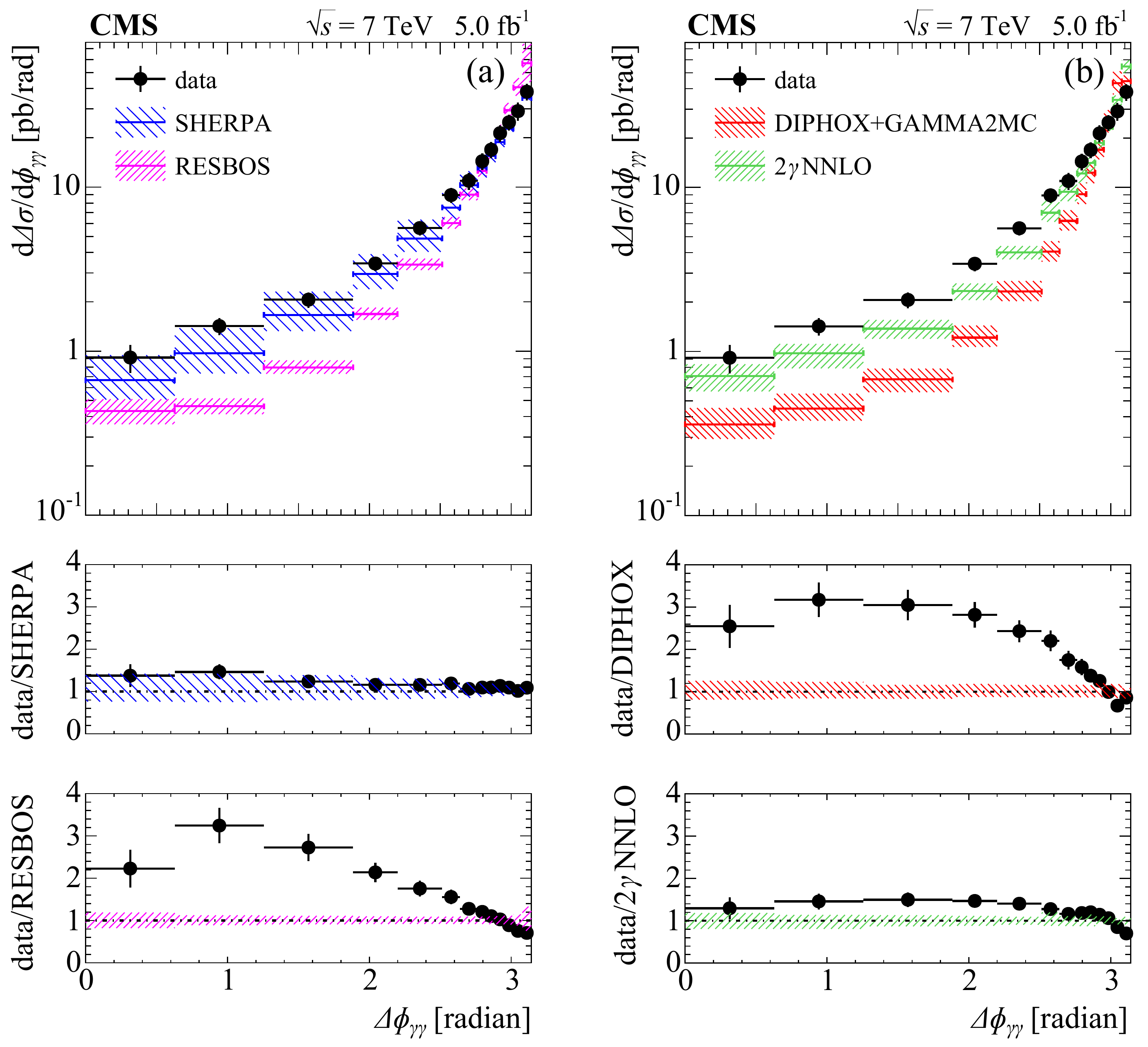}%
    \caption{Photon-pair production cross section at \unit{7}{\TeV}
      centre-of-mass energy as a function of the diphoton azimuthal
      angular separation $\Delta\phi_{\gamma\gamma}$ as measured by
      CMS~\cite{Chatrchyan:2014fsa}.
      (a) Comparison of the data to \SHERPA using tree-level
      matrix elements matched to the parton shower and to NLO
      results from \RESBOS that also include soft-gluon resummation.
      (b) Comparison to full NLO calculations including contributions
      of photons produced in jet fragmentation from \DIPHOX +
      \GAMMATWOMC as well as to NNLO results from \TWOGAMMANNLO.
      \textit{(Adapted from Ref.~\cite{Chatrchyan:2014fsa}.)}
    \label{fig:qcd:cms_diphoton}}
  \end{center}
\end{figure}

Four kinematic variables are usually chosen to investigate the
differential photon-pair production cross section: the invariant mass
$m_{\gamma\gamma}$, the transverse momentum of the photon pair
\pTgamgam, the azimuthal angular separation
$\Delta\phi_{\gamma\gamma}$, and the cosine of the polar angle in the
Collins--Soper reference frame~\cite{Collins:1977iv} of the diphoton
system, $\cos\theta^*_{\gamma\gamma}$. The invariant mass is of
obvious interest for resonance searches, where
$\cos\theta^*_{\gamma\gamma}$ is useful to examine the spin of
diphoton resonances. The transverse momentum $\pTgamgam$ and the
quantity $\Delta\phi_{\gamma\gamma}$ are well-suited for the
comparison of specific aspects of the theoretically very challenging
description of photon-pair production. In particular, at low \pTgamgam
or for well-balanced photons at $\Delta\phi_{\gamma\gamma}\approx\pi$,
where multiple soft-gluon emission becomes important, fixed-order
calculations are not expected to work. As an example,
\fig{\ref{fig:qcd:atlas_diphoton}} shows a comparison between data and
parton-shower improved LO predictions by \PYTHIASIX and \SHERPA for
\pTgamgam and $\Delta\phi_{\gamma\gamma}$ from
ATLAS~\cite{Aad:2012tba}. Since both MC event generators fail to
accurately predict the total cross section, the distributions have
been rescaled by a factor of 1.2 to match the total cross section in
data. In both cases \SHERPA provides a good description of the whole
distribution, while \PYTHIASIX exhibits some discrepancy at low
\pTgamgam.

Figure~\ref{fig:qcd:cms_diphoton}(a) furthermore presents comparisons
for the cross section differential in $\Delta\phi_{\gamma\gamma}$ for
\SHERPA and to NLO from \RESBOS including soft-gluon
resummation. Again \SHERPA gives a good overall description, while
\RESBOS predictions are too low by factors of 2--3 at small separation
angles. A similar behaviour can be seen in the comparison to NLO from
\DIPHOX + \GAMMATWOMC in \fig{\ref{fig:qcd:cms_diphoton}}(b), although
due to resummation \RESBOS performs somewhat better around
$\Delta\phi_{\gamma\gamma}\approx\pi$. The best description of the
data is given by the NNLO result from \TWOGAMMANNLO
(\fig{\ref{fig:qcd:cms_diphoton}}(b)), even though the fragmentation
contribution is not included. ATLAS observes similar results when
comparing their data to the predictions from \DIPHOX + \GAMMATWOMC and
\TWOGAMMANNLO~\cite{Aad:2012tba}.

%% file: wincl.tex
\subsection{Inclusive Vector-Boson Production}
\label{sec:qcd:vinclusive}

Due to the relatively simple
colour-singlet production mechanism and an experimental signature
based on leptons that can be measured very accurately,
the production of \Wb and \Zb bosons is well suited for precision tests of
QCD and electroweak interactions.

At hadron colliders,
massive electroweak bosons are dominantly produced via
quark-antiquark annihilation.
The production of \Zb bosons is
commonly referred to as Drell--Yan process, since in
1970 S.~Drell and T.M.~Yan applied the developing parton model to
predict the scaling of the $ \pp \to \gamma/\Zb \to \mpmm$ cross sections
$\dif\sigma/\dif M^2 \propto K \; L(M^2/s)$, with $M$ the invariant
boson mass, $L$ the parton luminosity
and $K$ a constant that depends on the parton spin and
coupling~\cite{Drell:1970wh}. In modern
QCD, the cross section can be expressed as
$\dif\sigma/\dif M^2 = (4 \pi \alpEM\squared)/(3 \NC M\squared s) L_{\qqbar}{(M\squared/s)}$,
where $L_{\qqbar}$ is the quark-antiquark luminosity, $\alpEM$ the
electromagnetic coupling, and $\NC$ the number of colours.

Vector-boson processes are relatively simple to calculate, since
there is no QCD interaction involving the final-state particles. In fact,
the Drell--Yan process was the first hadron-collider channel
known to NNLO accuracy~\cite{Anastasiou:2003yy,Melnikov:2006kv,Catani:2009sm}.
The NNLO Drell-Yan production cross sections can be calculated with a precision
of $1\%$
for the renormalisation-scale and factorisation-scale uncertainty.
The calculations also include the decay of \Wb and \Zb bosons to leptons.
For inclusive \Zb-boson production at the LHC, the NLO-to-LO
correction is sizeable (about 20\%), while
the NNLO-to-NLO correction amounts to 2\% only .
Two public codes implement the NNLO QCD calculations
for inclusive \Zb-boson and \Wb-boson production:
\DYNNLO~\cite{Catani:2009sm} and \FEWZ~\cite{Gavin:2010az,Li:2012wna}.
Furthermore, the dominant electroweak corrections have been determined~\cite{Baur:1997wa,Baur:2001ze,Dittmaier:2001ay,CarloniCalame:2007cd,Arbuzov:2007db,Dittmaier:2009cr}.

The precise measurement of the inclusive \Wb-boson and \Zb-boson
production cross sections at LHC and their successful comparison to precise
QCD calculations was one of the first and very important confirmations
of pQCD at very high energies and momentum transfers.
Measurements
at $\sqrt{s}=\unit{7}{\TeV}$ have been performed with the 2010 data set
corresponding to an integrated luminosity of \unit{36}{\invpb} by
ATLAS~\cite{Aad:2011dm} and CMS~\cite{CMS:2011aa}.
CMS also measured the cross sections at $\sqrt{s}=\unit{8}{\TeV}$
using a data set of \unit{18.2}{\invfb}~\cite{Chatrchyan:2014mua}.

The cross sections are measured by counting the number of events in
the detector acceptance and
subtracting the estimated
background contributions.
The efficiency $\epsilon$ is estimated from Monte Carlo simulations and
corrected for differences between these simulations and the data.
To better separate experimental and modelling uncertainties
(in particular from the extrapolation of the detector acceptance
to the total phase space), the efficiency is decomposed in a
fiducial acceptance, $A$, and a correction for detector effects, $C$,
i.e.\ $\epsilon = A \cdot C$.

The fiducial acceptance $A$ is given by the ratio of the number of
events passing the kinematic cuts applied on particle-level
(i.e.\ without detector simulation) over the total number of generated events.
The kinematic selection cuts are set close to the requirements on the
reconstructed objects typically for leptons: $\pT>\unit{20}{\GeV}$ and $|\eta| < 2.5$.
The detector correction factor $C$ is defined as the number of selected
events in the sample including the detector simulation
over the number of events passing the fiducial acceptance $A$.
Typically $A$ is 0.45--0.50 and $C$ is 0.7--0.8, depending on the analysis.

A summary of the inclusive Drell--Yan cross-section measurements is shown in
\tab{\ref{tab:inclusiveWZcrosssection}}. The statistical uncertainty amounts
to about 0.3\% for \Wb-boson production and 0.7\% for \Zb-boson production.
The systematic uncertainty is dominated by the knowledge of the integrated
luminosity, which at $\sqrt{s}= \unit{7}{\TeV}$ results in an uncertainty of
3.4\% (ATLAS) and 4.0\% (CMS) and in 2.5\% at $\sqrt{s}= \unit{8}{\TeV}$.
The measurements of ATLAS and CMS agree within their uncertainties, and both
agree with the NNLO prediction that has an
uncertainty of about 5\% (dominated by the PDF uncertainty).

\begin{table}
\begin{center}
\begin{tabular}{lccc}
\hline
            & $\Wp\rightarrow l^+\nu$           & $\Wm\rightarrow l^-\nu$             &   $\Zb\rightarrow \lplm$ \\
\hline
 \unit{7}{\TeV}   &                                  &                                     &                \\
ATLAS       & $6.05\pm0.12\pm 0.21$ nb $\;\;$&  $4.16\pm0.10\pm0.14$ nb  $\;\;$ & $0.937\pm0.02\pm0.03$ nb \\
CMS         & $6.04\pm0.10\pm0.24$ nb  $\;\;$ &  $4.26\pm0.08\pm0.17$ nb  $\;\;$ & $0.974\pm0.02\pm0.04$ nb \\
NNLO QCD    & $5.98\pm0.3$ nb          $\;\;$ &   $4.2\pm0.2$ nb   $\;\;$       & $0.991\pm0.05$ nb        \\
\hline
 \unit{8}{\TeV}   &                                  &                                     &                \\
CMS         & $7.11\pm0.14\pm 0.18$ nb $\;\;$ & $5.09\pm0.12\pm 0.13$  nb  $\;\;$ & $1.15\pm0.02\pm0.03$  nb  \\
NNLO QCD    & $7.12\pm0.2$ nb      $\;\;$      & $5.06\pm0.13$ nb        $\;\;$      & $1.13\pm0.04$ nb           \\
\hline
\end{tabular}
\caption{Summary of the cross-section results of the inclusive \Wb-boson and \Zb-boson
analyses of ATLAS and CMS. The total uncertainty (statistical, experimental and
acceptance modelling added in quadrature) and the luminosity uncertainties are also given.
The theory cross sections are calculated with \FEWZ~\cite{Gavin:2010az,Li:2012wna}
with the MSTW2008-NNLO PDF set~\cite{Martin:2009iq} and contain renormalisation-scale and
factorisation-scale uncertainties and PDF uncertainties.
\label{tab:inclusiveWZcrosssection}}
\end{center}
\end{table}

%% file: zpt.tex
\subsection{Differential Vector-Boson Cross Sections}

Apart from the total \Wb and \Zb production cross sections, also some differential
cross sections are known to order ${\mathcal{O}}(\alpS\squared)$; this includes
the dilepton invariant-mass spectrum, the vector-boson's transverse momentum
and its rapidity distribution.

\begin{figure}[htbp]
  \begin{center}
    \includegraphics[width=0.95\textwidth]{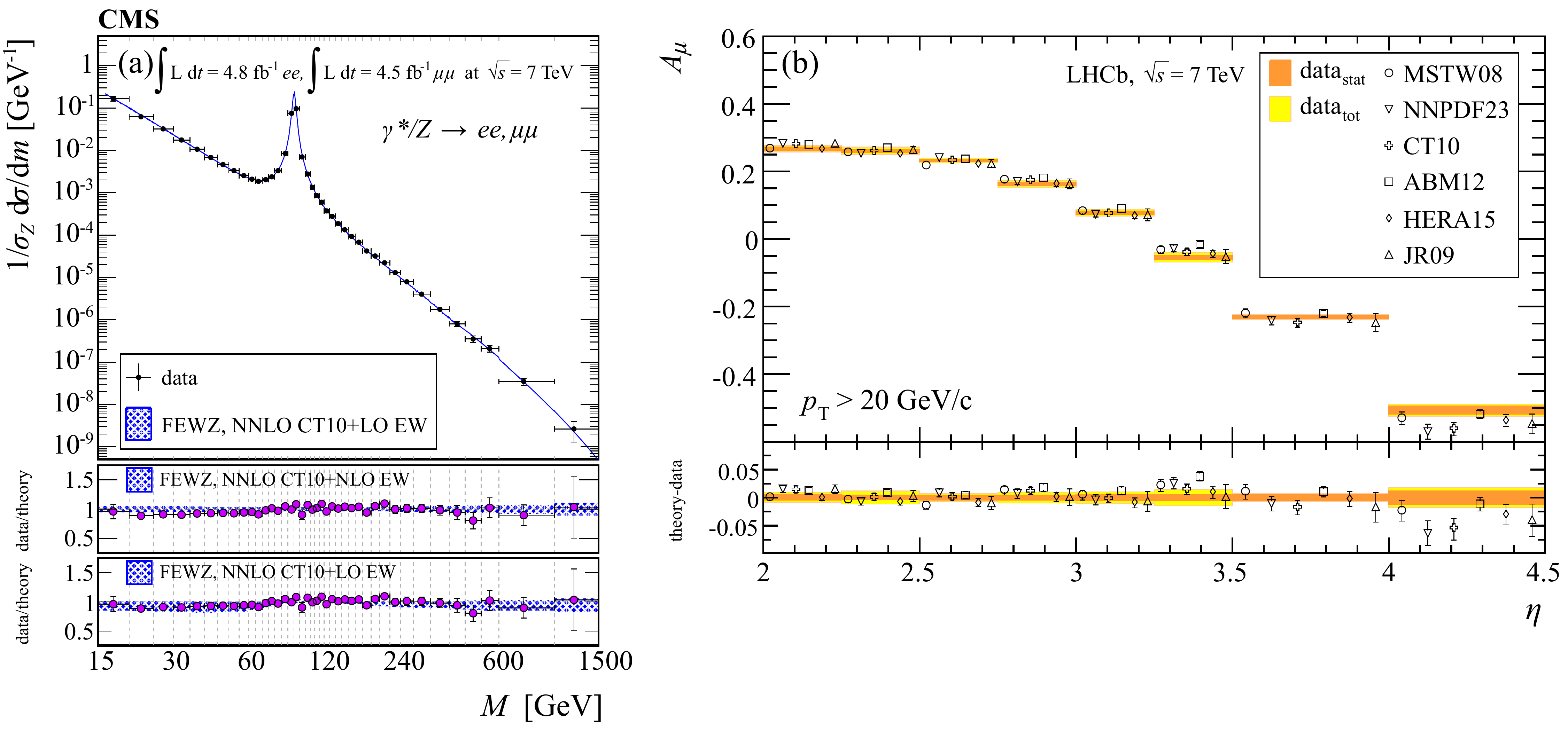}%
    \caption{(a) Drell--Yan cross sections at $\sqrt{s} = \unit{7}{\TeV}$ as a function of
      the dilepton invariant mass $M$.
      (b) \Wb-boson charge asymmetry as a function of the lepton
      pseudo-rapidity.  Data are compared to NNLO QCD
      predictions using various PDF sets.
      \textit{(Adapted from Refs.~\cite{Chatrchyan:2013tia,Aaij:2014wba}.)}
    \label{fig:qcd:cms_drellyan}}
  \end{center}
\end{figure}

As an example, \fig{\ref{fig:qcd:cms_drellyan}} presents a CMS measurement
of the \Zb-boson production cross section
as a function of the invariant mass $M$ of the two leptons from the \Zb decay
at $\sqrt{s}=\unit{7}{\TeV}$~\cite{Chatrchyan:2013tia}.
The cross section falls over
eight orders of magnitudes
in the mass range of
$15 < M < \unit{1500}{\GeV}$
and clearly shows the \Zb-boson resonance at \unit{90}{\GeV}.
The NNLO calculation using the CT10 PDF set describes
the data well. Similar measurements were performed
by ATLAS~\cite{Aad:2013iua}.

The boson's transverse momentum
is an interesting observable to
test various regimes in the strong-interaction dynamics. The high-\pT part of the
spectrum can be modelled by perturbative QCD calculations considering
processes with additional jets
($\Wb/\Zb+n$-jet processes). Presently calculations can be performed with up to $n=5$ additional partons.
When the transverse momentum of the vector boson is much smaller than
its invariant mass reconstructed from the two leptons,
soft QCD radiation is enhanced and fixed-order perturbation theory has to be
supplemented by the resummation of large logarithmic corrections of the form
$\alpS^n \ln^m(\pT/M)$ to all orders of \alpS. The resummed cross sections also
include a non-perturbative component at momentum scales below \unit{1}{\GeV}. This region
can also be modelled using Monte Carlo generators implementing parton
showers and models for hadronisation.

To lowest order, \Wb and \Zb bosons are produced via quark-antiquark
annihilation, i.e.\ $\qqbar\to \Zb$. Indeed, this process dominates for inclusive \Zb production
within a rapidity range of $|y|<2.1$.
However, already for \Zb transverse momenta around \unit{25}{\GeV}
quark-gluon scattering $\qq\glue \to \Zb\qq$ is of similar size,
and around $\pT = \unit{100}{\GeV}$ the latter process constitutes 80\% of
the total cross section~\cite{Brandt:2013hoa}.
For \Zb-boson transverse momenta of \unit{180}{\GeV}, parton-momentum fractions
$x_{1/2} = \left(M/\sqrt{s}\right) \exp(\pm y)$ of about $x=0.05$ are probed.

The program \RESBOS~\cite{Balazs:1997xd} implements soft-gluon resummation at
low \pT at next-to-next-to-leading-logarithm (NNLL) accuracy and matches to a matrix-element calculation of order
$\mathcal{O}(\alpS\squared)$ at high \pT. It provides fully differential cross sections
as a function of the rapidity, the invariant mass and the transverse momentum
of the vector boson. It also attempts to estimate non-perturbative
corrections~\cite{Guzzi:2013aja}. Recently, first calculations for Drell--Yan lepton-pair
production at NNLO accuracy including parton-shower effects have been
presented~\cite{Hoeche:2014aia,Karlberg:2014qua}.

\begin{figure}[htbp]
  \begin{center}
    \includegraphics[width=0.95\textwidth]{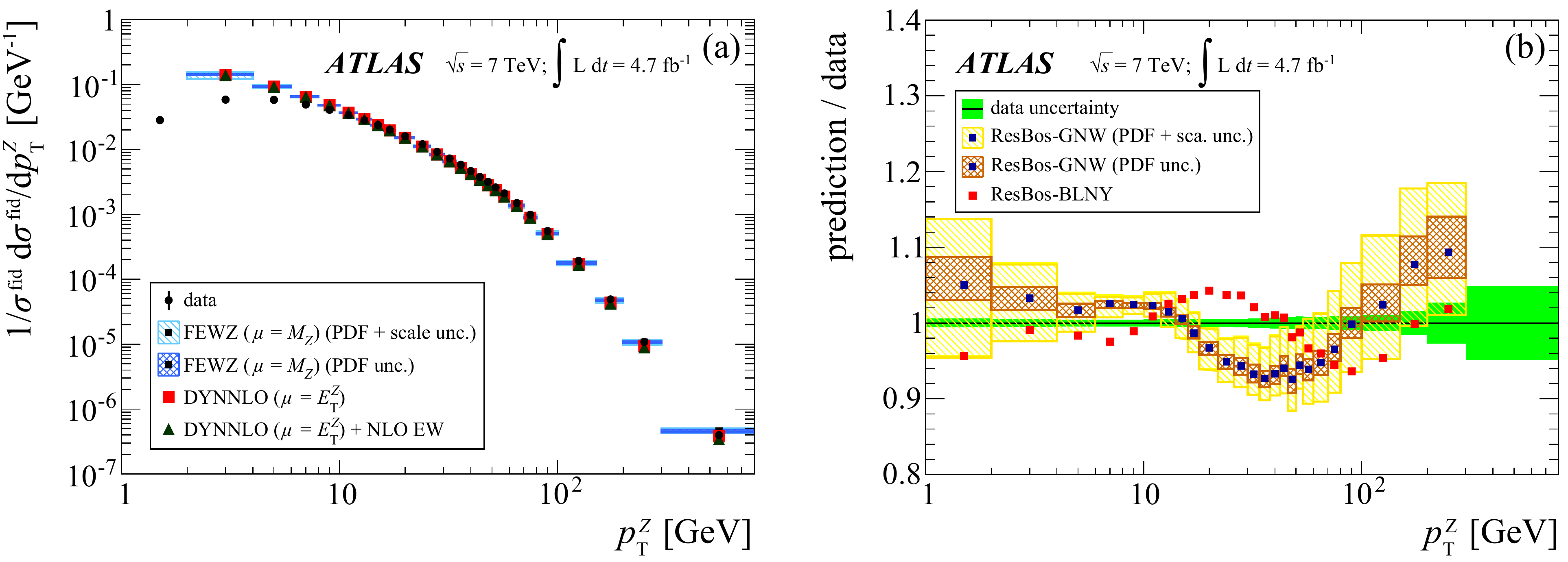}%
    \caption{Normalised transverse momentum of the \Zb boson
      measured within the detector acceptance at $\sqrt{s}= \unit{7}{\TeV}$.
      (a) ATLAS data compared to NNLO QCD calculations based on
      \DYNNLO~\cite{Catani:2009sm} and
      \FEWZ~\cite{Gavin:2010az,Li:2012wna}) with different
      renormalisation-scale and factorisation-scale settings.
      (b) Ratio of an NNLO QCD prediction including NNLL resummation to the
      data. Shown are calculations that differ in the parametrisation
      of non-perturbative effects. The bands indicate the respective
      theory uncertainty estimate.
      The band around 1 denotes the data measurement uncertainty.
      \textit{(Adapted from Ref.~\cite{Aad:2014xaa}.)}
    \label{fig:smqcd:zpt}}
  \end{center}
\end{figure}

Early measurements of differential cross sections at $\sqrt{s} = \unit{7}{\TeV}$
for \Zb~\cite{Aad:2011gj} and \Wb~\cite{Aad:2011fp,Chatrchyan:2011wt} bosons
were based on an integrated luminosity of \unit{35-40}{\invpb}.
Recently, the ATLAS experiment updated this measurement to a larger data set
corresponding to an integrated luminosity of \unit{4.7}{\invfb}.
This measurement reaches \Zb-boson transverse momenta of \unit{800}{\GeV}
for rapidities up to 2.4~\cite{Aad:2014xaa}.
The measurement uncertainty amounts to 1\% for $\pT < \unit{150}{\GeV}$ and rises to
about 5\% for the highest \pT. The leptonic decays of the \Zb boson to electrons
or muons are used in the analysis, and
electrons or muons
with $\pT>\unit{20}{\GeV}$ and rapidity $|y|<2.4$ are required. The invariant mass of the dilepton pair must lie between
$66 < M < \unit{116}{\GeV}$.
Figure~\ref{fig:smqcd:zpt} shows the comparison of the data to the NNLO QCD predictions
based on \DYNNLO~\cite{Catani:2009sm} and \FEWZ~\cite{Gavin:2010az,Li:2012wna}
using the CT10 PDF set.
Shown as a band are theory uncertainties that are due to the
renormalisation-scale and factorisation-scale variations and associated with the choice of
PDFs for both calculations.
The measurement uncertainties are below a percent in most bins and are smaller than the theory
uncertainties that reach 10\% around \unit{50}{\GeV}.
The two calculations are in excellent
agreement with each other for $\pT<\unit{100}{\GeV}$, but differ at large \pT by
about 10\%.
In the region $10 < \pT < \unit{100}{\GeV}$ the predictions are 10\% lower
than the data. For $\pT<\unit{10}{\GeV}$ the fixed-order calculations diverge and clearly
disagree with the data.

The resummed calculation based on \RESBOS using the CTEQ6M PDF~\cite{Pumplin:2002vw}
describes the data within 10\%. The observed deviations are covered by the
theory uncertainties of the standard non-perturbative parametrisation
(denoted ``BLNY''). A recent new parameterisation denoted ``GNW''~\cite{Guzzi:2013aja} predicts a somewhat different shape.

An NLO QCD calculation based on \POWHEG interfaced with \PYTHIA for parton showers
and hadronisation describes the data within 5\% for $\pT<\unit{60}{\GeV}$, but deviates
by up to 20\% over the full \pT range. An NLO QCD calculation based on \MCATNLO
interfaced with \HERWIG for parton shower and hadronisation deviates by up to
40\% at high \pT. Leading order Monte Carlo generators based on multi-leg matrix
elements, like \SHERPA or \ALPGEN, agree with data within 5\% except for the
highest \pT bin. A dedicated tuning of parton-shower parameters in \PYTHIAEIGHT
interfaced to \POWHEG~\cite{Aad:2014xaa} achieves agreement
within 2\% for $\pT < \unit{50}{\GeV}$.

%% file: charge.tex
Since the dominant \Wpm production mechanisms at the LHC are
$\dq \ubarq \to \Wm$ and $\uq \dbarq \to \Wp$, measurements of \Wpm-boson
production cross sections provide a powerful tool to study the parton density functions
of \dq and \uq quarks in the proton.
In particular, the measurement of the lepton charge asymmetry
$A_\ell = (\sigma_{\Wp \to \ell^+ \nu} - \sigma_{\Wm \to \ell^- \bar{\nu}})
/(\sigma_{\Wp \to \ell^+ \nu} + \sigma_{\Wm \to \ell^- \bar{\nu}})$
gives valuable information on the \dq-to-\uq ratio
and also on the sea-quark and sea-antiquark distributions, in particular of the strange quark.
By measuring differential cross sections, for instance
as a function of the lepton rapidity, different parton-momentum fraction
values $x$ can be probed, since the parton-momentum fraction depends on the
vector-boson mass and on the rapidity $y$:
$
x_{1/2} = (M_{\Wb,\Zb} / \sqrt{s}) \exp\left({\pm y}\right)
$.

ATLAS published such differential cross-section measurements
$\dif \sigma_{\Wpm/\Zb}/\dif y_\ell$
for electrons and muons up to lepton rapidities of $y_\ell=4.9$ based on the
2010 data set at $\sqrt{s} = \unit{7}{\TeV}$~\cite{Aad:2011dm}.
Leptons with a $\pT > \unit{20-25}{\GeV}$ are selected.
For \Zb bosons, the accuracy reaches about 2\% in the central region
of the \Zb-boson rapidity and 10\% at $y_Z = 3.2$.
A precision of about 2\% is obtained for \Wb-boson cross sections
measured within $|y_\ell |<2.5$.
For the
$A_\ell$ measurement the accuracy ranges between 4 and 8\%.
The data can be described by an NNLO calculation using the
HERAPDF1.5 and the MSTW08-NNLO PDF sets.
Further measurements by ATLAS and CMS can be found in Refs.~\cite{Aad:2011yna,Chatrchyan:2011wt,Chatrchyan:2011jz,CMS:2011aa,Aad:2010yt,Khachatryan:2010xn}.

The LHCb experiment measured \Wb-boson production in the forward region covering
lepton rapidities in the range $2 < |y_\ell | <4.5$. A measurement was performed with the 2011 data set
at $\sqrt{s} = \unit{7}{\TeV}$ using an integrated luminosity of \unit{1}{\invfb}.
Figure~\ref{fig:qcd:cms_drellyan}(b) shows the measured lepton charge asymmetry for muons
with $\pT>\unit{20}{\GeV}$ as a function of the
muon rapidity compared to NNLO calculations using various PDF sets.
The band shows the statistical (inner, dark) and total uncertainty (light)
on the data. The experimental accuracy is similar to the theoretical
uncertainty estimate, and the predictions are in good agreement with the measurement.

%% file: nlops.tex
\section{Multi-Jets: Precision Meets Multiplicity}\label{sec:qcd:nlops}

Understanding the QCD dynamics governing the emergence
of multi-jet final states is of utmost importance at
the LHC. Given that typically jet objects
with minimal values of their transverse momentum
of order \unit{30}{\GeV} are considered, the phase space for
producing multiple jets is huge. Accordingly, these
jet-production processes need to be described as well
as possible, as they constitute severe backgrounds to
almost every search for new physics and furthermore
have an impact on the appearance of the signals
themselves. The modelling of inclusive processes that
receive contributions from final states with potentially
many jets is the realm of so-called ``matching and merging''
techniques that combine (multi-parton) fixed-order
calculations with parton-shower simulations.

On the one hand, fixed-order calculations provide a
well-defined estimate for inclusive production cross sections
and reliably account for hard, well-separated parton configurations.
On the other hand, parton-shower simulations capture the dominant
terms driving the emission of additional soft and/or collinear
partons and thus provide the means to perturbatively account for
the internal structure of jets and inter-jet energy flows. When
combining fixed-order matrix elements with parton-shower
resummation beyond leading-order 2-to-2 scattering processes,
the obstacle of double-counting configurations that appear in both
approaches needs to be resolved. Furthermore, any consistent
scheme should neither spoil the logarithmic accuracy of the
inherent parton-shower resummation nor destroy the formal
precision of the fixed-order part.

There are two basic strategies to distinguish. Tree-level merging
techniques correct the hardest emissions of the parton shower
off a given core process through exact leading-order QCD matrix
elements~\cite{Catani:2001cc,Lonnblad:2001iq,Krauss:2002up}. This is
achieved through a slicing of the real-emission phase space in terms
of a hardness-measure that regulates any infrared singularities and
allows to consistently combine matrix elements of varying parton
multiplicity dressed with parton showers into an inclusive sample.
Several variants of such leading-order merging techniques exist
and are widely used in LHC analyses, see
Refs.~\cite{Mangano:2006rw,Alwall:2007fs,Hoeche:2009rj,Lonnblad:2011xx,Platzer:2012bs}.

The second ansatz relies on the exact matching of an NLO QCD
calculation with a parton-shower cascade off the underlying
Born process.
Accordingly, the real-emission correction as part
of the fixed-order calculation has to be properly synchronised with the
first, i.e.\ hardest, shower splitting. Furthermore, the NLO accuracy
with respect to the inclusive production process considered needs to
be preserved. Two basic solutions exist to this problem, which are known
as \MCATNLO~\cite{Frixione:2002ik} and \POWHEG~\cite{Nason:2004rx}.
Over the last years there has been tremendous development in
implementing these techniques for a wide range of processes and
ultimately their automation, see
Refs.~\cite{Frixione:2007vw,Alioli:2010xd,Platzer:2011bc,Hoeche:2011fd,Alwall:2014hca}.

Most recently hybrid solutions emerged that combine
NLO plus parton-shower calculations with higher-order tree-level
QCD corrections~\cite{Hamilton:2010wh,Hoche:2010kg} or that even
combine parton-shower-matched NLO calculations of varying jet
multiplicity~\cite{Hoeche:2012yf,Frederix:2012ps,Lonnblad:2012ix,Hamilton:2012rf,Hoeche:2014qda}.

In the following section, the main focus shall be on a class of
processes that constitutes a prototypical Standard Model
background and serves as a test bed for the various types
of advanced QCD calculations outlined above: $\Wb/\Zb+$jets
production.

\subsection{Weak Bosons and Jets}

\begin{figure}[htbp]
  \begin{center}
    \includegraphics[width=0.95\textwidth]{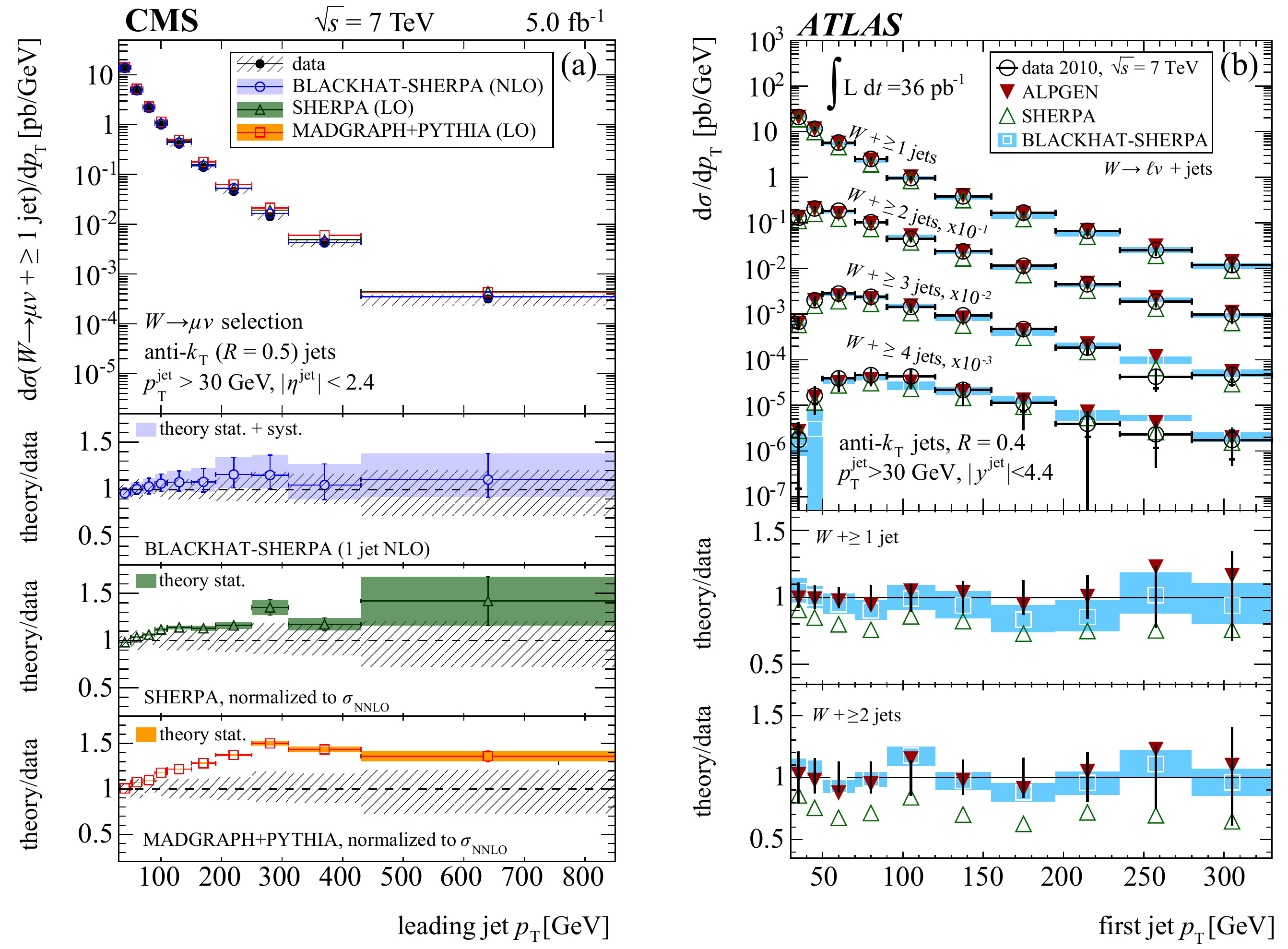}
    \caption{The $\Wb+$jet cross section as a function of the transverse momentum
    of the leading jet as measured by CMS (a) and
      ATLAS (b). The data are compared to various QCD
      calculations. For the ATLAS measurement, jet multiplicities up
      to four jets are shown. The results for higher jet
      multiplicities are scaled by the factors given in the
      figure. The leading-order multi-leg calculations are normalised to the inclusive NNLO cross section.
      \textit{(Adapted from Refs.~\cite{Khachatryan:2014uva,Aad:2012en}.)}
    \label{fig:qcd:wbosonjet}}
  \end{center}
\end{figure}

In the production of \Wb and \Zb bosons in association with jets
($\Wb+$jets), very high jet multiplicities can be reached at the LHC. Cross sections
for the associated production with up to seven jets with
transverse momentum \pT larger than \unit{30}{\GeV} have been measured by
ATLAS and CMS. Already
the leading-order prediction for this process is of order $\alpS^7$. However,
this class of processes is extremely important as it constitutes a major background
to many other processes with complex final states such as top-quark pair or diboson
production or rare signatures from the Standard Model or beyond.

The $\Wb+$jets processes allow an assessment of the validity of
Monte Carlo generators like \ALPGEN~\cite{Mangano:2002ea},
\SHERPA~\cite{Gleisberg:2003xi,Gleisberg:2008ta},
or \MADGRAPH~\cite{Alwall:2011uj}
(together with recent progress in \MADGRAPHFIVE~\cite{Alwall:2014hca})
that merge tree-level matrix-element
calculations with up to five additional partons interfaced with parton
showers and hadronisation.
Presently, many experimental analyses are in a transition phase
adopting NLO matrix-element calculations plus parton showers as
baseline for MC event generation instead of the former LO based ones
(see Section~\ref{sec:qcd:nlo}).

Figure~\ref{fig:qcd:wbosonjet} shows the $\Wb+$jets cross section as a function of the
leading-jet \pT as measured by CMS~\cite{Khachatryan:2014uva}
and ATLAS~\cite{Aad:2012en} using proton-proton collision data
recorded at $\sqrt{s}= \unit{7}{\TeV}$. In the ATLAS measurement, jets are
defined by the anti-\ktalgo jet algorithm using a radius parameter of $R=0.4$ and measured within
$30 \le \pT < \unit{300}{\GeV}$ and $|y|<4.4$. The results for up to four additional jets
are superimposed and are compared to the predictions of \SHERPA and \ALPGEN.
The measured shape of the jet \pT distribution is described
by the Monte Carlo simulation within the uncertainties.

In the CMS measurement, jets defined by the anti-\ktalgo jet algorithm with $R=0.5$
are selected with $\pT > \unit{30}{\GeV}$ and $|y|<2.4$.
The jet spectrum reaches up to \pT values of \unit{800}{\GeV}. The data fall below
the leading-order QCD predictions towards high \pT. The \MADGRAPH (\SHERPA) estimate
using up to four additional partons is almost 50\% (20\%) higher than the data for
$\pT>\unit{200}{\GeV}$.

At the time of the LHC start-up in 2008, NLO QCD calculations with maximally two associated jets
were available~\cite{Campbell:2003hd}.
Since then, in particular,
with the \BLACKHATSHERPA~\cite{Berger:2010gf,Berger:2010zx} program significant progress was made in NLO QCD
calculations for \Wb or \Zb production in association with up to five additional
partons~\cite{Bern:2013gka}. Fixed-order calculations, by nature, are not matched to parton
showers and hadronisation, so that corrections to account for these
effects are derived
from generators like \PYTHIA or \HERWIG are needed. The resulting predictions from
\BLACKHATSHERPA are in good agreement with the measured distributions within the
systematic uncertainties as shown in \fig{\ref{fig:qcd:wbosonjet}}.

Recently, automated computations of NLO QCD cross sections matched to parton-shower
simulations for a large variety of processes (with up to four partons) have been developed.
The programs \SHERPA (in the \MEPSATNLO option~\cite{Gleisberg:2008fv,Hoeche:2012yf})
and \MADGRAPHFIVE~\cite{Alwall:2014hca} presently provide
the most accurate predictions for multi-jet final states at particle level.

\begin{figure}[htbp]
  \begin{center}
    \includegraphics[width=0.95\textwidth]{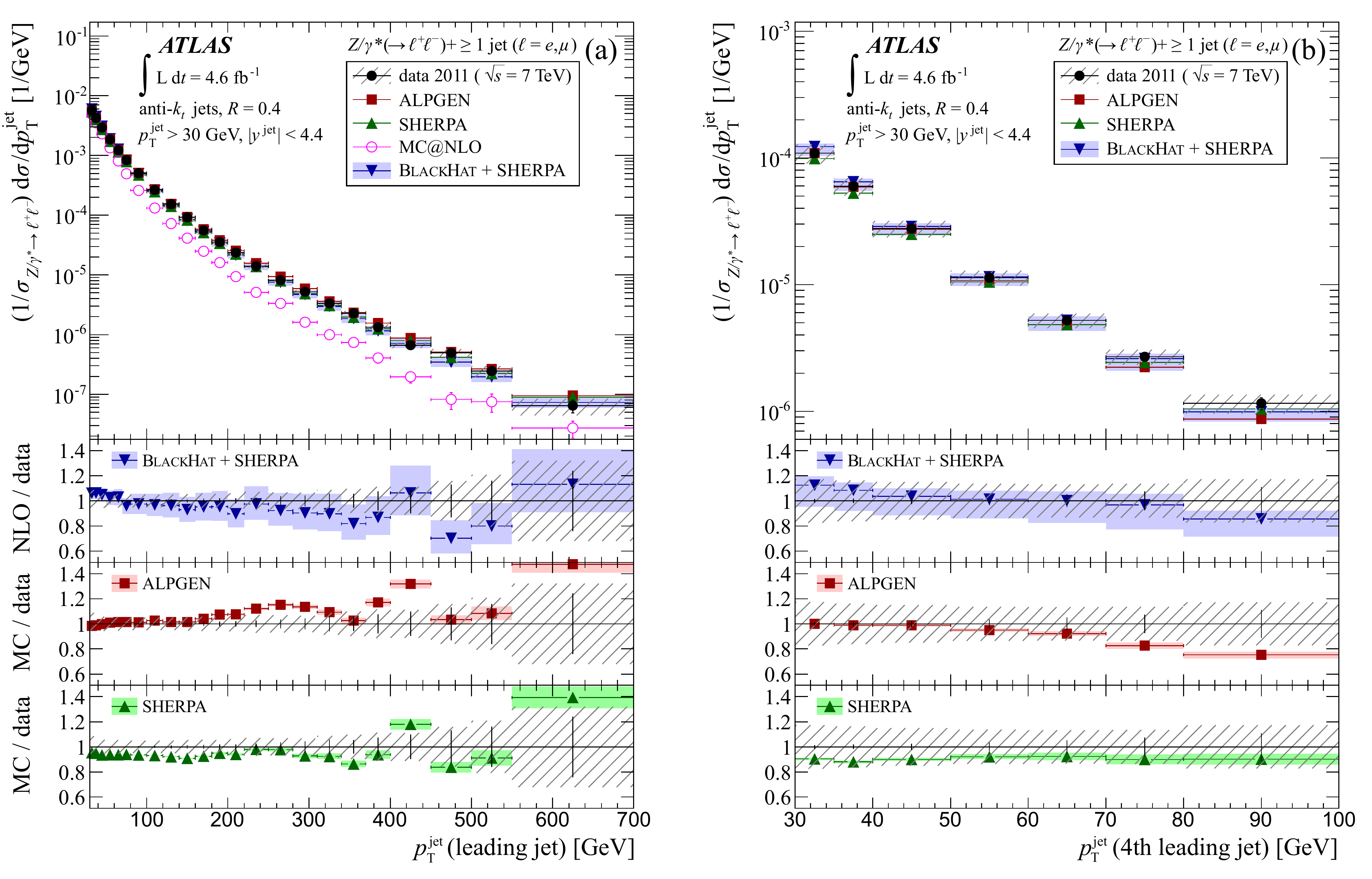}
    \caption{$\Zb+$jets production cross section as a
      function of the leading and fourth-leading jet transverse
      momentum as measured by ATLAS. The data are compared to
      various QCD calculations.
      \textit{(Adapted from Ref.~\cite{Aad:2013ysa}.)}
    \label{fig:qcd:atlas_zbosonjet}}
  \end{center}
\end{figure}

Similar measurements~\cite{Aad:2013ysa} for the $\Zb+$jets processes using the
full data set at $\sqrt{s} = \unit{7}{\TeV}$ are presented in
\fig{\ref{fig:qcd:atlas_zbosonjet}} as a function of the leading-jet \pT
and the fourth-leading jet \pT. Jets are defined using the anti-\ktalgo jet
algorithm with $R=0.4$ and are selected if $\pT > \unit{30}{\GeV}$ and $|y|<4.4$.
The cross sections are normalised to the inclusive \Zb cross section.
Again, the \BLACKHATSHERPA predictions are in good agreement with both
measured distributions.

An NLO QCD calculation for the inclusive \Zb cross section merged with
parton showers and hadronisation as incorporated in \MCATNLO is
capable of describing the data for events with one or two low-\pT
jets, but falls far below the data towards high values of the
leading-jet \pT. This indicates that the fraction of events exhibiting
a second jet increases considerably with the \pT of the leading jet
and that the parton shower approximation, which is used by \MCATNLO for
the second jet, fails to describe the data.

The results are also compared to two multi-leg Monte Carlo simulations
produced with up to five additional partons. The \ALPGEN simulation
overestimates the data at high \pT for the leading jets and underestimates
the data for the fourth-leading jets. \SHERPA is in good agreement with
the data considering the experimental uncertainties.
The measurements of the $\Zb+$jets cross section are limited
by the experimental uncertainty on the jet energy measurements.
By considering jet-multiplicity ratios, many experimental and theoretical
uncertainties cancel, allowing for an even more precise data-to-theory comparison.

\begin{figure}[htbp]
  \begin{center}
    \includegraphics[width=0.95\textwidth]{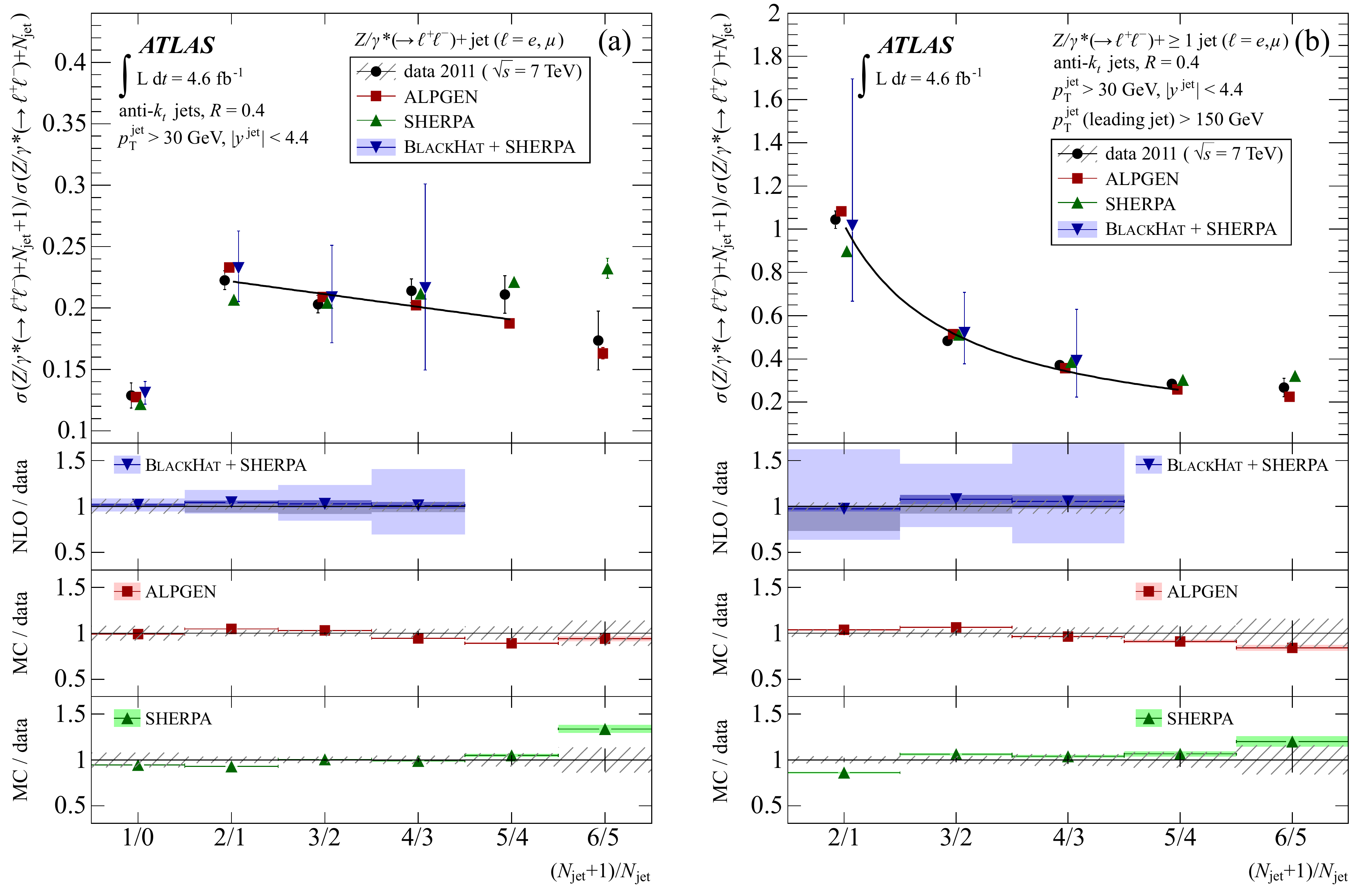}
    \caption{Ratios of $\Zb$+jets cross sections for successive
      exclusive jet multiplicities
      for (a) inclusive events
      and (b) for events with at least one jet
      with $\pT > \unit{150}{\GeV}$ and $|y_{\rm jet}|<4.4$.  The data are
      compared to various QCD calculations.
      \textit{(Adapted from Ref.~\cite{Aad:2013ysa}.)}
    \label{fig:qcd:atlas_zbosonjetscaling}}
  \end{center}
\end{figure}

The ratio of the $\Zb+$jet cross sections for $N_{\text{jet}}+1$ over $N_{\text{jet}}$
($R_{(N+1)/N} = \sigma(\Zb+N_{\text{jet}}+1)/\sigma(\Zb+N_{\text{jet}})$) was measured
by ATLAS~\cite{Aad:2013ysa}. An earlier measurement was
performed by the CMS experiment on a smaller data set~\cite{Chatrchyan:2011ne}.
Figure~\ref{fig:qcd:atlas_zbosonjetscaling} shows the ratio $R_{(N+1)/N}$ for successive
exclusive jet multiplicities for all events (``inclusive'') and for events with at least
one jet with $\pT >\unit{150}{\GeV}$.
For the inclusive case, the ratio shows a rather flat, approximately linear
dependence on the jet multiplicity with a small slope only, while for events
with $\pT > \unit{150}{\GeV}$ the measured distribution steeply rises towards low jet
multiplicities. These measurements illustrate two limiting cases
of scaling patterns, the ``staircase scaling''~\cite{Berends:1989cf} with
$R_{(N+1)/N}$ constant and the ``Poisson scaling''~\cite{Gerwick:2012hq}
with $R_{(N+1)/N} = \av{N}/N$, where $\av{N}$ is the average
number
of jets.
These scaling patterns can be qualitatively understood from the expected
Poisson-distributed jet multiplicity.
For large jet-\pT differences,
the Poisson-scaling is directly observed (see \fig{\ref{fig:qcd:atlas_zbosonjetscaling}}(b)).
However, for low \pT and small $N_{\text{jet}}$, the emission
of additional partons is suppressed by the PDFs, and for high $N_{\text{jet}}$ the
emission of additional partons no longer follows a Poisson distribution due to
the non-Abelian nature of QCD, leading to a proliferation of the
number of jets
originating from gluon splitting (see \fig{\ref{fig:qcd:atlas_zbosonjetscaling}}(a)).

The scaling patterns observed in data are described by the Monte Carlo
simulations \BLACKHATSHERPA, \ALPGEN and \SHERPA. The scale uncertainty is
only shown for the NLO QCD calculation based on \BLACKHATSHERPA. The
different grey shades indicate the scale uncertainty when it is considered
uncorrelated (as proposed in Ref.~\cite{Stewart:2011cf}) or correlated across
the various jet multiplicities. Assuming the predicted and now observed
scaling patterns to be valid, it becomes possible to give
estimates for even higher jet multiplicities where there exists no
complete NLO prediction yet~\cite{Bern:2014voa}.

%% file: Vplusb.tex
\subsection{Weak Bosons and Jets with Flavour}\label{sec:qcd:Vplusb}

\begin{figure}[htbp]
  \begin{center}
    \includegraphics[width=0.7\textwidth]{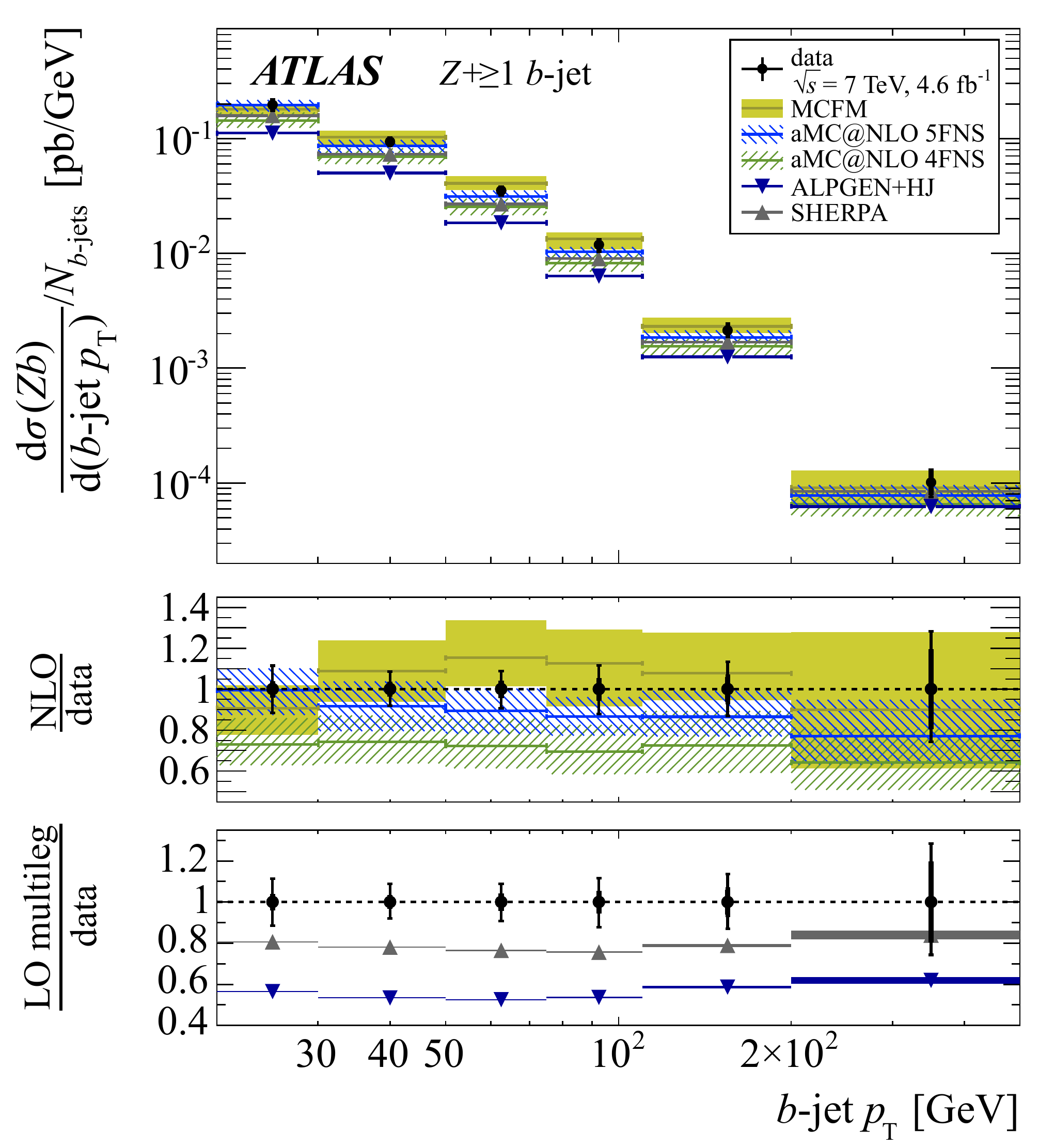}
    \caption{Production cross sections for \Zb bosons produced in
      association with jets with heavy-flavour content as a function
      of the jet \pT.  Superimposed are various QCD calculations.
      \textit{(Adapted from Ref.~\cite{Aad:2014dvb}.)}
    \label{fig:qcd:atlas_ZHF}}
  \end{center}
\end{figure}

Several key processes in the Standard Model and beyond---such as the production of
single top quarks, top-quark pairs or Higgs bosons---involve heavy quarks. The
understanding of their production mechanisms is therefore an important part of
the LHC physics programme. For a detailed discussion of flavour physics
at the LHC see Chap.~8 of Ref.~\cite{thisbook}.

The predictions of processes with heavy-flavour content are more difficult to
accomplish than those for the inclusive case. Bottom ($\bq$) quarks are significantly heavier
than the proton, and in strong interactions at high scales they can
only be created in pairs. The calculations can be classified in two types: In the
4-flavour (4F) scheme heavy quarks appear only in the final state and are typically
considered massive, while 5-flavour (5F) scheme calculations include heavy
quarks in the initial state, i.e.\ as present in the proton. For this purpose, \cq-quark
and \bq-quark PDFs are used.

The production of \Wb and \Zb bosons in association with heavy flavour is
an important background for many processes. Several NLO QCD calculations
are available~\cite{Giele:1995kr,Ellis:1998fv,Campbell:2000bg,Campbell:2003dd,
Maltoni:2005wd,Campbell:2005zv,Campbell:2008hh,Cordero:2009kv,Badger:2010mg,
Oleari:2011ey,Frederix:2011qg}. Recently, the $\Wb+\bbbar$ and $\Zb+\bbbar$
processes (with and without additional jets) have been included in
automated NLO calculations matched with parton showers and
hadronisation~\cite{Alwall:2014hca}.

\Zb bosons provide a particularly clean experimental signature that allows
for precise measurements. Figure~\ref{fig:qcd:atlas_ZHF} presents a recent
ATLAS measurement  of the cross section
of \Zb bosons produced in association with \bq jets as a function of the \pT
of the leading jet containing a \Bmeson hadron.
The measurements uses the full 2011 data set.
Transverse jet momenta up to
\unit{500}{\GeV} are reached. The data are compared to various QCD calculations.

The \MCFM prediction is based on a 5F NLO QCD calculation of
$\Zb+1$\bq jet~\cite{Campbell:2003dd,Campbell:2005zv}, corrected for hadronisation and
effects of multi-parton interactions
(see \sect{\ref{sec:MPI}}).
Full particle-level predictions with
NLO QCD matrix element calculations are also obtained using
\MADGRAPHFIVE~\cite{Alwall:2014hca} in both the 4F and 5F schemes.
For these calculations, the renormalisation and factorisation scales are set
to the transverse \Zb-boson mass
and varied up and down independently by a
factor of two to assess the residual dependence on this scale choice.
The data are described within the experimental and theoretical uncertainties.

Also shown are the predictions for multi-leg LO QCD calculations based on
\ALPGEN using the 4F scheme with up to five partons in the matrix element
and \SHERPA using the 5F scheme with up to four partons.
The shape of these predictions is in good agreement with the data. However,
the LO normalisation of these predictions undershoots the measurement,
\SHERPA reproducing the normalisation better than \ALPGEN.

Further measurements by ATLAS and CMS of vector-boson production in association with heavy-flavour
jets can be found in Refs.~\cite{Aad:2011jn,Chatrchyan:2014dha,Chatrchyan:2013zja}.

%% file: resummation-fragmentation.tex
\section{Resummation: The Realm of Large Logarithms}\label{sec:qcd:resumfrag}

When discussing the process of vector-boson production before, the
necessity to incorporate effects of multiple emissions of initial-state and
final-state partons to appropriately model certain observables has
been touched upon already. In fact, phase-space regions sensitive to the
emission of soft and/or collinear QCD quanta often resist a
satisfactory description in terms of fixed-order perturbation theory.
Rather, one is forced to identify the relevant enhanced contributions,
namely large logarithmic terms, and to reorganise the perturbative expansion
for the observable at hand. The aim is to account for the
enhanced terms to all orders in perturbation theory. This procedure is
referred to as ``resummation'' and allows an appropriate treatment
in a wide range of perturbative QCD phenomena.

Examples for hadron-hadron collider observables sensitive to
multiple-emission effects include the transverse-momentum distribution of
lepton pairs or gauge bosons at
low \pT~\cite{Balazs:1997xd,Becher:2011xn,Banfi:2012du}, event-shape
variables~\cite{Banfi:2004nk,Banfi:2010xy,Stewart:2010tn}, or certain
jet-substructure observables~\cite{Feige:2012vc,Gerwick:2012fw,Larkoski:2013eya}.
Furthermore, resummation techniques become important when the
real-emission phase space is constrained, leaving behind
finite but possibly large uncancelled QCD corrections. Examples
thereof are observables that exhibit
thresholds~\cite{Idilbi:2005ky,Becher:2007ty,Bonvini:2010tp,Cacciari:2011hy}
or that rely on an explicit veto on QCD
activity~\cite{Forshaw:2009fz,DuranDelgado:2011tp,
  Banfi:2012yh,Banfi:2012jm,Tackmann:2012bt}. Resummed expressions for
this type of observables might enable us to extract QCD
parameters such as the strong coupling, quark masses and the parton
distribution functions much more accurately than when having to rely on
fixed-order perturbation theory alone. In the following, two concrete examples shall
be considered: the application of a veto on additional jet activity
and the jet-mass distribution in $\Zb+$jet production.

\subsection{Jet Vetos and Gap Fractions}

At first, so called ``gap fractions''
will be
discussed.  For this class of observables, used also to select the
weak-boson-fusion production channel of Higgs bosons (see Chap.~6 of Ref.~\cite{thisbook}),
one considers a specific kinematic structure of two hard and
widely separated jets.  Analysing the QCD activity in-between these
hard jets, events are vetoed in case there are additional jets with
transverse momentum above a certain veto scale $Q_0$, filtering out
those events that exhibit a gap between the two boundary jets.
On the one hand, the presence of the potentially soft jet-veto
scale $Q_0$ induces large logarithms with argument $Q/Q_0$, where $Q$
denotes the typical hard process scale, e.g.\ the boundary jets'
transverse momenta. This demands for the all-orders treatment of
wide-angle soft-gluon radiation~\cite{Forshaw:2009fz,DuranDelgado:2011tp}.
On the other hand, when considering large rapidity separations $\Delta y$ between
the boundary jets, manifestations of BFKL-like
dynamics~\cite{Kuraev:1977fs,Balitsky:1978ic} are expected, requiring to sum terms
proportional to $\alpS^n(\Delta y)^n$ to all orders
$n$~\cite{Forshaw:2005sx,Andersen:2009nu}.
In Refs.~\cite{Aad:2011jz,Aad:2014pua}
ATLAS presented measurements of dijet production with a veto on additional jets.
In order to gain further insight into the QCD dynamics determining the gap-fraction
and gap-jet measurements, Ref.~\cite{Aad:2014pua} considers additional
azimuthal-decorrelation observables. In this latter analysis,
the data are compared to
predictions from \POWHEG~\cite{Alioli:2010xd,Alioli:2010xa}, interfaced
to the \PYTHIAEIGHT and \HERWIG event generators, as well as to the program
\HEJ~\cite{Andersen:2011hs} with and without invoking the \ARIADNE
shower model~\cite{Andersen:2011zd}. The two \POWHEG simulations
provide NLO accuracy for the inclusive dijet production process
invoking leading-logarithmic DGLAP
resummation through the parton
showers attached.
The \HEJ approach provides a resummation of small-$x$,
BFKL-type logarithmic terms that can be supplemented with DGLAP
resummation through, in this case, the \ARIADNE parton cascade.

\begin{figure}[htbp]
  \begin{center}
    \includegraphics[width=0.95\textwidth]{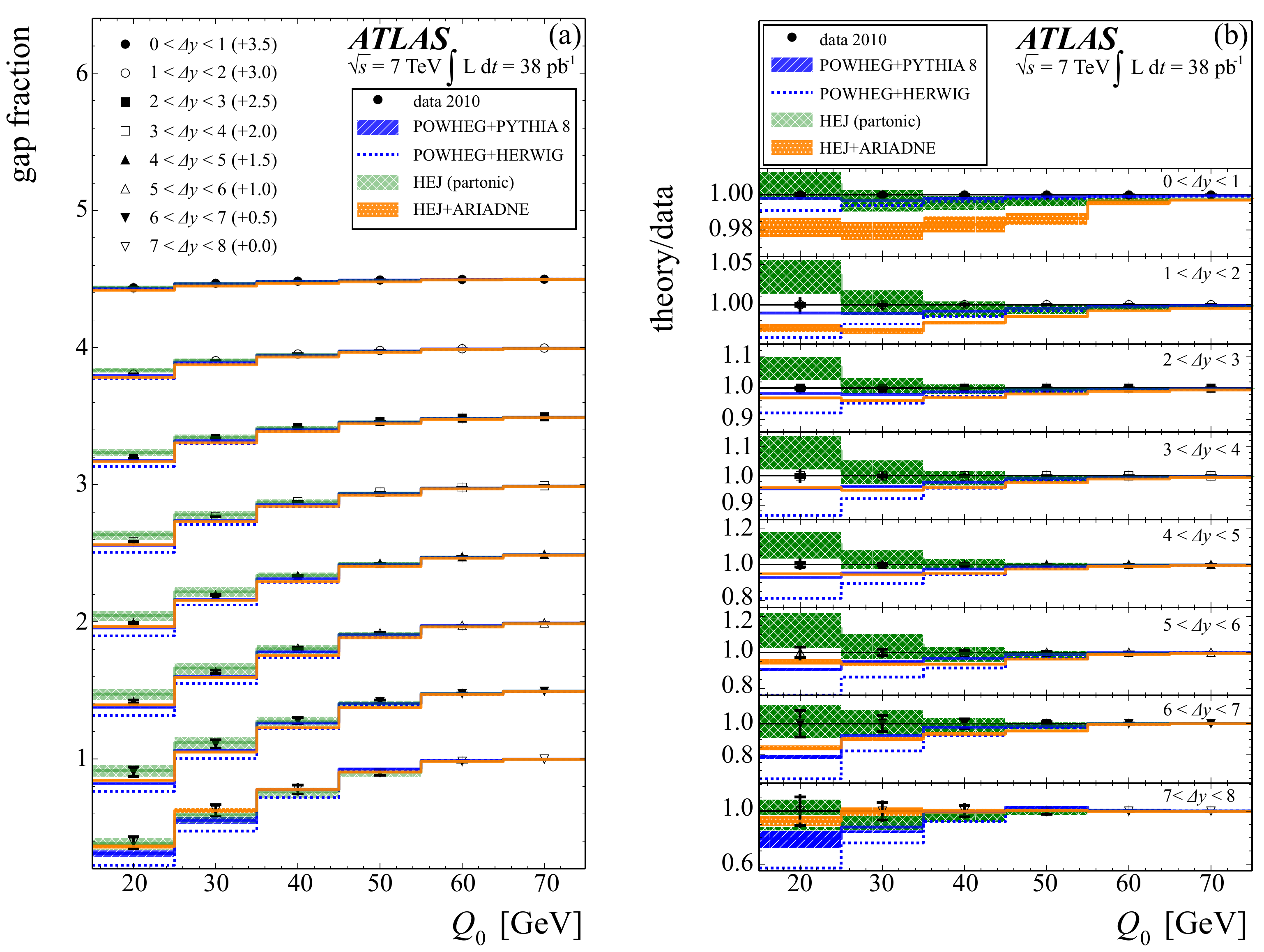}
%
    \caption{Gap fraction for dijet events as a function of the jet-veto scale $Q_0$
    for various $\Delta y$ ranges. The data are
      compared to predictions from \POWHEG+\PYTHIAEIGHT, \POWHEG+\HERWIG and
      from the generator \HEJ with and without inclusion of the \ARIADNE
      shower algorithm. (a) Direct comparison with the unfolded data
      (for illustrative purposes the various results are shifted by arbitrary constants);
      (b) ratio of theory over data.
      \textit{(Adapted from auxiliary material provided with Ref.~\cite{Aad:2014pua}.)}
          \label{fig:qcd:atlas_gapfractions}}
  \end{center}
\end{figure}

In \fig{\ref{fig:qcd:atlas_gapfractions}} the gap fraction as a
function of the veto scale $Q_0$ is presented for various slices of
the boundary jets' rapidity separation $\Delta y$. The fraction of
events exhibiting a rapidity gap decreases as the veto scale is
lowered or when increasing $\Delta y$. This corresponds to the
expectation of an increase of the jet activity when considering lower
jet transverse-momentum thresholds together with the rise in jet
multiplicity when considering larger rapidity intervals. The
\POWHEG+\PYTHIAEIGHT prediction yields a reasonable description of the
data. Employing the \HERWIG parton shower the agreement worsens, an effect that
originates from a prediction of too many jets above the veto
scale. For the two predictions from \HEJ the agreement with data
improves when considering large rapidity separations.  Given the high
precision of the measured data, no single theoretical prediction is
capable of describing the data in all phase-space regions. Considering
more observables, including angular variables, Ref.~\cite{Aad:2014pua}
establishes the DGLAP-based \POWHEG+\PYTHIAEIGHT and BFKL-like \HEJ+\ARIADNE
simulations to yield the best overall description of the data. So far
no clear-cut statements about the evidence for BFKL effects can be
established. However, the small experimental uncertainties allow
theoretical models for QCD radiation between widely separated and high-transverse momentum jets to be further constrained and improved~\cite{Deak:2011ga,Alioli:2012tp,Hoche:2012wh,Gerwick:2012hq,Hatta:2013qj}.

\subsection{The Jet-Mass Distribution}

Many phenomenological studies over the past years have investigated
and highlighted the potential of jet-substructure techniques to
identify the hadronic decays of boosted heavy objects or to
discriminate scenarios of new physics from Standard Model (i.e.\ QCD)
backgrounds~\cite{Abdesselam:2010pt,Altheimer:2012mn,Altheimer:2013yza}. The
successful application of jet-substructure methods requires a more
detailed and even finer understanding and modelling of the inner
structure of QCD jets. The new experimental challenges to address
include the consideration of various jet algorithms and radii that
need to be calibrated, the quantification and possible removal of
non-perturbative contributions from underlying event or pile-up
activity, or the application of tracking and flavour-tagging methods
in busy environments.
On the theoretical side, precise and reliable
predictions for the new observables at hand are required. This, in
particular, asks for resummed calculations, in the form of Monte Carlo
parton-shower simulations or as dedicated analytical results.

\begin{figure}[htbp]
  \begin{center}
    \includegraphics[width=0.9\textwidth]{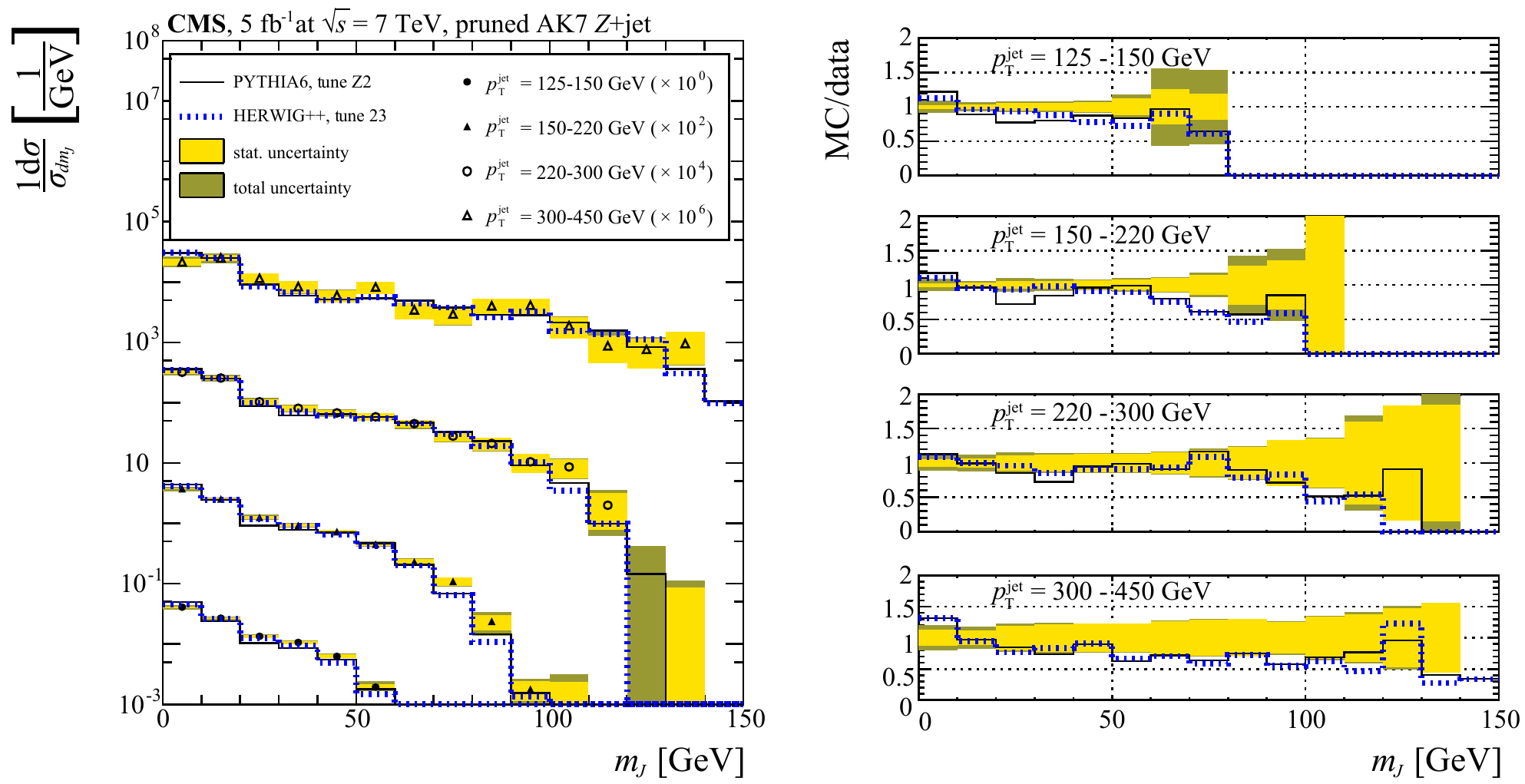}
    \caption{Unfolded jet-mass distribution in $\Zb(\ell\ell)+$jet
      events. Jets are reconstructed using the anti-\ktalgo jet algorithm with
      distance parameter $R=0.7$ and are groomed using the pruning
      technique~\cite{Ellis:2009me}. The data from CMS are compared to
      Monte Carlo predictions from \MADGRAPH+\PYTHIASIX (solid) and
      \HERWIGPP (dotted). The right panel shows the ratio of Monte
      Carlo over data.
      \textit{(Adapted from Ref.~\cite{Chatrchyan:2013vbb}.)}
    \label{fig:qcd:cms_ak7pr_jetmass}}
  \end{center}
\end{figure}

ATLAS and CMS have presented several analyses that address these
questions and compare their data to theoretical
models~\cite{Aad:2012meb, ATLAS:2012am, Aad:2013ueu, Aad:2014haa,
  Chatrchyan:2012ku, Chatrchyan:2013vbb, Khachatryan:2014hpa,
  Khachatryan:2014gha}.
One particular and rather simple observable shall be considered as an
example here: the invariant mass of jets---the jet-mass
distribution.
The jet-mass variable
probes the composition of jet objects and thus depends on the
production process, the jet algorithm and its parameters used to
cluster the jets, and on additional selection or filtering
steps applied to the final states.  The aim of
``jet grooming''
procedures is to reduce the impact
of underlying-event and pile-up contributions on the jet properties
and sometimes also to remove certain types of intra-jet parton
splittings, all that in order to maximise the sensitivity to the
desired signal. Examples of jet-grooming procedures include
trimming~\cite{Krohn:2009th},
pruning~\cite{Ellis:2009me}, mass-drop \cite{Butterworth:2008iy,Plehn:2009rk}
or soft-drop~\cite{Larkoski:2014wba} filtering.
Dedicated
theoretical predictions for the jet-mass distribution exist for a
variety of production processes, including dijet final states,
vector-boson plus jet production and Higgs-boson plus jet
production~\cite{Li:2011hy,Li:2012bw,Dasgupta:2012hg,Jouttenus:2013hs}.
The mass distribution of QCD jets after the application of grooming
techniques has recently been discussed in
Refs.~\cite{Dasgupta:2013ihk,Dasgupta:2013via}.

In Ref.~\cite{Chatrchyan:2013vbb} the CMS collaboration has presented an
analysis of the jet-mass observable in dijet and $\Wb/\Zb+$jet events,
focusing on the effects of various grooming techniques and comparing
data to parton-shower Monte Carlo predictions. For an integrated
luminosity of \unit{5}{\invfb} of \unit{7}{\TeV} collision data they
considered both the anti-\ktalgo and Cambridge--Aachen~\cite{Dokshitzer:1997in,Wobisch:1998wt,Wobisch:2000dk}
jet algorithms
using various distance parameters, namely $R=0.7$ for anti-\ktalgo and
$R=0.8$ and $R=1.2$ for Cambridge--Aachen. The groomed and ungroomed
data are compared to Monte Carlo predictions from \PYTHIASIX and
\HERWIGPP. As an example, the unfolded jet-mass distribution of
$\Zb(\ell\ell)+$jet events is shown in \fig{\ref{fig:qcd:cms_ak7pr_jetmass}}.
Depicted there is the jet-mass
variable for anti-\ktalgo jets after a pruning procedure, analysed for
four slices in jet transverse momentum (note the arbitrary scaling
factors). For all the jet-\pT slices both generators describe
the data well. A similar level of congruence is observed for the
alternative grooming techniques and combinations of jet algorithm and
distance parameter. Somewhat larger deviations between data and
Monte Carlo estimates are observed for regions of small jet mass.
However, it is in particular this region of phase space that is most
affected by non-perturbative corrections from pile-up, underlying
event and hadronisation. Attempts to analytically estimate the effects
of the underlying event and parton-to-hadron fragmentation on the
jet-mass distribution have been presented in Ref.~\cite{Dasgupta:2007wa}.

From the analysis presented by CMS and a similar one presented by
ATLAS~\cite{ATLAS:2012am}, agreement of jet-mass
and substructure observables with modern parton-shower Monte Carlo
tools is established. This paves the way to successfully use these techniques in
future analyses and particularly in searches for new phenomena.

%% file: beyond-pqcd.tex
\section{Beyond Perturbative QCD}\label{sec:qcd:npqcd}

So far the focus has been mainly on measurements that can reliably be
addressed using QCD perturbative techniques.  Experimentally, however,
hadrons are observed in the detectors, and accordingly the transition
of partons into hadrons needs to be accounted for. Furthermore, there
is a wealth of effects that require the use of non-perturbative
methods and models---most prominently the so-called ``underlying
event''
that accompanies a hard scattering.
As an example, the underlying event includes contributions from multiple interactions of the
colliding protons' constituents (``multi-parton interactions'') that give rise to the production of
many additional final-state particles.

In this section these aspects shall be exemplified by presenting LHC
measurements of jet-shape variables, of jet-radius ratios and of
observables sensitive to soft-particle production and double and
multi-parton scattering.

\subsection{Jet Shapes}

\begin{figure}[htbp]
  \begin{center}
    \includegraphics[width=1.00\textwidth]{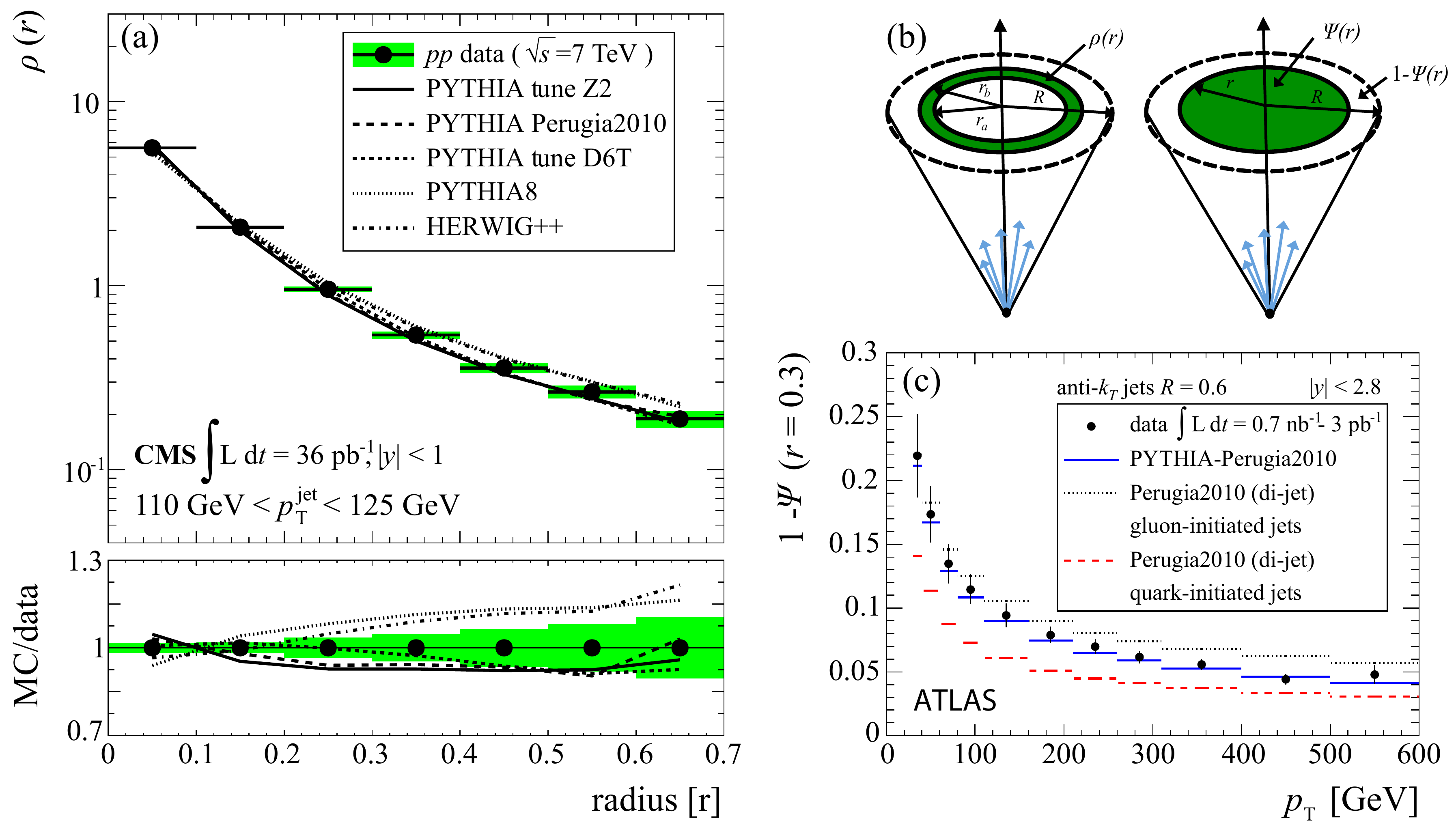}
  \caption{(a) Differential jet shape $\rho(r)$ as a function of the distance $r$ from the jet
    axis for jets with $110 < \pT < \unit{125}{\GeV}$ from
    CMS. The data are compared to various tunes of
    the \PYTHIASIX and to the \PYTHIAEIGHT and \HERWIGPP event generators.
    (b) Illustration of the jet shape observables.
    (c) Integrated jet shape $1-\Psi(r=0.3)$ as a function of the
    jet \pT from ATLAS~\cite{Aad:2011kq}. The data are compared to the prediction
    for gluon-initiated and quark-initiated jets using the \PYTHIASIX MC event
    generator with tune Perugia2010.
    \textit{(Adapted from Refs.~\cite{Chatrchyan:2012mec,Aad:2011kq}.)}
    \label{fig:qcd:jetshapes}}
\end{center}
\end{figure}

Focusing on the internal structure of jets, the profile of the
distribution of transverse momentum within a jet can be examined.  For
this type of observable the term ``jet shape'' has been
coined~\cite{Ellis:1992qq,Abe:1992wv}. The differential jet shape
$\rho(r)$ as a function of the distance
$r_i=\sqrt{(\Delta_{i,\text{jet}}y)\squared+(\Delta_{i,\text{jet}}\phi)\squared}$
to the jet axis is defined as the average fraction of jet \pT
contained inside an annulus of inner radius $r_a = r-\Delta r/2$ and
outer radius $r_b = r+\Delta r/2$ for an ensemble of $N$ jets:

  \begin{equation*}
    \rho(r) = \frac{1}{N}\sum_{\text{jets}}
    \frac{1}{\Delta r} \frac{\sum_{r_a \leq r_i < r_b}\pTi}
    {\sum_{r_i \leq R}\pTi}\, ,
  \end{equation*}

where the second sum runs over all jet constituents $i$. The
integrated jet shape $\Psi(r)$ is then given by the integral of the
differential jet shape up to a radius $r$ (see the illustrations in
\fig{\ref{fig:qcd:jetshapes}}(b)). Conventionally, measurements are
presented in terms of $1-\Psi(r=r_\text{core})$, where $r_\text{core}$
is usually taken to be 0.3~\cite{Acosta:2005ix}.

Perturbatively, the internal structure of jets is determined by
multiple emissions of gluons and depends on the type of initiating
parton, i.e.\ a quark or gluon.  Perturbative QCD predicts that
gluon-initiated jets are broader in shape with a higher particle
multiplicity on average~\cite{9780521581899}. To some extent it is
therefore possible to differentiate statistically between gluon and
quark jets by looking into the jet substructure. Jet shapes, however,
are also sensitive to non-perturbative effects of the parton
fragmentation and the underlying event. Corresponding measurements are
valuable inputs to improve the modelling of such soft contributions in
MC event generators like \PYTHIA, \HERWIGPP, or \SHERPA.

Figure~\ref{fig:qcd:jetshapes}(a) shows a CMS
measurement~\cite{Chatrchyan:2012mec} of the differential jet shape,
for the example of the jet \pT interval $110 < \pT <
\unit{125}{\GeV}$, in comparison to several LO MC event generators at
particle level.  While \HERWIGPP and \PYTHIAEIGHT predict somewhat
broader jet shapes than measured, the \PYTHIASIX tunes D6T and Z2
deviate in the opposite direction. The best description of the data is
given by \PYTHIASIX with the tune Perugia2010, something, which is
also observed in a similar investigation by the ATLAS
collaboration~\cite{Aad:2011kq}.  By employing this tune for the
integrated jet shape, shown in the form of $1-\Psi(r=0.3)$ as a
function of the jet \pT in \fig{\ref{fig:qcd:jetshapes}}(c), ATLAS
demonstrates the sensitivity of this observable with respect to the
jet-initiating parton. At small jet \pT the examined inclusive jet
sample is predominantly composed of gluon jets, while with increasing
jet \pT the quark-initiated component grows. In the figure, the change
in the fraction of quark-initiated to gluon-initiated jets is averaged
over the jet rapidity up to $|y| < 2.8$. If studied
double-differentially, a mild dependence on $|y|$ of this fraction is
observed as expected. The sensitivity of jet shapes to
non-perturbative effects and modelling uncertainties, however,
prevents an extraction of the quark-gluon jet fraction.  Instead, they
provide valuable input to the tuning of these effects in MC event
generators.

In the context of jets initiated by heavy boosted objects,
the investigation of jet substructure has
attracted much attention lately. Potential new particles like \Zprime
bosons or heavy $\tq^\prime$ quarks that decay hadronically may be at
the origin of jets with two or three, respectively, high-energetic
partons fragmenting into only one jet. The required boosts for these
hypothetical particles are attainable at the LHC, in particular at
\unit{13}{\TeV} centre-of-mass energy as foreseen for the restart in
2015\@. Searches for these particles profit from a better
understanding of the QCD background in the form of quark-initiated or
gluon-initiated jets. For some of the investigated observables, e.g.\
jet mass, resummed predictions are available (see
\sect{\ref{sec:qcd:resumfrag}}).

\subsection{Jet-Radius Ratio}

As described in \sect{\ref{sec:qcd:nlo}}, the inclusive jet cross
section can precisely be compared to NLO predictions, and the strong
coupling constant can be extracted. A small correction for
non-perturbative effects, which is \pT-dependent and increases at
small \pT values, has to be taken into account. Looking more closely
into the results presented by ATLAS for anti-\ktalgo jet sizes of $R=0.4$
and $0.6$~\cite{Aad:2011fc} or by CMS for anti-\ktalgo jet sizes of
$R=0.5$ and $0.7$~\cite{Chatrchyan:2011ab, Chatrchyan:2012bja}, it is
observed that the relative normalisation of the measured cross
sections to the theoretical predictions exhibits a dependence on $R$.
Theoretically, this $R$ dependence has been examined in
Refs.~\cite{Dasgupta:2007wa, Cacciari:2008gd}, where in a collinear
approximation it was found that the impact of perturbative radiation
and of the non-perturbative effects of hadronisation and the
underlying event on jet transverse momenta scales roughly with $\ln
R$, $-1/R$, and $R^2$ for small $R$, respectively. By choosing the jet
size parameter $R$, one can therefore steer which aspects of jet
formation are emphasised in a jet analysis. In
Ref.~\cite{Dasgupta:2007wa} it is suggested, in order to gain insight
into the interplay of these effects, to study the relative difference
between inclusive jet cross sections that emerge from two different
jet definitions:

  \begin{equation*}
    \left. \left( \frac{\dif \sigma^\text{alt}}{\dif \pT} -
      \frac{\dif\sigma^\text{ref}}{\dif\pT} \right)
    \middle/ \left( \frac{\dif\sigma^\text{ref}}{\dif\pT} \right) \right. =
    \mathcal{R}(\text{alt},\text{ref}) - 1\, .
    \label{qcd:eq:rdef}
  \end{equation*}

Provided that partons in opposite hemispheres are not clustered
together (a condition that is usually fulfilled), the LO two-parton cross sections are identical for
arbitrary jet algorithms. Therefore, only partonic final states with
three or more partons lead to a numerator different from zero.  Hence,
\eqn{\eqref{qcd:eq:rdef}} defines a three-jet observable,
$\mathcal{R}$, for which it was shown in Ref.~\cite{Soyez:2011np} that
it is calculable to NLO with terms up to $\alpS^4$ with
\NLOJETPP~\cite{Nagy:2001fj,Nagy:2003tz}.

\begin{figure}[htbp]
  \begin{center}
    \includegraphics[width=0.95\textwidth]{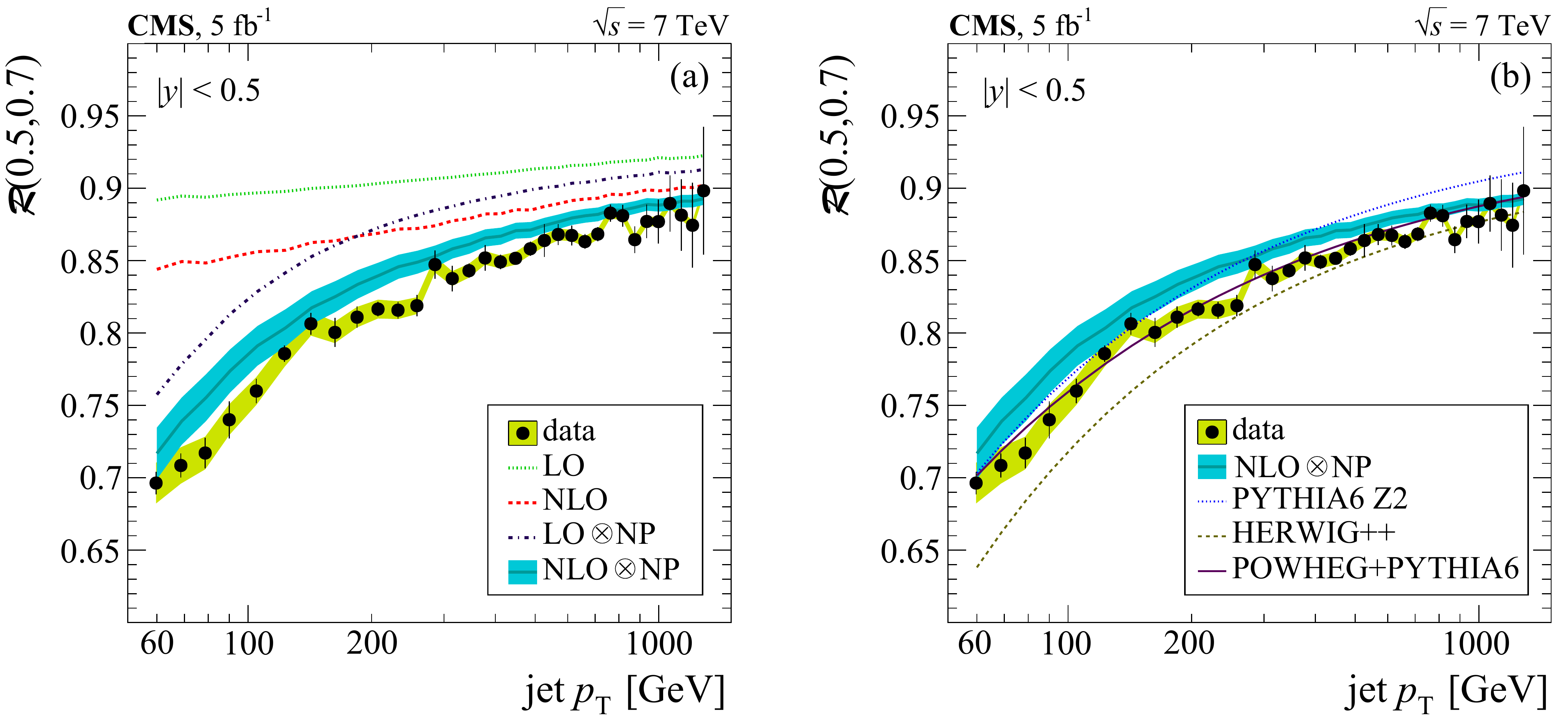}%
    \caption{Jet-radius ratio $\mathcal{R}$ as measured by
      CMS in comparison to (a) fixed-order
      predictions with and without non-perturbative corrections and to (b)
      particle-level predictions of LO and NLO event generators with
      matched parton showers and modelling of hadronisation and the
      underlying event at central rapidity $|y| < 0.5$.
      \textit{(Adapted from auxiliary material provided with Ref.~\cite{Chatrchyan:2014gia}.)}
      \label{fig:qcd:radiusratio}}
  \end{center}
\end{figure}

Studies in that direction have been performed at the HERA collider by
the ZEUS collaboration for two different jet
algorithms~\cite{Abramowicz:2010ke} and by the ALICE experiment for
the two different anti-\ktalgo jet sizes of $R=0.2$ and
$0.4$~\cite{Abelev:2013fn}. The results of a CMS analysis for their
two jet sizes of 0.5 and 0.7 is presented in
Ref.~\cite{Chatrchyan:2014gia} as a function of jet \pT and rapidity
$y$. It is expected that QCD radiation reduces this jet-radius ratio
$\mathcal{R}$ below unity and that this effect disappears with
increasing collimation of jets at high \pT.
Figure~\ref{fig:qcd:radiusratio} confirms this assumption with a
comparison of the measured ratio $\mathcal{R}$ as a function of the
jet \pT up to the \TeV\ scale to theoretical predictions at central
rapidity $|y| < 0.5$. Figure~\ref{fig:qcd:radiusratio}(a) clearly
demonstrates that fixed-order calculations up to NLO, even when
combined with non-perturbative corrections, are systematically above
the data. The LO event generators \PYTHIASIX and \HERWIGPP lie
systematically on either side of the data, as can be seen in
\fig{\ref{fig:qcd:radiusratio}}(b). Presumably, this can be improved
by including this observable into the tuning of these generators. The
best description is given by \POWHEG that matches the dijet production
process evaluated at NLO with the parton showers and non-perturbative
models of \PYTHIASIX, emphasising the importance of parton showers for
the given choice of jet sizes.

\input{totalxsection}

\subsection{The Underlying Event and Multi-Parton Interactions}
\label{sec:MPI}

An important aspect of hadron-hadron collisions, neglected in our
discussion so far, is the observation that additional mostly
low-energetic (soft) particles are produced. The simple picture of a
hard scattering between one parton from each hadron with subsequent
showering and hadronisation at least needs to be complemented by a
treatment of the beam-beam remnants, which is beyond the capabilities
of pQCD\@. In addition, the primary interaction may be accompanied by
the production of further particles, a circumstance which is usually modelled in the
form of multi-parton interactions (MPI). The concepts employed for
MPI by \PYTHIA and \HERWIGPP, for example, are described in
Refs.~\cite{Sjostrand:1987su, Sjostrand:2004pf, Sjostrand:2004ef}
and~\cite{Bahr:2008dy}, respectively. A very useful overview of
general-purpose MC event generators including all aspects of event
generation for the LHC is given in Ref.~\cite{Buckley:2011ms}.

An unambiguous association of a specific particle to the reaction from
which it originates is impossible. Therefore, the extra activity in a
hadron-hadron collision cannot be uniquely separated from effects like
initial-state or final-state radiation. One thus conventionally
defines the ``underlying event'' (UE) as everything except for the
hard process of interest. Two strategies are applied to measure the
UE: The first evaluates events that are collected requiring only
minimalistic trigger conditions such that even very soft collisions
are registered. These so-called ``minimum bias'' (MB) measurements are
supposed to be dominated by underlying-event physics.

\begin{figure}[htbp]
  \begin{center}
    \includegraphics[width=0.95\textwidth]{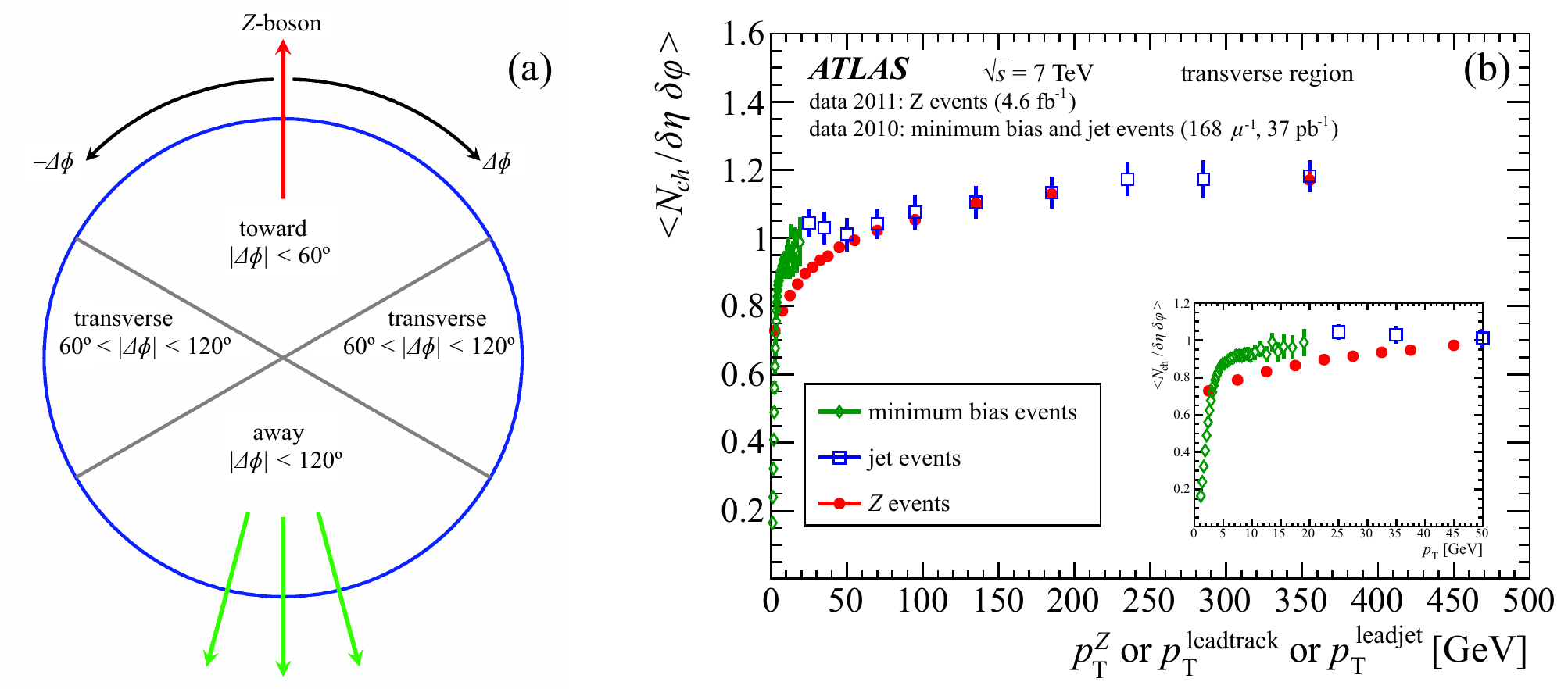}
    \caption{(a) Sketch of the geometrical subdivision of events in
      azimuthal angle for UE measurements.
      (b) Average density
      of charged particles in the transverse region as a function of
      the leading-jet or leading-track \pT, or the \pT of the reconstructed \Zb
      boson.
      \textit{(Adapted from Ref.~\cite{Aad:2014jgf}.)}
      \label{fig:qcd:ue}}
  \end{center}
\end{figure}

In the second approach, reactions are considered that have a high rate
for the production of a leading particle or jet with reasonably high
\pT (for triggering purposes). The event is then separated into hard
and soft contributions through a physically motivated definition of
phase-space regions that are dominated by either the hard or soft
component of a collision. Traditionally, this is achieved by
geometrically subdividing the event into different regions
(``towards'', ``away'', and ``transverse'') with respect to the
direction of the triggering particle or jet
(\fig{\ref{fig:qcd:ue}}(a)). Reactions that are typically chosen for
this second approach are dijet or Drell--Yan lepton-pair production.
Implicitly, it is assumed that the soft particle production is
uncorrelated to details of the hard process---an assumption that can
be tested experimentally.

As an example for such a strategy \fig{\ref{fig:qcd:ue}}(b) shows a
measurement by ATLAS~\cite{Aad:2014jgf} of the charged-particle
density in the transverse region as a function of the transverse
momentum that is chosen as scale of the event hardness. Three event
types are compared: minimum bias, dijet production, and inclusive \Zb
production, with event scales chosen to be the \pT of the leading
track, the leading jet, and the \Zb boson, respectively.  An
approximately continuous transition from MB to jet events is visible
in the inset plot. In the comparison between jet and \Zb events of the
soft-particle density in the transverse region, which is attributed to
the UE, differences are exhibited, which are even more pronounced in
the average summed \pT density of all tracks (not shown). These can
be explained in terms of a bias between the definitions of the event
scale. In the case of the jet events further jets with transverse
momenta larger than the \pT of the leading jet are forbidden by
definition, while for the inclusive \Zb events additional jet activity
with $\pTjet > \pTZ$ is possible. Taking this bias into account, the
conjecture that the UE activity is independent of the hard process is
supported by the ATLAS measurement and by similar measurements from
CMS~\cite{Chatrchyan:2012tb}.

Another quantity that is not predictable by perturbative methods is
the average density of charged particles produced in MB events for
example at central rapidity. Corresponding measurements usually must
be performed early on at the beginning of a data-taking period, when
the instantaneous luminosity is small and the probability of multiple
hadron-hadron collisions in the same or adjacent bunch crossings, the
so-called pile-up, is negligible. It should be emphasised that such
measurements are of utmost importance for a precise understanding of
LHC data as they are used to model individual pile-up collisions.  At
later stages of the data-taking
and with high instantaneous luminosity, the occurrence of up to 10--100
of such pile-up collisions simultaneously to the hard reaction of
interest demands that these MB-like collisions are very well
understood and modelled in simulations. Otherwise it becomes very
difficult to precisely calibrate jet energies (see
Chap.~3 of Ref.~\cite{thisbook})
or to estimate the background to searches
for new physics. The charged-particle density is also of much interest
for heavy-ion physics, and its dependence on the centre-of-mass energy
is shown in Chap.~9, Fig.~9.1, of Ref.~\cite{thisbook}
for various combinations of colliding
hadron-hadron beams.

\subsection{Double-Parton Scattering}

The basis for all perturbative results in the previous sections (see
in particular \sect{\ref{sec:qcd:pqcd}}) is the assumption that
collinear factorisation holds and that the cross section of interest
can be calculated following \eqn{\eqref{eq:qcd:QCDxsec}}.
To some extent it is a great fortune that such a simple picture with only one
interacting parton from each colliding hadron and only longitudinal
degrees of freedom describes such a huge amount of measurements.
Partially this is due to the fact that one is mostly interested in
inclusive processes at large \pT. Looking into the regime of small
transverse momenta, however, it becomes apparent that this picture
must fail, since otherwise the relevant cross sections would grow
beyond all limits for $\pT \to 0$. They would thus violate unitarity,
i.e.\ become even larger than the measured total cross sections. So,
somehow, the cross sections have to be regularised.
Realising that at \pT values of the
order of a few \GeV, protons are probed at fractional momenta of
\power{10}{-3} or below, one can interpret the consequently large
parton densities in such a way that more than one parton per hadron
participates in a scattering process. In terms of the MPI, as
explained above, this concept has already been successfully applied to
model the part of the soft-particle production that manifests itself
in the form of the underlying event. The possibility of two hard or
semi-hard reactions with two partons from each of the colliding
hadrons participating in their own parton-parton scatter has not yet
been considered, though.

This type of process is called ``double-parton scattering'' (DPS), in
contrast to the single parton scattering (SPS) discussed previously.
Double-parton scattering is one possibility to dampen the rise of the
low-\pT SPS cross sections. In a simplified form, the impact of
double-parton scattering on the production rate for a final state
$A+B$ can be parametrised by one effective cross section that can be
written as

  \begin{equation}
    \sigma_{\text{eff}} = \frac{m}{2}\frac{\sigma_\text{A}\cdot\sigma_\text{B}}
    {\sigma^\text{DPS}_\text{A+B}}\, ,
    \label{qcd:eq:sigeff}
  \end{equation}

where $\sigma_\text{A}$ and $\sigma_\text{B}$ are the cross sections
for the independent processes $A$ and $B$,
$\sigma^\text{DPS}_\text{A+B}$ is the cross section for processes $A$
and $B$ to occur simultaneously, and $m$ is a symmetry factor that is
equal to unity for indistinguishable final states and two otherwise.
In this simple picture, $\sigma_{\text{eff}}$ is a measure of the
overlap in transverse size of the parton distributions in the
colliding hadrons. A naive geometrical interpretation leads to an
estimate of $\sigma_{\text{eff}} \approx \pi\R\squared_\text{p}
\approx \unit{50}{\milli\barn}$, where $R_\text{p}$ is the proton
radius.

\begin{figure}[htbp]
  \begin{center}
    \includegraphics[width=0.95\textwidth]{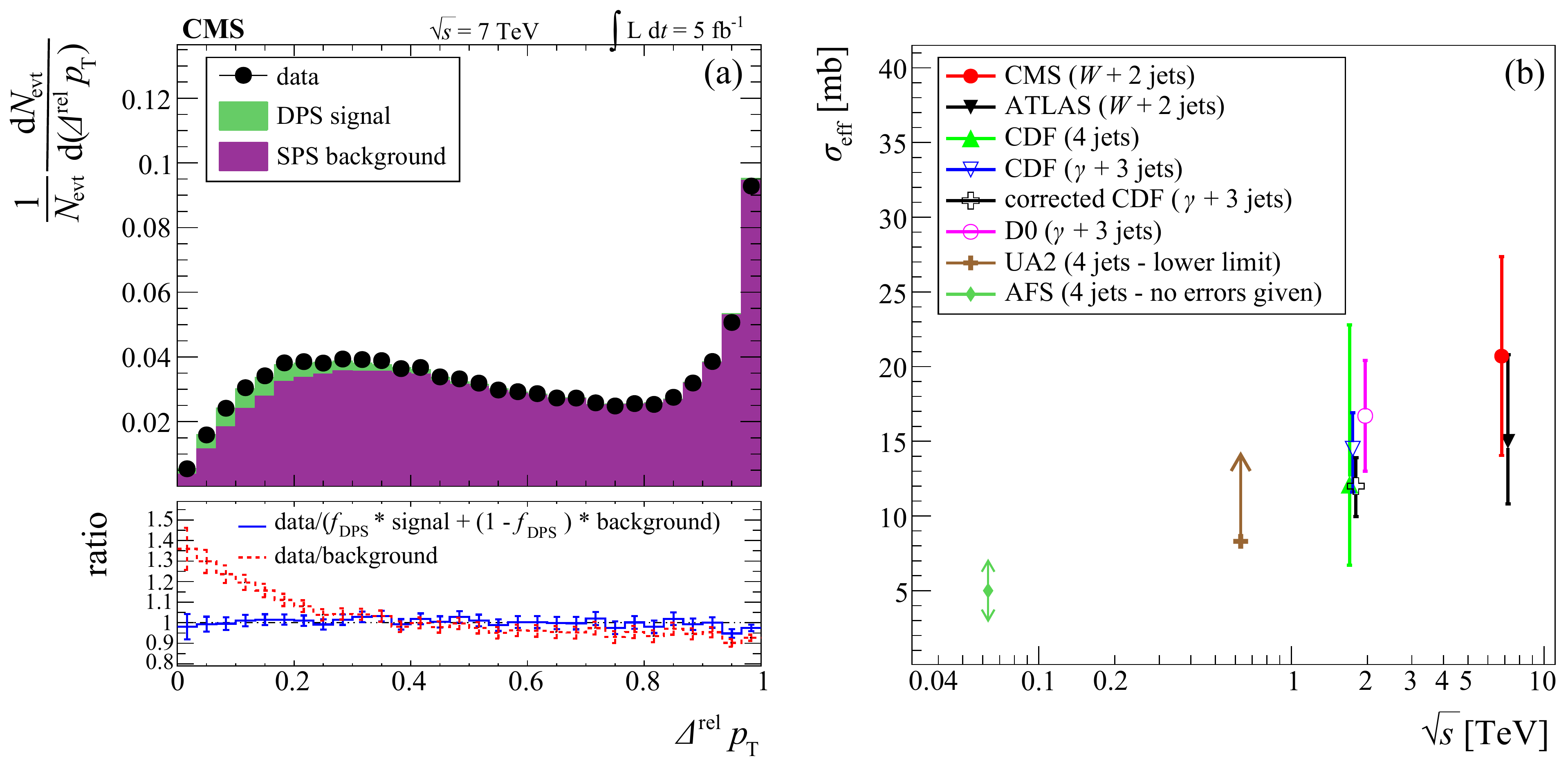}
    \caption{(a) Template fit of SPS background and DPS signal to data for the
      $\Delta^{\text{rel}}\pT$ distribution in $\Wb+$2-jet events as measured by CMS.
      (b) Extracted effective cross section $\sigma_{\text{eff}}$ as a function of the centre-of-mass energy
      as measured by various experiments.
      \textit{(Adapted from Ref.~\cite{Chatrchyan:2013xxa}.)}
      \label{fig:qcd:dps}}
  \end{center}
\end{figure}

Measurements have been performed for a number of processes $A$ and
$B$: 4-jet~\cite{Abe:1993rv, Chatrchyan:2013qza},
$\gamma+$3-jet~\cite{Abe:1997xk, Abazov:2009gc, Abazov:2014fha}, and
$\Wb+$2-jet~\cite{Aad:2013bjm,Chatrchyan:2013xxa} production. In each
case, the second independent scattering reaction is chosen to have a
dijet final state, because dijet production occurs at by far the
highest rates. To differentiate double-parton scattering from normal
high-multiplicity single parton scattering events, the particular
properties of DPS events are exploited: the \pT of the jet pair
(both jet pairs in 4-jet production) should be balanced and the
azimuthal angle between the vectors of the final-state objects of the
leading scattering and the lower-\pT jet pair should be uncorrelated, i.e.\
randomly oriented. Specialising to $\Wb+$2-jet events, observables
sensitive to DPS can then be defined as

  \begin{eqnarray*}
    \Delta^{\text{rel}}\pT & = &\frac{\left|\vec{p}_\mathrm{T}(j_1)+\vec{p}_\mathrm{T}(j_2)\right|}
    {\left|\vec{p}_\mathrm{T}(j_1)\right|+\left|\vec{p}_\mathrm{T}(j_2)\right|}\, ,\\
    \Delta S & = &\arccos\left(
      \frac{\vec{p}_\mathrm{T}(\mu,\ETmiss)+\vec{p}_\mathrm{T}(j_1,j_2)}
      {\left|\vec{p}_\mathrm{T}(\mu,\ETmiss)\right|+\left|\vec{p}_\mathrm{T}(j_1,j_2)\right|}\right)\, ,
  \end{eqnarray*}

where $\vec{p}_\mathrm{T}$ are the transverse momentum vectors of the
jets $j_1$, $j_2$, or of the jet pair $(j_1,j_2)$ and the decay
products of the \Wb boson, one muon and missing transverse energy
\ETmiss representing the undetected neutrino.
Figure~\ref{fig:qcd:dps}(a) shows, as an example, the measurement of
$\Delta^{\text{rel}}\pT$ by CMS~\cite{Chatrchyan:2013xxa} together
with the composition of single parton scattering background and
double-parton scattering signal, which peaks at small \pT differences,
as determined in a template fit.  Figure~\ref{fig:qcd:dps}(b)
summarises determinations of $\sigma_{\text{eff}}$ from various
experiments as a function of the centre-of-mass energy. The
``Corrected CDF'' point includes an improved
estimate~\cite{Treleani:2007gi, Bahr:2013gkj} for the fact that
\eqn{\eqref{qcd:eq:sigeff}} implies processes $A$ and $B$ to be
inclusive such that also higher terms than double scatterings are
parametrised by $\sigma_{\text{eff}}$. The bias of exclusive event
selections, when applying \eqn{\eqref{qcd:eq:sigeffw2jet}} in the
$\Wb+$2-jet case,
  \begin{equation}
    \sigma_{\text{eff}} =
    \frac{\sigma_\text{\Wb+0-jet}}{\sigma^\text{DPS}_\text{\Wb+2-jet}}
    \cdot\sigma_\text{2-jet}\, ,
    \label{qcd:eq:sigeffw2jet}
  \end{equation}
has been considered in the presented analysis from
CMS~\cite{Chatrchyan:2013xxa} and a similar measurement by
ATLAS~\cite{Aad:2013bjm}. Otherwise, $\sigma_{\text{eff}}$ would not
be independent any more of the choice of scattering processes $A$ and
$B$.

The validity of the presented simplified theoretical approach anyway
relies on a couple of additional assumptions. For a recent critical
overview see Ref.~\cite{Diehl:2011yj}.

%% file: totalxsection.tex
\subsection{Soft Hadron-Hadron Collisions}

When two hadrons collide at high energy, their entire volumes or at least a significant part
of them participates in the collision. Most of these collisions are caused by so-called
``soft'' hadronic interactions.

In a simple geometrical model,
the total hadron-hadron collision cross section
is related to the transverse extension of the scattering system.
Therefore, it is expected to be of the order of
$\sigma_{\text{tot}} \approx {\mathcal{O}}( \unit{2}{\fm\squared}) \approx {\mathcal{O}}(\unit{40}{\invmb})$.
Since the effective range of soft hadron interactions does not change much with energy, only
a weak energy dependence is expected.
Moreover, if the collision is soft, the scattering cross section at high energy should not
depend on the quark structure of the colliding hadrons.

In 1952 Heisenberg used a geometrical model where hadrons are modelled as
Lorentz-contracted disks with a given energy density. The interaction
takes place in the disk-overlap region, and only if there is sufficient
energy to create two mesons. In this model the total cross section increases like
$\ln^2{s}$. A consequence of this model is that
the interaction cross sections of all hadrons are similar at high energy.
For a recent review
on the model giving a $\ln^2{s}$ high-energy behaviour
see Ref.~\cite{Dosch:2002pg}.

General arguments based on unitarity, analyticity and factorisation imply a bound
on the high-energy behaviour of total hadronic cross sections.
This bound is independent of the details of
the strong-interaction dynamics and states that the total
cross section can not rise faster than $\ln^2{s}$~\cite{Froissart:1961ux}.

\begin{figure}[htbp]
  \begin{center}
    \includegraphics[width=0.85\textwidth]{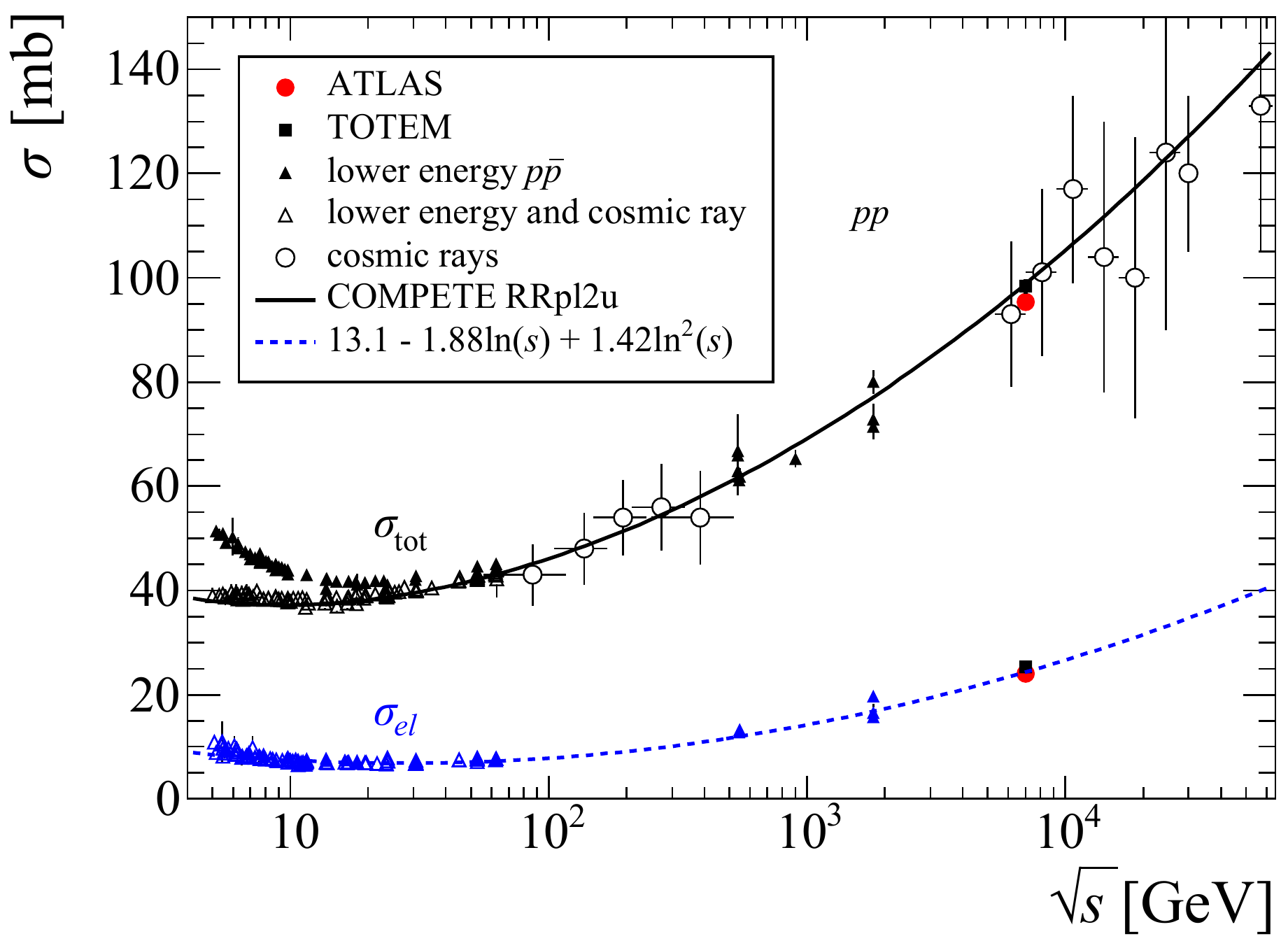}
    \caption{Comparison of total and elastic cross-section
      measurements in proton-proton and antiproton-proton
      collisions and in cosmic-ray interactions as a function of the centre-of-mass
      energy.
      \textit{(Adapted from Ref.~\cite{Aad:2014dca}.)}
    \label{fig:qcd:atlas_totalpp}}
  \end{center}
\end{figure}

Soft hadron collisions can not be calculated in perturbative QCD.
Instead, they are usually described by phenomenological models where the
force between hadrons is assumed to be mediated by mesons, for instance by pions~\cite{Yukawa:1935xg}. The basic idea
is that the exchanged meson interacts
with the colour charges confined in the scattered hadrons.
Since many mesons exists, the strong force between hadrons can be
quite complicated in such models. The typical range of the
strong force is anti-proportional to the mass of the exchanged meson.
Consequently, the largest range is provided by the lightest meson which has
the appropriate quantum numbers.
Such models are usually called ``Regge theory''.
For reviews see Refs.~\cite{Kaidalov:2001db,Collins:1977jy}.
Regge theory
aimed at explaining the spectrum of the various hadrons,
the forces between them, and the high-energy behaviour
of the hadron-hadron scattering cross section.
It is based on simple assumptions like that the scattering matrix is analytic and unitary.
Mesons are understood as composite quark-antiquark states bound by an effective strong
potential that depends on the distance between the quark-antiquark pair $r$
and the orbital angular momentum $J$.
Elastic hadron-hadron scattering
is viewed as the exchange of a meson in the $s$-channel or $t$-channel.
In 1961, Chew and Frautschi~\cite{Chew:1961yz} made the observation that
for $\pi^- \proton \to \pi^0 \neutron$ scattering there is a linear
relation between the mass of the exchanged hadron ($m$) and its angular momentum $J$.
Each integer value of the slope $\alpha$---called ``Regge trajectory''---corresponds to one
particle that can be experimentally identified (``Regge-pole'').
In the time-like region where $t>0$ ($s$-channel meson exchange)
the momentum transfer $t$ corresponds to the hadron mass $m$.
If one continues the linear relation between $m$ and $J$ using the slope $\alpha$
for $t<0$, i.e.\ the space-like regime,
a good description of the experimental measurements for $\alpha$ is obtained.
In this case, however, $t$ can take any value.
This trajectory is called ``Pomeron trajectory''.
The exchanged object carrying the quantum numbers of the vacuum ($\CCC=\PARITY=1$)
is called ``Pomeron''.

The meson exchange leads to a simple power behaviour $s^{\alpha -1}$
of total hadron-hadron scattering cross sections.
In 1992 Donnachie and Landshoff~\cite{Donnachie:1992ny} made the observation that
hadron-hadron total cross sections can be described by
two powers, one $\alpha \approx 1.1$ corresponding to Pomeron exchange and
one $\alpha = 0.5$ corresponding to the exchange of mesons
like $\rho$, $\omega$, $f_2$ and $a_2$.
Therefore at low-energies, where the Reggeon exchange dominates, a behaviour
$\sigma \sim s^{-0.5}$ is expected, while at high energies $\sigma \sim s^{0.1}$
should prevail.

Such a behaviour was indeed experimentally observed. Figure~\ref{fig:qcd:atlas_totalpp}
shows the total inclusive proton-proton cross section
as a function of the centre-of-mass energy $\sqrt{s}$ for various experiments.
In the range \unit{2-20}{\GeV} the proton-proton cross section decreases
as measured by fixed-target experiments, while at higher energies
it increases towards larger $\sqrt{s}$.
The difference between proton-proton and proton-antiproton
cross section decreases towards high energy; at about
$\sqrt{s} \approx \unit{30}{\GeV}$, the two cross sections are
very similar.
The rise of the proton-antiproton cross section for $\sqrt{s} \gtap \unit{30}{\GeV}$
was first observed at the \sps, confirmed by \Tevatron and LHC, where
the ATLAS~\cite{Aad:2011eu}, CMS~\cite{Chatrchyan:2012nj} and
TOTEM~\cite{Antchev:2013gaa} experiments measured at $\sqrt{s} = \unit{7}{\TeV}$
(TOTEM measured also at $\sqrt{s} = \unit{8}{\TeV}$~\cite{Antchev:2013paa}).
Data from cosmic-ray experiments~\cite{Aielli:2009ca,Honda:1992kv,Baltrusaitis:1984ka} are able to provide useful measurements
at energies up to \unit{60}{\TeV} that are  not covered by colliders.
Overlayed in \fig{\ref{fig:qcd:atlas_totalpp}}
as curve is a parameterisation~\cite{Cudell:2002xe} based on
the $\ln^2{s}$ behaviour that describes the data well at high energy.
The power-law behaviour can also describe the present data~\cite{Donnachie:2013xia}.
Also shown in \fig{\ref{fig:qcd:atlas_totalpp}} are recent measurements of the
elastic proton-proton cross section~\cite{Aad:2014dca,Antchev:2013gaa}.

%% file: summary.tex
\section{Summary and Outlook}\label{sec:qcd:summary}

The excellent performance of the LHC machine and of the four major
detectors around the ring made it possible to carry out precise
measurements at very high particle, i.e.\ jet transverse momenta, high
invariant masses, and for very complex event topologies with large
particle multiplicities. The high luminosities and the centre-of-mass
energies of \unit{7 and 8}{\TeV} reached for proton-proton collisions
are unprecedented. These conditions allowed to assess the validity of
the QCD theory in the \TeV\ regime and to stress-test our current
understanding of particle-production processes at such high
energies. Many measurements in particular of vector-boson and
jet-production processes reached an experimental accuracy of a few
percent uncertainty only---true QCD precision tests.

To meet the experimental challenges, theoretical predictions are needed
that match the precision and complexity requirements. Accordingly,
enormous efforts have been undertaken to provide better and better
theoretical predictions accompanied and driven by exciting new
theoretical developments and insights. This includes largely automated
NLO and many new NNLO calculations, improved analytical and
parton-shower resummation techniques, the merging (i.e.\ matching) of
parton showers and exact matrix elements, and attempts to gain
a better understanding of the low-energy (i.e.\ non-perturbative)
regime of the strong interaction.
A major role is played by
the Monte Carlo event generators that have been developed and
optimised for the analysis and interpretation of LHC measurements.

Calculations at NNLO in the strong coupling provide predictions for
inclusive production cross sections with an estimated uncertainty,
e.g.\ due to missing higher orders, of a few percent only. Such
precision predictions were successfully tested against the LHC data.
Often the dominant theoretical uncertainty originates from
the knowledge of the proton PDFs. Also in this field LHC measurements
have already contributed to a refined understanding of the proton
structure.

Around the year 2002, two theoretical break-throughs have
revolutionised the field of Monte Carlo generators: first, the exact
matching of $2\to 2$ production processes evaluated at NLO accuracy
with parton showers, and second, the consistent merging of tree-level
matrix elements with varying final-state parton multiplicity with
parton showers.  These two innovative concepts enable fully exclusive
predictions at particle-level with an inclusive production rate
accurate to NLO or the first few hardest emissions being modelled
through exact real-emission matrix elements. Consequently, Monte-Carlo
simulations based on matching or merging of matrix elements and parton
showers have become the standard tools in any LHC analysis.
Driven by the enormous progress in the evaluation of NLO cross sections
(which are largely automated by now), there appeared many multi-leg NLO
calculations that can directly be matched with parton showers.

The currently emerging new standard are simulations based on exact NLO
calculations with increasing parton multiplicity all matched with
parton showers and combined into an inclusive description of the
considered production process. However, this is not the end of the
development: first prototypes of exact NNLO calculations matched with
parton showers have recently been presented. These exciting
theoretical developments pave the way to a further scrutiny of our
understanding of QCD in the upcoming high-energy and high-luminosity
LHC runs. In consequence, they will significantly boost the sensitivity
and prospects to find potential new physics at the LHC.